\documentclass[longauth]{aa} % for the long lists of affiliations 
\graphicspath{{img}}
\usepackage{graphicx}
\usepackage{caption}
\usepackage{subcaption}
\usepackage{float}
\usepackage{comment}
\usepackage{natbib}
\usepackage{txfonts}
\usepackage{xcolor}
\usepackage[symbol]{footmisc}
\renewcommand{\thefootnote}{\fnsymbol{footnote}}
\usepackage{hyperref}
\hypersetup{%
    colorlinks=true, 
    citecolor=blue,
    filecolor=blue,
    linkcolor=blue,
    urlcolor=black
    }

\begin{document} 

   \title{A search for the afterglows, kilonovae, and host galaxies of two short GRBs: GRB\,211106A and GRB\,211227A}

   %\subtitle{**Subtitle**}

   \author{M. Ferro\inst{1,2}
          \and
          R. Brivio\inst{1,2}
          \and
          P. D'Avanzo\inst{2}
          \and
          A. Rossi\inst{3}
          \and
          L. Izzo\inst{4,5}
          \and
          S. Campana\inst{2}
          \and
          L. Christensen\inst{6,7}
          \and
          M. Dinatolo\inst{8}
          \and
          S. Hussein\inst{9,10}
          \and
          A. J. Levan\inst{11,12}
          \and
          A. Melandri\inst{13}
          \and
          M. G. Bernardini\inst{2}
          \and
          S. Covino\inst{2}
          \and
          V. D'Elia\inst{13,14}
          \and
          M. Della Valle\inst{4}
          \and
          M. De Pasquale\inst{15}
          \and
          B. P. Gompertz\inst{16}
          \and
          D. Hartmann\inst{17}
          \and
          K. E. Heintz\inst{6,7}
          \and
          P. Jakobsson\inst{18}
          \and
          C. Kouveliotou\inst{19}
          \and
          D. B. Malesani\inst{11,6,7}
          \and
          A. Martin-Carrillo\inst{20}
          \and
          L. Nava\inst{2}
          \and
          A. Nicuesa Guelbenzu\inst{21}
          \and
          G. Pugliese\inst{22}
          \and
          C. Salvaggio\inst{2}
          \and
          R. Salvaterra\inst{23}
          \and
          S. Savaglio\inst{24,3,25}
          \and
          T. Sbarrato\inst{2}
          \and
          N. R. Tanvir\inst{26}
          \and
          R. A. M. J. Wijers\inst{27}
          \and
          T. Zafar\inst{28}
          }
    
    \institute{
            Università degli Studi dell’Insubria, Dipartimento di Scienza e Alta Tecnologia, Via Valleggio 11, 22100 Como, Italy 
        \and
             INAF–Osservatorio Astronomico di Brera, Via E. Bianchi 46, 23807 Merate (LC), Italy
        \and
            INAF–Osservatorio di Astrofisica e Scienza dello Spazio, Via Piero Gobetti 93/3, 40129 Bologna, Italy
        \and
            INAF–Osservatorio Astronomico di Capodimonte, Salita Moiariello 16, I-80131 Napoli, Italy
        \and
            DARK, Niels Bohr Institute, University of Copenhagen, Jagtvej 128, 2200 Copenhagen N, Denmark
        \and
            Cosmic Dawn Center (DAWN), Copenhagen, Denmark
        \and
            Niels Bohr Institute, University of Copenhagen, Jagtvej 128, 2200 Copenhagen N, Denmark
        \and
            Università degli Studi di Milano, Dipartimento di Fisica, via Celoria 16, 20133 Milano, Italy
        \and
            Université Paris-Saclay, CNRS/IN2P3, IJCLab, F-91405 Orsay, France
        \and
            An-Najah National University, Nablus, Palestine
        \and
            Department of Astrophysics/IMAPP, Radboud University, 6525 AJ Nijmegen, The Netherlands
        \and
            Department of Physics, University of Warwick, Coventry, CV4 7AL, UK
        \and
            INAF-Osservatorio Astronomico di Roma, Via di Frascati 33, I-00040, Monte Porzio Catone (RM), Italy
        \and
            ASI, Italian Space Agency, Space Science Data Centre,Via del Politecnico snc, I-00133 Rome, Italy
        \and
            University of Messina, Mathematics, Informatics, Physics and Earth Science Department, Via F.D. D’Alcontres 31, Polo Papardo, 98166, Messina, Italy
        \and
            School of Physics and Astronomy and Institute for Gravitational Wave Astronomy, University of Birmingham, B15 2TT, UK
        \and
            Department of Physics and Astronomy, Clemson University, Clemson, SC 29634, USA
        \and
            Centre for Astrophysics and Cosmology, Science Institute, University of Iceland, Dunhagi 5, 107, Reykjavik, Iceland
        \and
            The George Washington University, Physics Department, 725 21st street NW, Washington DC, 20052, USA
        \and
            School of Physics and Centre for Space Research, University College Dublin, Belfield, D04 V1W8 Dublin, Ireland
        \and
            Thüringer Landessternwarte Tautenburg, Sternwarte 5, 07778 Tautenburg, Germany
        \and
            Astronomical Institute Anton Pannekoek, University of Amsterdam, 1090 GE Amsterdam, The Netherlands
        \and
            INAF–IASF Milano, Via Corti 12, 20133 Milano, Italy
        \and
            Physics Department, University of Calabria, 87036 Arcavacata di
            Rende, CS, Italy
        \and
            INFN – Laboratori Nazionali di Frascati, Frascati, Italy
        \and
            School of Physics and Astronomy, University of Leicester, University Road, Leicester, LE1 7RH, UK
        \and 
            Anton Pannekoek Institute for Astronomy, University of Amsterdam, Science Park 904, NL-1098 XH Amsterdam, The Netherlands
        \and
            School of Mathematical and Physical Sciences, Macquarie University, NSW 2109, Australia
            }
    \date{Received xxx; accepted yyy}

% \abstract{}{}{}{}{} 
% 5 {} token are mandatory

\abstract
  % context heading (optional), leave it empty if necessary  
  {GRB\,211106A and GRB\,211227A are two recent gamma-ray bursts (GRBs) whose initial X-ray position enabled us to possibly associate them with bright, low-redshift galaxies ($z$<0.7). The prompt emission properties suggest that GRB\,211106A is a genuine short-duration GRB and GRB\,211227A is a short GRB with extended emission. Therefore, they are likely to be produced by a compact binary merger. However, a classification based solely on the prompt emission properties can be misleading.}
   {The possibility of having two short GRBs occurring in the local Universe makes them ideal targets for the search of associated kilonova (KN) emission and for detailed studies of the host galaxy properties.}   
   {We carried out deep optical and near-infrared (NIR) follow-up with the ESO-VLT FORS2, HAWK-I, and MUSE instruments for GRB\,211106A and with ESO-VLT FORS2 and X-Shooter for GRB\,211227A, starting from hours after the X-ray afterglow discovery up to days later. We performed photometric analysis to look for afterglow and KN emissions associated with the bursts, together with imaging and spectroscopic observations of the host galaxy candidates. We compared the results obtained from the optical/NIR observations with the available \textit{Swift} X-Ray Telescope (XRT) and others high-energy data of both events.}
   {For both GRBs we placed deep limits to the optical/NIR afterglow and KN emission. We identified their associated host galaxies, GRB\,211106A at a photometric redshift $z$=0.64, GRB\,211227A at a spectroscopic $z$=0.228. From MUSE and X-Shooter spectra we derived the host galaxy properties, which turned out to be consistent with short GRBs typical hosts. We also compared the properties of GRB\,211106A and GRB\,211227A with those of the short GRBs belonging to the S-BAT4 sample, here extended up to December 2021, in order to further investigate the nature of these two bursts.} 
   {Our study of the prompt and afterglow phase of the two GRBs, together with the analysis of their associated host galaxies, allows us to confirm the classification of GRB\,211106A as a short GRB, and GRB\,211227A as a short GRB with extended emission. The absence of an optical/NIR counterpart down to deep magnitude limits is likely due to high local extinction for GRB\,211106A and a peculiarly faint kilonova for GRB\,211227A.}


    %\titlerunning{GRB 211106A and GRB 211227A}

   \maketitle

%
%-------------------------------------------------------------------
\section{Introduction}\label{intro}

Two classes of gamma-ray bursts (GRBs), short and long, were identified  based on the distribution of their prompt gamma-ray emission duration. Short GRBs (SGRBs) are those with duration shorter than about two seconds and with harder high-energy spectra with respect to long bursts \citep{Kouveliotouetal93}. While it was firmly established that long GRBs (LGRBs) originate in core-collapse supernova (SN) explosions \citep{Hjorth&Bloom12}, according to the most popular model, short GRBs are produced by the merger of compact objects, namely neutron stars (NSs) and black holes (BHs). The knowledge of short GRBs experienced an impressive boost in the past two decades. The discovery of short GRB afterglows in 2005 with the \textit{Neil Gehrels Swift Observatory} (\textit{Swift}) and the \textit{HETE-II} satellites was a watershed moment in the study of these sources, providing the key to unravel their distance, energy scale, environments, and host galaxies \citep{Gehrelsetal05, Foxetal05, Villasenoretal05, Hjorthetal05, Barthelmyetal05, Bergeretal05, Covinoetal06}.\\
To date, more than 150 short GRBs were found with \textit{Swift} ($\sim10$/yr). A sizeable fraction of them have X-ray and optical afterglow detections, with a few having been detected also in radio \citep{Fongetal15, Fongetal22}. Short GRB afterglows are fainter on average than those of long GRBs and the great majority of short GRB redshifts are obtained through optical spectroscopy of their associated host galaxies. To this end, the precise localisation with \textit{Swift} is a crucial asset to achieve a firm short GRB-host galaxy association \citep{Davanzo2014}. Properties like the absence of associated SNe, the afterglow faintness, the occurrence also in early-type galaxies, and the offset and redshift distribution definitely point towards a compact merger origin, at variance with what was observed for long GRBs \citep{Berger14, Davanzo15}. All these findings are in agreement with the compact object binary progenitor model \citep{Eichleretal89, Narayanetal92, Nakar07}. These progenitors are also expected to be sources of high-frequency gravitational waves (GWs).\\
Another key signature of a NS-NS or NS-BH binary merger is the production of a so-called kilonova (KN), whose electromagnetic emission is powered by the decay of heavy radioactive species produced by rapid neutron capture (r-process) and ejected during the merger process \citep{Li&Paczynski98, Rosswog05, Metzgeretal10}. Excess infrared light seen in the light curves of several SGRBs between 3 and 10 days post-burst were argued as being likely due to such kilonova emission (e.g., \citealt{Tan13}). The compact object binary progenitor model for short GRBs was spectacularly confirmed on August 17, 2017, when a GW event associated with a NS-NS merger, GW 170817, was detected for the first time by the Advanced Laser Interferometer Gravitational wave Observatory (aLIGO, \citealt{Aasietal2015}) and localized also thanks to Advanced Virgo \citep{Acerneseetal2015}. The subsequent detections of the weak short GRB\,170817A and the bright kilonova AT2017gfo \citep{Abbottetal17} were associated with the GW event.  The emergence, weeks after the event, of an X-ray and radio counterpart, suggested the possibility of off-axis GRB afterglow emission \citep{Hallinanetal17,Trojaetal17}. Besides providing the long-sought "smoking gun" of short GRB progenitors, the case of GRB\,170817A demonstrated that the GRB emission geometry differs (in this case, at least) from a simple uniform jet (\citealt{Mooleyetal18,Ghirlandaetal19}, for a review, see \citealt{Margutti&Chornock21}).\\
The observations of GRB\,060614 (\citealt{2006EE}) and, more recently, of GRB\,211211A (\citealt{Rastinejad_211211A}) suggested the existence of a subclass of short GRBs that mimic the behavior of long-duration GRBs, regarding the duration of their gamma-ray emission that largely exceeds the classical $\sim$2s separation threshold for GRB families. This fraction of short GRBs exhibits the presence of a prolonged $\gamma$-ray emission that is softer than the prompt spike \citep{Lazzatietal01}, lasts tens of seconds, and may rise with a delayed onset \citep{Norris06, Norrisetal10}. Such an emission component can dominate (in terms of fluence) the prompt spike emission \citep{Perleyetal09}. 
Moreover, the short and long classification is further confused by the recent observation of peculiar GRBs of short duration associated with a supernova (e.g., GRB\,200826A, \citealt{Rossietal22}), which was until then only associated with long GRBs. Besides the spectacular case of GRB\,170817A and AT2017gfo, in recent years, other convincing evidence of KNe associated to short GRBs were reported \citep[for a review, see][]{ROSSI20-KN}. A peculiar case was the photometric detection of KN signatures in the optical and near-infrared (NIR) light curve of the long duration GRB\,211211A, and more recently, the detection of heavy neutron capture elements associated with the long GRB\,230307A \citep{230307AArx}. These events emphasised the need of caution in GRB classification, providing evidence that also long duration GRBs may orginate from compact binary merger progenitors. On the other hand, the characteristics of the prompt emission of GRB 2111211A may indicate that this event can be actually classified as a Short GRB with extended emission \citep{Rastinejad_211211A, Troja_211211A, Gompertz_211211A}.
\\ 
In this work we present the results of an extensive multi-band study of two GRBs, possibly originating from the merger of compact objects, namely GRB\,211106A (a short-duration GRB) and GRB\,211227A (a short GRB with extended emission). The properties of the GRBs and host galaxies are compared to those of the short GRBs belonging to the S-BAT4 sample \citep{Davanzo2014}, for which we provide here an extension including events up to 2021. In \S \ref{GRB211106A} and \S \ref{GRB211227A} we present observations of the two GRBs. In \S \ref{Obs&analysis211106A} and \S \ref{sec:211227An} data analysis procedure and the first results are described, and in \S \ref{discussion} a full discussion is presented. Our main conclusions are summarized in \S \ref{conclusions}.\\
Throughout the paper, we adopt $\Lambda$CDM model with cosmological parameters $\Omega_M$ = 0.308, $\Omega_\Lambda$ = 0.692, and $H_0$ = 67.8 km s$\textsuperscript{-1}$ Mpc$\textsuperscript{-1}$ \citep{PLANCK16}. All magnitudes presented in this work are given in AB system, unless otherwise specified. Errors are at the 68\% confidence level (c.l.), unless stated otherwise. All filter curves considered in this work are taken from the the SVO Filter Profile Service (\citealt{SVO1}; \citealt{SVO2}).

%--------------------------------------------------------------------
\section{GRB\,211106A}\label{GRB211106A}
\subsection{Prompt phase}\label{sec:211106_prompt}

\begin{figure}
    \centering
    \includegraphics[width=9cm]{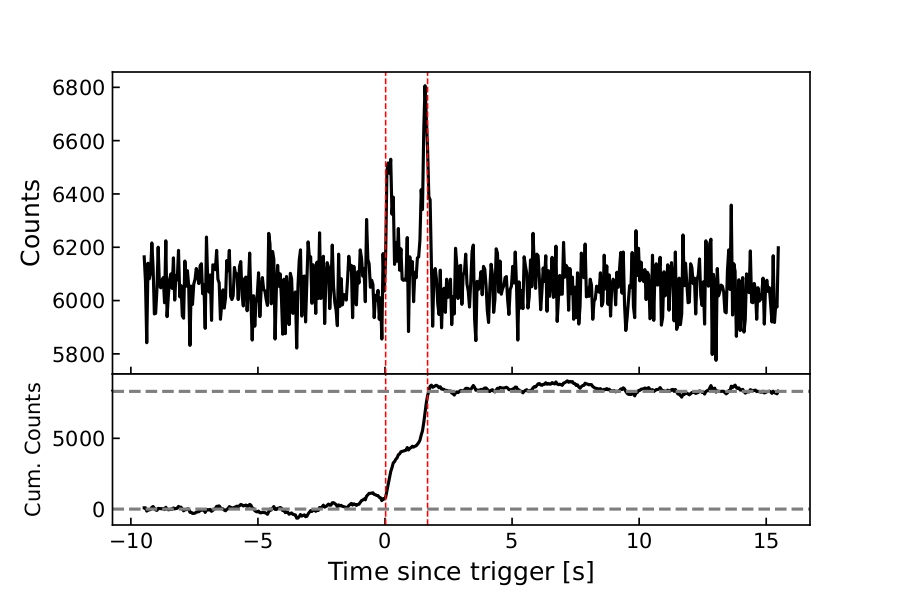}\caption{ \textit{INTEGRAL}/SPI-ACS light curve of the prompt emission of GRB\,211106A with 50 ms resolution (from \url{https://www.isdc.unige.ch/integral/science/grb\#ACS}). The red dashed lines represent the T$_{90}$, the grey dashed line in the bottom panel corresponds to the T$_{100}$.}
    \label{GRB211106A_IGR}
\end{figure}

GRB\,211106A was first detected with the \textit{INTEGRAL} gamma-ray satellite \citep{INTEGRAL} on 2021 Nov 6 at T$_0$: 04:37:31.2 UTC. The INTEGRAL notice (\#9504) was distributed in near real-time and triggered the Swift Mission Operations Center Gamma-ray Urgent Archiver for Novel Opportunities \citep[GUANO,][]{GUANO}. Consequently, GUANO sent a command to the \textit{Swift} Burst Alert Telescope \citep[BAT,][]{BAT} to save 90 seconds of BAT event-mode data from [-45,+45] s around the time of the burst. Then, the BAT likelihood search, NITRATES \citep{NITRATES_paper}, confidently detected the burst with an estimated T$_{90}$ equal to 1.75 $\pm$ 0.05 s in the 15-350 keV energy band \citep{GUANO_GCN}, making it a candidate for the short GRB class. From the \textit{Swift}/BAT data, it has also been possible to derive the GRB position with coordinates (J2000) R.A. = 22:54:34.32, Dec. = $-$53:14:0.9, with an uncertainty of 7 arcmin (radius, 90\% c.l.; \citealt{GUANO_GCN}).\\
GRB\,211106A also triggered the Konus-\textit{Wind} satellite on 2021 Nov 6 at T$_0^{'}$: 04:37:32.485 UTC \citep{KONUS_GCN}. The burst light curve showed a single pulse starting at T$_0^{'}-$0.3 s with a total duration of $\sim$ 0.5 s. The Konus-\textit{Wind} trigger time is consistent with the time of the second peak in the \textit{INTEGRAL} and BAT light curve \citep{Laskar22}. The GRB prompt emission has also been detected at a sub-threshold level by \textit{Fermi}/GBM, through a targeted search \citep{FERMI_GCN}. \\
The \textit{INTEGRAL} light curve (Fig. \ref{GRB211106A_IGR}) shows two pulses, with T$_{90} = 1.65 \pm 0.95$ s. The same shape and duration is also observed in the \textit{Swift}/BAT and \textit{Fermi}/GBM light curves \citep{Laskar22}. 

\subsection{Afterglow phase}\label{sec:211106_aftgl}
Following the \textit{Swift}/BAT-GUANO detection of GRB\,211106A, the \textit{Swift} X-Ray Telescope \citep[XRT,][]{XRT} performed follow-up observations, starting at T$_0$+39.2 ks \citep{AAGCN_XRT01}. An uncatalogued, fading source was detected at the coordinates (J2000) R.A. = 22:54:20.45, Dec. = $-$53:13:49.0, with an uncertainty of $3.4''$ (radius, 90\% c.l.) and was proposed as the GRB X-ray afterglow \citep{AG_GCN}. The source was followed-up in the optical from the ground using the ESO-VLT UT1 equipped with the FORS2 camera, and two nearby galaxies were found \citep{FORS_GCN}. The closest (in projection) of them is 4.8$''$ away from the center of the \textit{Swift}/XRT error circle at a redshift $z$ = 0.097, as inferred by the detection of the [OIII] doublet and H$_{\beta}$ in the emission \citep{MUSE_GCN}. \\
Follow-up observations of the X-ray afterglow were also performed with the \textit{Chandra X-ray Observatory} and with the \textit{XMM-Newton}, between about 10.4 and 59.8 days after the burst \citep{CHANDRA_GCN, XMM_GCN}. These observations enabled a more precise localisation of the afterglow. In particular, the position obtained with \textit{Chandra} is (J2000) R.A. = 22:54:20.518, Dec. = $-$53:13:50.59, with an uncertainty of $0.18''$ and an angular offset of 3.76$''$ with respect to the galaxy at $z$ = 0.097 \citep{CHANDRA_GCN}.\\
Further optical/NIR observations were taken with the \textit{Hubble} Space Telescope (\textit{HST}) in three different epochs, at 19.2, 25.3 and 48.2 days after the trigger, respectively. A single, possibly extended, source within the \textit{Chandra} error circle was detected at (J2000): R.A. = 22:54:20.54, Dec. = $-$53:13:50.6, with an uncertainty of $\sim$ 0.05$''$. Photometry indicates magnitudes of $m_{F814W}=$ 25.791 $\pm$ 0.069 and $m_{F110W}=$ 25.709 $\pm$ 0.016 \citep{HST_GCN,Laskar22}, and no evidence for fading during the 29-day baseline. \\
Finally, GRB\,211106A was followed-up in the millimeter (97.5 GHz) and radio (5.5, 9.0 and 18.0 GHz) bands with the Atacama Large Millimeter/Submillimeter Array (ALMA) and with the Australia Telescope Compact Array (ATCA), respectively. The observations were carried out between $\sim$ 13 and 117 days after the burst and led to a detection of mm and radio afterglows  \citep{Laskar22} at the following (ALMA) position (J2000): R.A. = 22:54:20.53056 ($\pm$0.011$''$), Dec. = $-$53:13:50.525 ($\pm$0.010$''$). This position is consistent with the \textit{Chandra} one and lies within the optical, non-variable, extended source detected with \textit{HST}, making it a likely host galaxy candidate for GRB\,211106A \citep{Laskar22}.

\section{GRB\,211227A}\label{GRB211227A}
\subsection{Prompt phase}

\begin{figure}
    \centering
    \includegraphics[width=9cm]{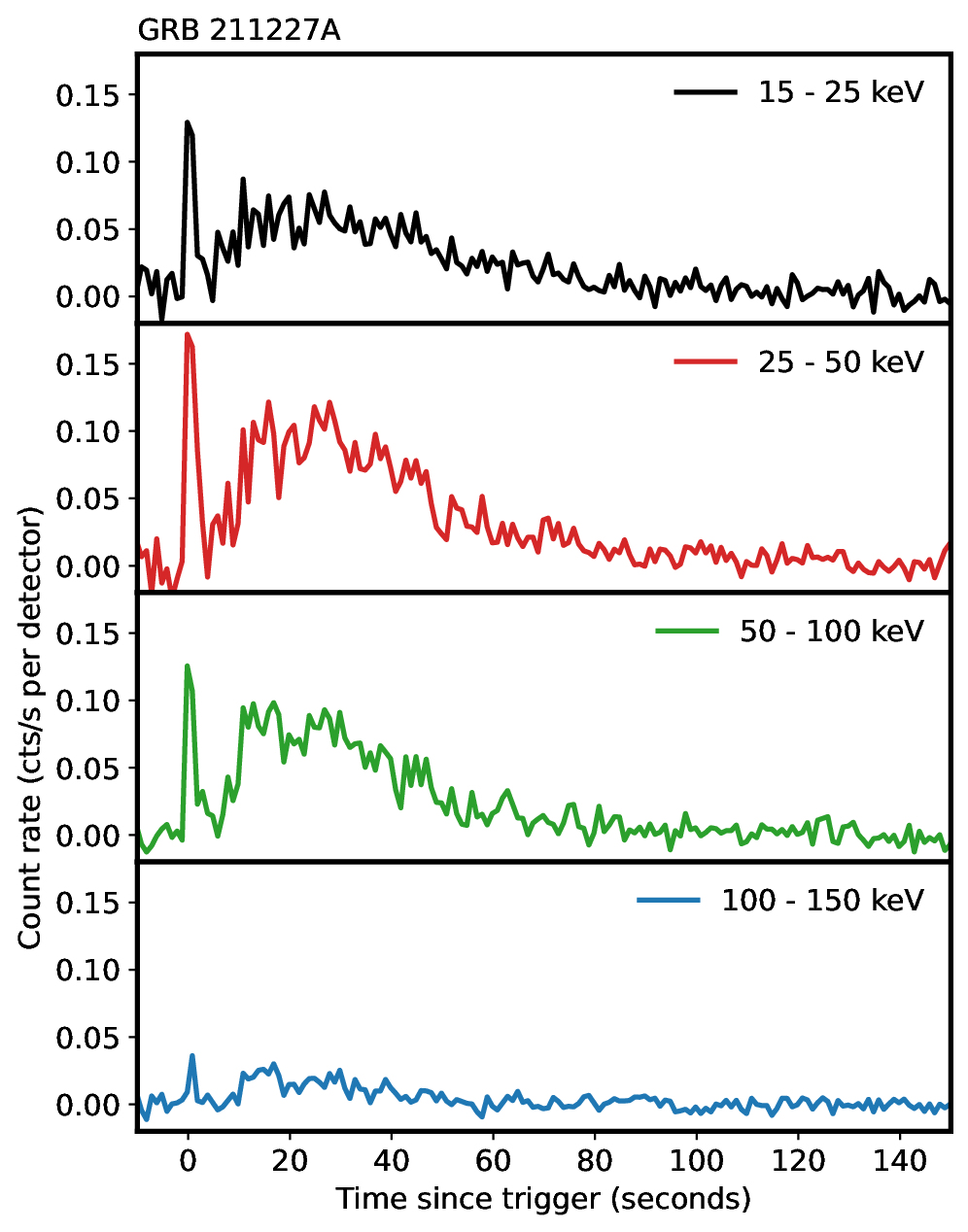}\caption{\textit{Swift}/BAT light curve of the prompt emission of GRB\,211227A with 1 s resolution (data collected from \url{https://gcn.gsfc.nasa.gov/notices_s/1091101/BA/}).}
    \label{GRB211227_BAT}
\end{figure}

GRB\,211227A was detected with \textit{Swift}/BAT on 2021 Dec 27 at T$_0$ = 23:32:06 UT \citep{GRB211227A_GCN_SWIFT}, at the position (J2000) R.A. = 08:48:34.5, Dec. = $-$02:44:10.9, with an uncertainty of 1.4 arcmin (radius, 90\% c.l.). T$_{90}$ in the 15-150 keV band is 83.8 $\pm$ 8.5 s. The GRB was detected with MAXI/GSC \citep{GRB211227A_GCN_MAXI}, \textit{CALET} \citep{GRB211227A_GCN_CALET} and Konus-\textit{Wind} \citep{GRB211227A_GCN_KONUS}. A joint analysis of the prompt emission data collected with Konus-\textit{Wind} and \textit{Swift}/BAT revealed a light curve with two main episodes, a narrow pulse peaking at $\sim$ T$_0$ and a second broader pulse peaking at $\sim$ T$_0$+11 s \citep{GRB211227A_GCN_KONUS}. The \textit{Swift}/BAT light curve morphology (Fig. \ref{GRB211227_BAT}) suggests that GRB\,211227A is a short GRB with extended emission \citep{GRB211227A_GCN_XSHOOTER}. Such a conclusion was reached also by \cite{Luetal22} and \cite{zhu22} from the joint analysis of the Konus-\textit{Wind} and \textit{Swift}/BAT prompt emission properties.

\subsection{Afterglow phase}\label{GRB211227Aaft}
Following the \textit{Swift}/BAT detection, the \textit{Swift}/XRT began observing the field from T$_0$+63 s and found a bright, uncatalogued fading source at the coordinates (J2000) R.A. = 08:48:35.73, Dec. = $-$02:44:07.1, with an uncertainty of 2.4$''$ (radius, 90\% c.l.).\\
Ground-based observations were performed using the ESO-VLT UT3 equipped with the X-Shooter spectrograph \citep{XShooter}. No optical counterpart was detected within the XRT error circle in the images obtained with the acquisition camera, but a fairly bright galaxy ($r$ $\sim$ 19.4 mag) was found $\sim$ 3.67$''$ away from its center, with its light extending within the XRT error circle \citep{GRB211227A_GCN_XSHOOTER}. X-Shooter spectroscopy of this galaxy revealed the presence of several emission lines ([O II], H$_{\beta}$, [O III] 5007, H$_{\alpha}$, [N II], [S II]), all at a common redshift $z$ = 0.228 \citep{GRB211227A_GCN_XSHOOTER}. Further deep optical observations were performed with the Liverpool telescope \citep{perley21}, CAHA 2.2m telescope \citep{kann2021}, and Gemini South telescope \citep{Oconnor2021}. No optical counterpart was detected within the XRT error circle in the three cases.

\section{Observations and data analysis: GRB\,211106A}\label{Obs&analysis211106A}
\renewcommand{\thefootnote}{\arabic{footnote}}

%----------------------------------------------------- LOG TABLE
\begin{table*}[]
    \centering
    \caption{Log of imaging and spectroscopic VLT ground-based observations of the GRB\,211106A field.}
    %\captionsetup{justification=centering,margin=2cm}
    \begin{tabular}{l|l|l|l|l|l}
    Date: $t_{\text{start}}$ (UT) & $t-t_0$ [d] & Instrument & $t_{\text{exp}}$ [$\#$ x    $s$] & Seeing [$''$] & Filter\\
    \noalign{\smallskip}
    \hline
    \hline
    \noalign{\smallskip}
    2021-11-09 02:35:46.701 & 2.92 & FORS2 & 5 x 240 & 0.70 & $R$ \\
    2021-11-11 00:25:12.736 & 4.83 & HAWK-I & 20 x (6 x 10) & 0.86 & $H$\\
    2021-11-12 00:36:16.074 & 5.85 & FORS2 & 16 x 180 & 0.83 & $R$\\
    2021-12-04 00:45:50.842 & 27.85 & FORS2 & 8 x 300 & 0.69 & $R$\\
    \noalign{\smallskip}
    \hline
    \noalign{\smallskip}
    2021-11-10 00:15:43.829 & 3.88 & MUSE & 4 x 700 & 0.8-0.9 & -\\
    \end{tabular}
    \label{log_table_211106}
\end{table*}

\begin{table}
    \centering
    \caption{Results of GRB\,211106A photometric analysis.}
    \begin{tabular}{ccccc}
    \hline
    \noalign{\smallskip}
    $t-t_0$ [d]& Instrument / Filter &Magnitude& Mag. limit \\
    \noalign{\smallskip}
    \hline
    \hline
    \noalign{\smallskip}
    2.92& FORS2 / $R$& 26.37 $\pm$ 0.23 & 26.7 (27.5)$^b$\\
    5.85& FORS2 / $R$& 26.69 $\pm$ 0.39$^a$ & 26.8 (27.5)$^b$ \\
    27.85& FORS2 / $R$&26.51 $\pm$ 0.23 & 26.6 \\
    \noalign{\smallskip}
    \hline
    \noalign{\smallskip}
    4.83& HAWK-I / $H$& $-$ & 23.6 \\
    \noalign{\smallskip}
    \hline
    \noalign{\smallskip}
    3.88& MUSE / $V$& 27.42 $\pm$ 0.08 & $-$ \\
    3.88& MUSE / $R$& 26.45 $\pm$ 0.10 & $-$ \\
    3.88& MUSE / $I$& 25.67 $\pm$ 0.26 & $-$ \\
    \noalign{\smallskip}
    \hline
    \end{tabular}
    \tablefoot{Results of PSF photometry analysis of GRB\,211106A field. Magnitudes of the source detected within the \textit{Chandra} error circle from FORS2 image, along with 3$\sigma$ detectability limits for each image, are reported.\\
    $^a$ In the second epoch the source is detected at more than 3$\sigma$ c.l. due to lower S/N.\\
    $^b$ Limits found from photometry on the background, source-free, region.}
    \label{tab-magnitudes}
\end{table}

\begin{figure}
    \centering
    \includegraphics[width=9cm]{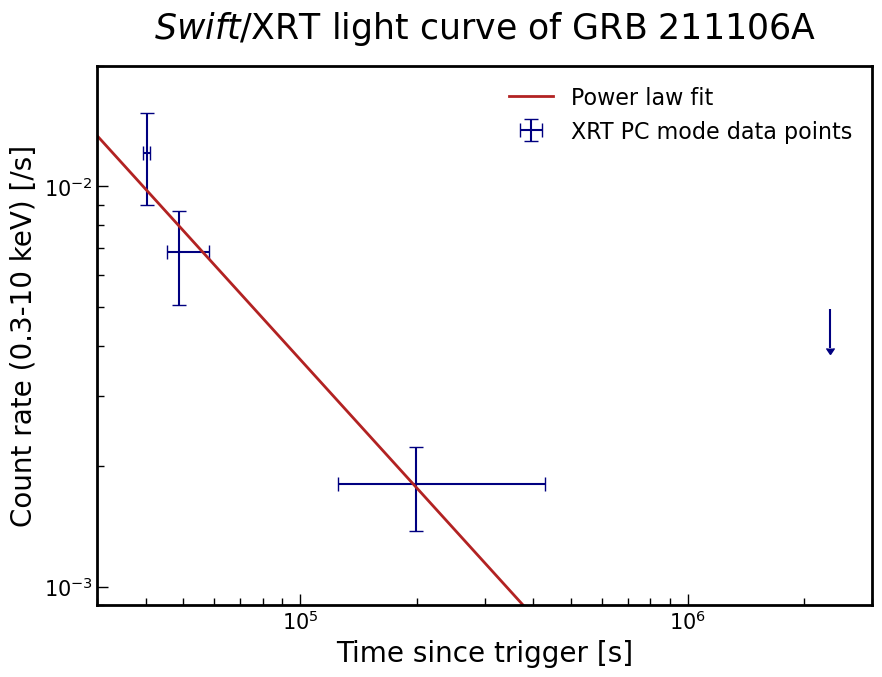}
    \caption{\textit{Swift}/XRT light curve of the GRB\,211106A afterglow. The best power-law fit (with index $\alpha$ = 1.07 $\pm$ 0.21) is represented by a red solid line. The downward pointing arrow on the right indicates an upper limit for the last measurement.}
    \label{211106_XRTlc}
\end{figure}

\begin{figure*}
    \centering
    \begin{subfigure}[b]{.49\textwidth}
        \includegraphics[width=1.\textwidth]{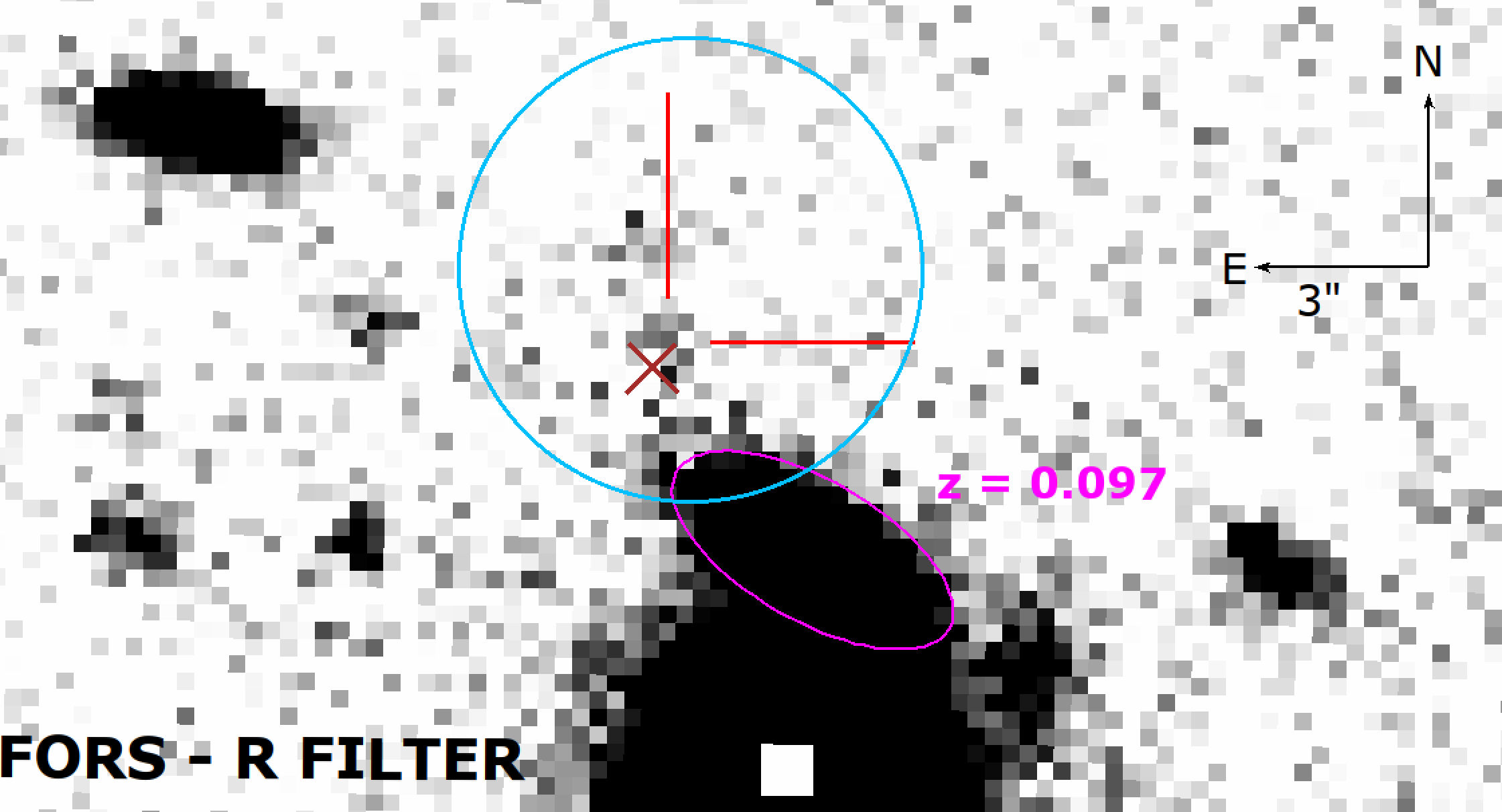}
    \end{subfigure}
    \hfill
    \begin{subfigure}[b]{.49\textwidth}
        \includegraphics[width=1.\textwidth]{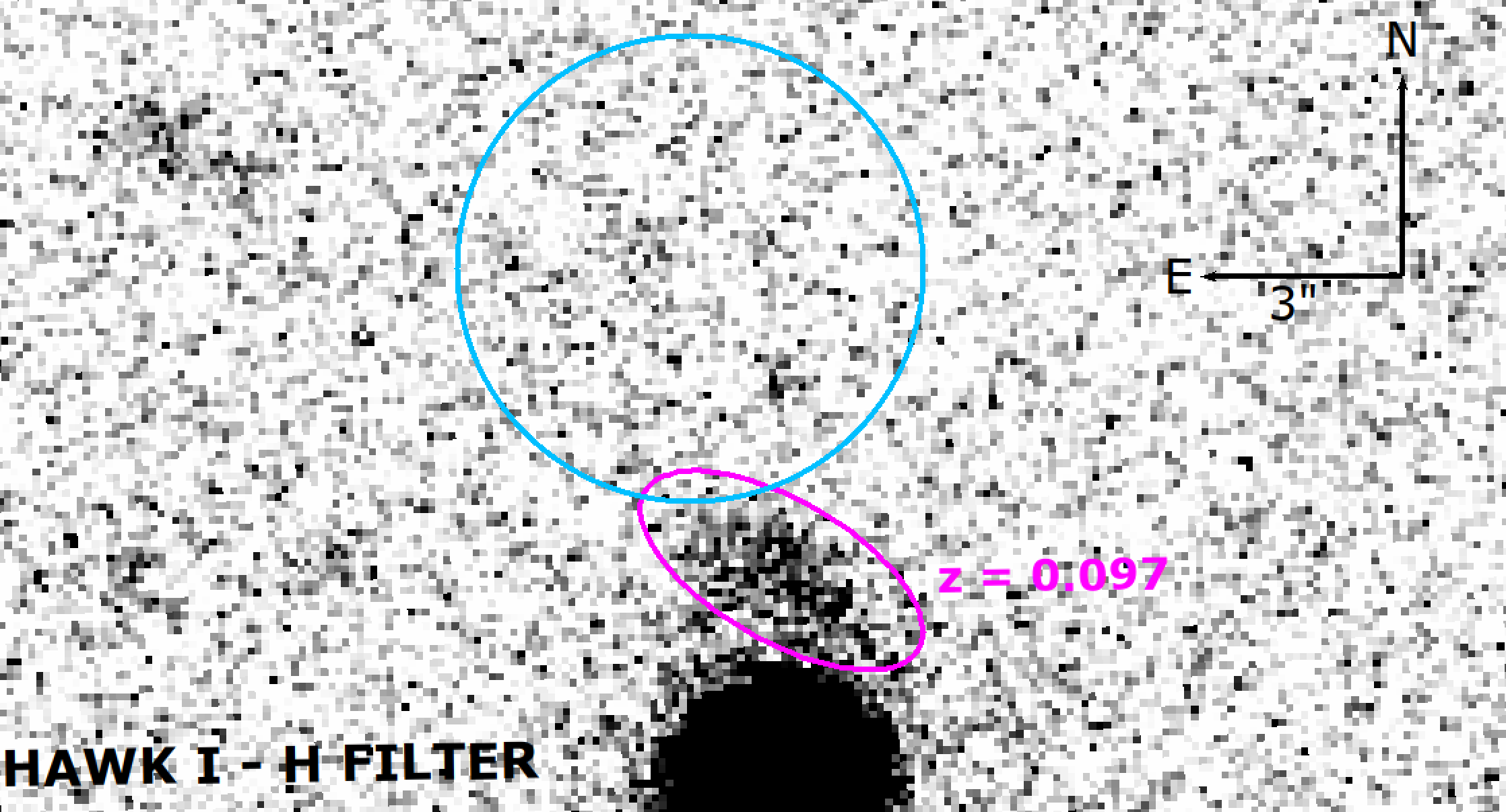}
    \end{subfigure}
    \caption{Finding chart showing the VLT/FORS2 $R$-band (left panel) and VLT/HAWK-I $H$-band (right panel) image of GRB\,211106A field taken on 2021 Dec 4 and 2021 Nov 11, respectively (see Table \ref{log_table_211106}). The \textit{Swift}/XRT position is marked with a blue circle. In the left figure,  the mm afterglow  is highlighted with a brown cross, and a potential host galaxy within the XRT error region with red bars. The possible host galaxy discussed in this work is marked with the magenta ellipse in both figures.}
    \label{211106A_fc}
\end{figure*}

\subsection{\textit{Swift}/XRT}\label{sec:211106A_XRT}
We retrieved the count-rate \textit{Swift}/XRT light curve of GRB\,211106A from the GRB XRT light curve repository\footnote{\url{https://www.swift.ac.uk/xrt_curves/00021466/}} \citep{Evans07, Evans09}. The light curve comprises 9.9 ks of Photon Counting (PC) mode data obtained between T$_0$+39.2 ks and T$_0$+428.8 ks. We found that the light curve can be satisfactorily fitted with a power-law ($F \propto t^{-\alpha}$), with a slope $\alpha$ = 1.07 $\pm$ 0.21 (Fig. \ref{211106_XRTlc}). The few photons collected, due to the late time pointing, prevented us from performing a useful spectral analysis. 

\subsection{Optical/NIR}
We observed the field of GRB\,211106A with ESO-VLT FORS2, HAWK-I and MUSE instruments. A complete log of the observations presented and analysed in this work is summarized in Table \ref{log_table_211106}, and finding charts for the observations carried out with FORS2 and HAWK-I are shown in Fig. \ref{211106A_fc}. The nights of observation were clear, with the seeing in the 0.7${''}$– 0.9${''}$ range.

\subsubsection{VLT/FORS2}\label{211106A/FORS}
We obtained three epochs of observations with the ESO-VLT FORS2 instrument in the $R$ band at about 2.9, 5.9 and 27.9 days after the trigger. Data reduction was carried out following the standard procedures: after bias subtraction, non-uniformities were corrected using a normalized flat-field frame using the \textit{jitter} tool under the ESO-Eclipse software package\footnote{\url{https://www.eso.org/sci/software/eclipse/}}. The photometric calibration was achieved using Stetson field standard stars\footnote{\url{https://www.canfar.net/storage/list/STETSON/Standards}} and by applying the computed offset for five more stars in the field. Astrometric solutions were computed against the USNO B-1 catalogue \citep{USNO}: the three images were found to have a manageable spatial offset $\leq$ 0.3$''$. We performed aperture and point spread function (PSF) photometry using the DAOPHOT package \citep{DAOPHOT}. We carried out image subtraction with the \texttt{HOTPANTS} \citep[High Order Transform of Psf ANd Template Subtraction code,][]{HOTPANTS} package to find and pinpoint variable sources in the field among the three FORS2 epochs, considering the latest epoch as the template image. The full width half maximum (FWHM) of each image is needed to properly model the PSF of both input images, and we computed average values of 3.34$''$, 3.67$''$, and 3.56$''$ for the three epochs, respectively.
\begin{figure*}
    \centering
    \begin{subfigure}[b]{0.3\textwidth}
        \includegraphics[width=1.05\textwidth]{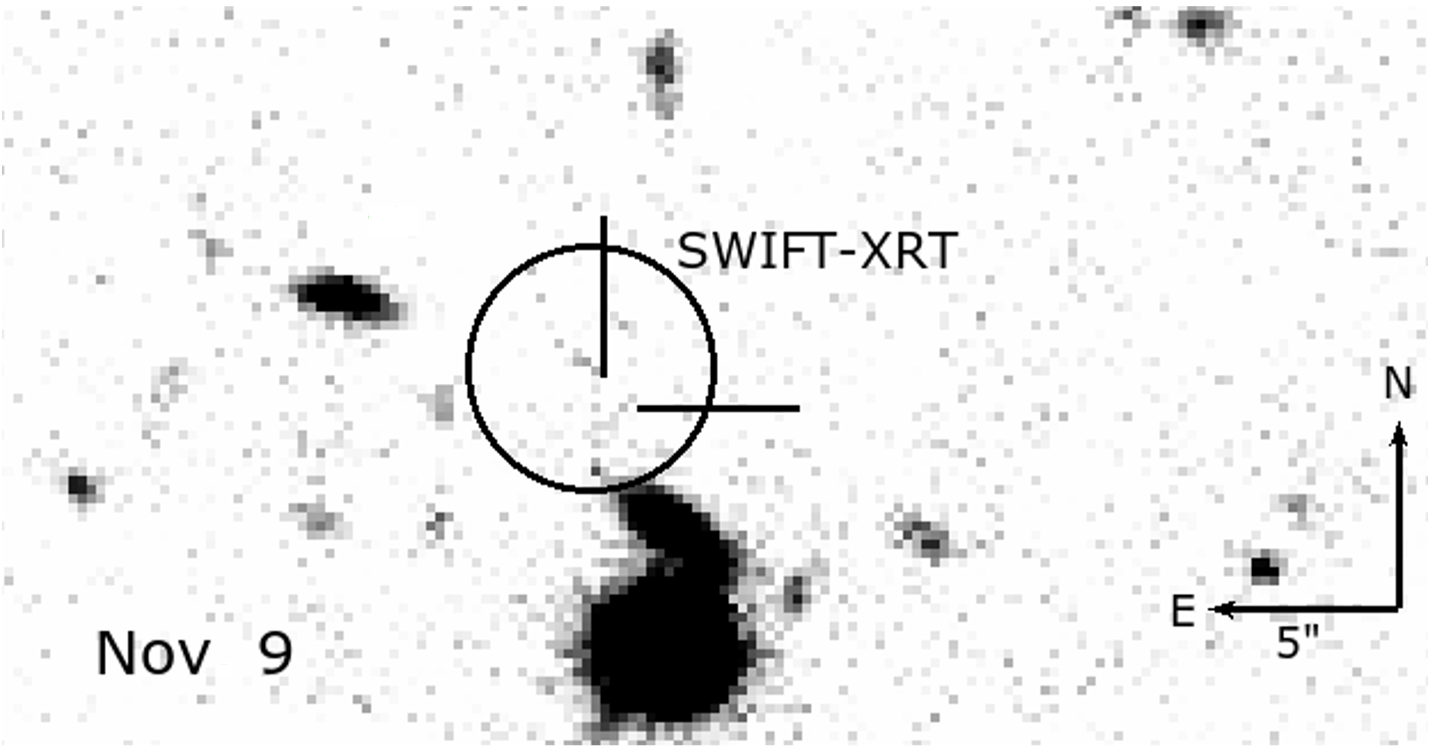}
    \end{subfigure}
    \hfill
    \begin{subfigure}[b]{0.3\textwidth}
        \includegraphics[width=1.05\textwidth]{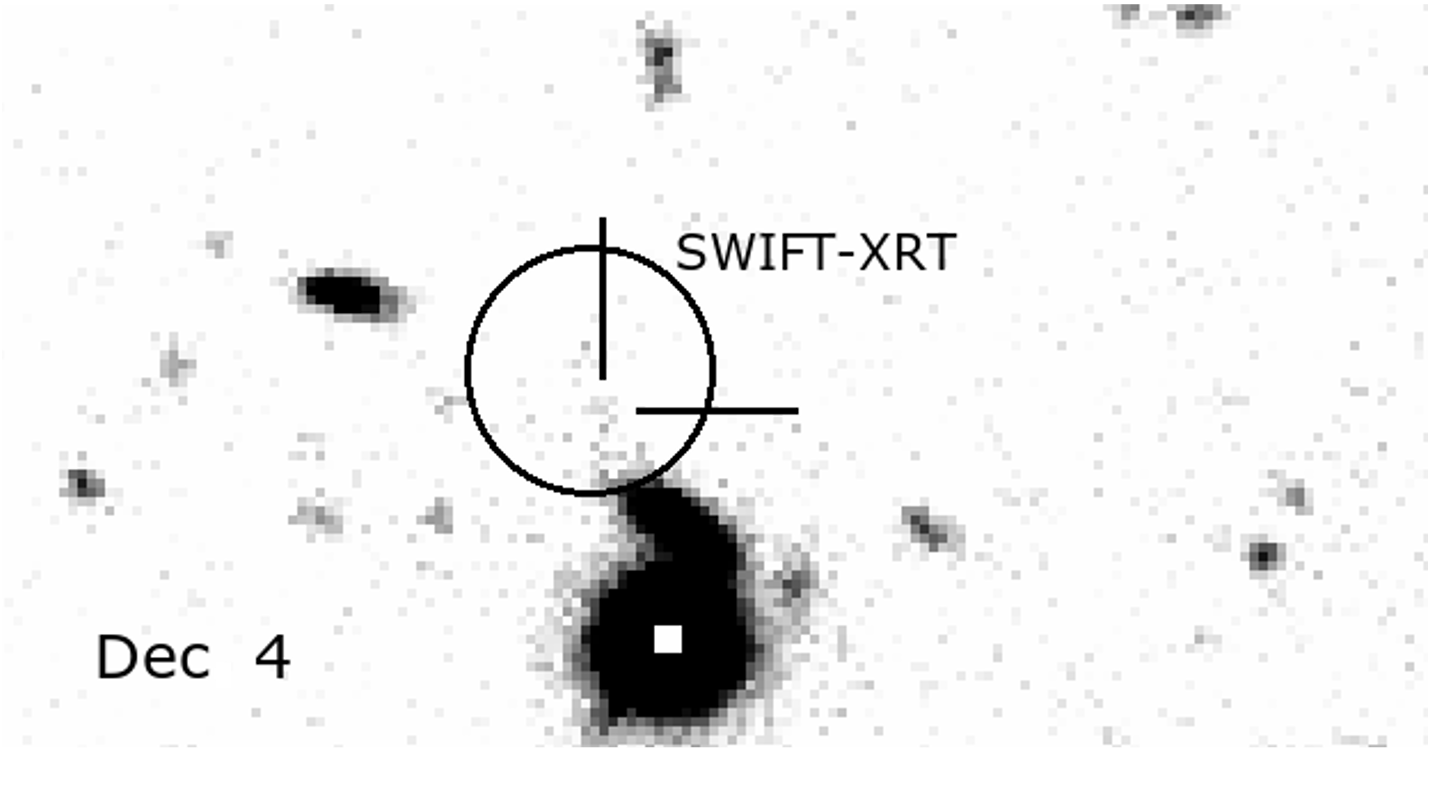}
    \end{subfigure}
    \hfill
    \begin{subfigure}[b]{0.3\textwidth}
        \includegraphics[width=1.05\textwidth]{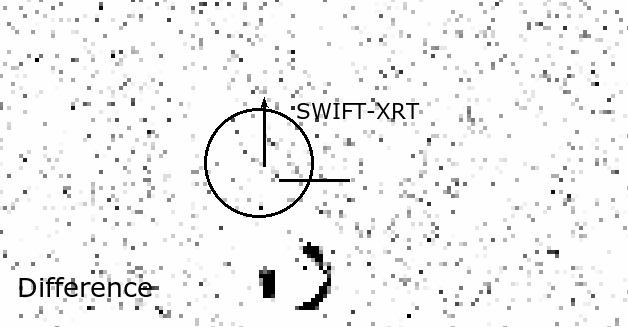}
    \end{subfigure}
    \begin{subfigure}[b]{0.3\textwidth}
        \includegraphics[width=1.05\textwidth]{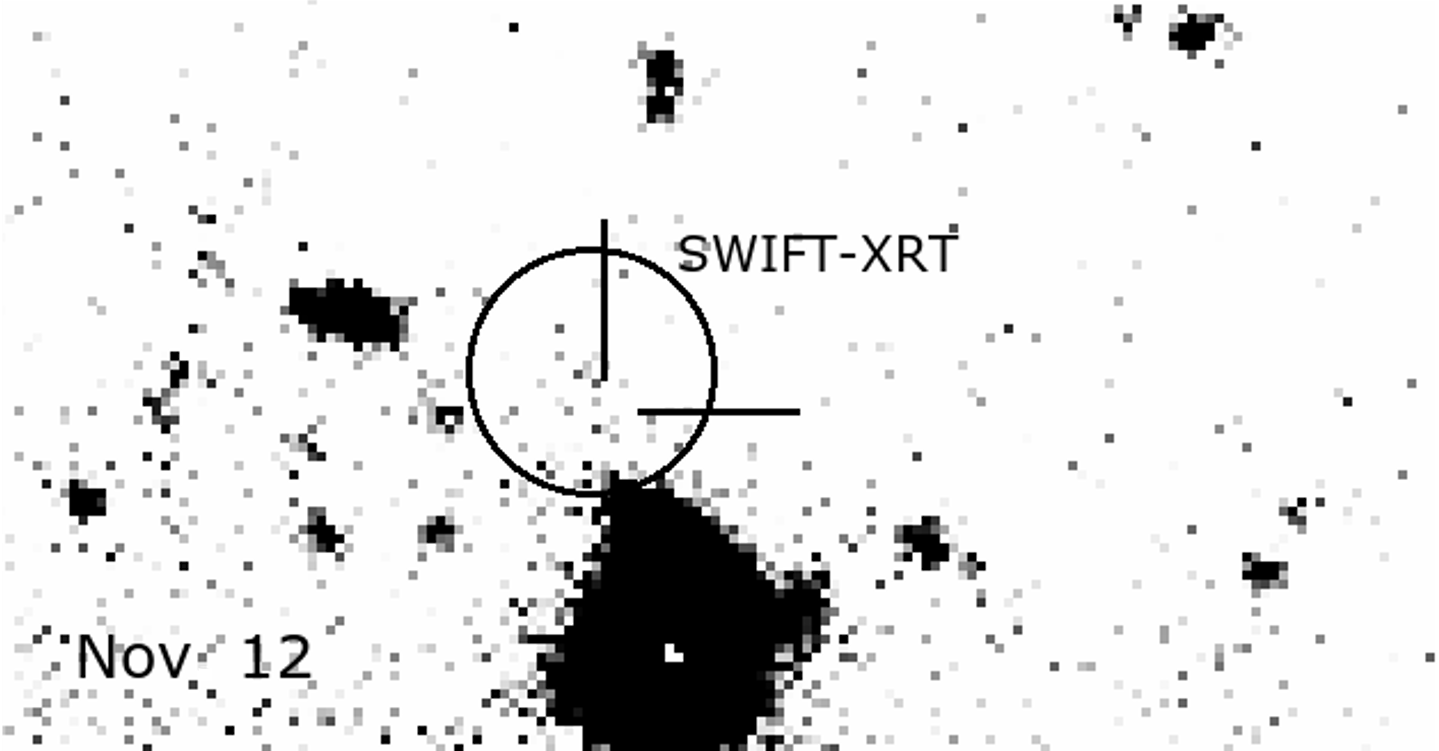}
    \end{subfigure}
    \hfill
    \begin{subfigure}[b]{0.3\textwidth}
        \includegraphics[width=1.05\textwidth]{211106_e3.png}
    \end{subfigure}
    \hfill
    \begin{subfigure}[b]{0.3\textwidth}
        \includegraphics[width=1.05\textwidth]{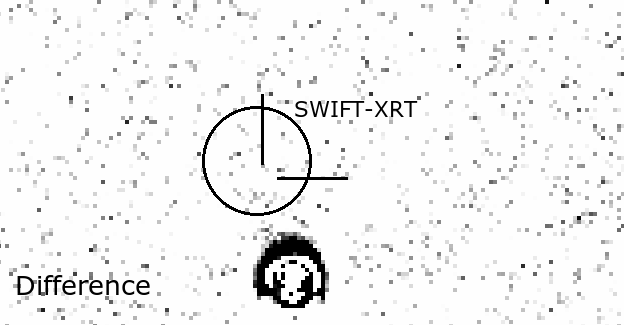}
    \end{subfigure}
    \caption{FORS2 $R$-band images of the field of GRB\,211106A.\\
    \textit{Top panels}.  Epoch 1 (Nov 9), 3 (Dec 4), and the corresponding subtraction. The circle represents the XRT 90\% error region, while the position of the mm afterglow \citep{Laskar22} is indicated by the two black lines.\\ \textit{Bottom panels}. Same as above with the comparison between epochs 2 (Nov 12) and 3 (Dec 4).}
    \label{subtracted}
\end{figure*}\\
In our first $R$ band image, taken $\sim$ 2.9 days after the trigger, we found a source inside the \textit{Swift}/XRT and \textit{Chandra} error circles. For this source, we derived the following position (J2000): R.A. = 22:54:20.5, Dec. = $-$53:13:50.5, with an uncertainty of 0.3$''$. This position is fully consistent with the one of the non-variable extended object detected in the \textit{HST} images of the field \citep{HST_GCN} and proposed as the GRB\,211106A host galaxy by \cite{Laskar22}. The first VLT/FORS2 observation was obtained with a CCD suffering fringing in the red part of the spectrum, while the second one showed the worst seeing among the three, therefore we considered the results obtained from the third epoch as the best value for the source: ${R}=$ 26.51 $\pm$ 0.23 mag. Lastly, the third epoch was taken sufficiently late to confidently rule out a significant possible contribution from the afterglow.\\
Image subtraction through the \texttt{HOTPANTS} software was performed to investigate the detectability of the optical afterglow. We placed in both epoch 1 and 2 images a dummy star of variable magnitude both at the position of the mm afterglow and in a nearby, source-free region, and we computed the difference with respect to the third epoch image. Then, we performed aperture photometry on the subtracted images at the two positions in order to find the deepest limit both for the afterglow detectability and the background at 3$\sigma$ confidence level. For the two subtractions in epoch 1 and 2 we found, within the candidate host galaxy region, limits of $R = 26.7$  mag and $R = 26.8$ mag, respectively; we also obtained $R = 27.5$ mag for the background in both images. 
For the third epoch, we computed $R$ = 26.6 as the minimum magnitude that the afterglow should have had, embedded in our detected source, in order to show a 3$\sigma$ variation. All images obtained for each epoch and the corresponding subtractions are shown in Fig. \ref{subtracted}.

\subsubsection{VLT/HAWK-I}
We carried out NIR imaging observations of the field of GRB\,211106A with the ESO-VLT HAWK-I instrument, in the $H$ band, at $t-t_0 \sim$ 4.8 d. Image reduction was carried out as above, while photometric calibration and astrometry were both performed against five 2MASS catalogue \citep{2MASS} stars present in the field. A $\simeq$ 0.2$''$ spatial offset was found.
Aperture and PSF Photometry were carried out with the DAOPHOT package; no credible source was detected within the XRT error circle down to a limiting magnitude of $H$ $\sim$ 23.6 mag.

\subsubsection{VLT/MUSE}\label{sec:MUSE}
We obtained a single-epoch observation, $\sim$ 3.9 days after the trigger, with the MUSE integral field spectrograph, consisting of 4 exposures of 700\,s each. Data reduction was carried out following standard procedures with a Python-based custom script. Given the faintness of the object detected in our FORS2 observations (\S \ref{211106A/FORS}), we were unable to retrieve a clean spectrum for this source from the MUSE datacube. Nevertheless, we obtained three images integrating the datacube for a wavelength coverage equivalent to $V$, $R$, and $I$ filter, respectively. We computed the corresponding magnitudes for the source using a circular aperture of 0.6$''$. All our photometric results are reported in Table \ref{tab-magnitudes}.

\section{Observation and data analysis: GRB\,211227A}\label{sec:211227An}
%----------------------------------------------------- LOG TABLE
\begin{table*}[]
    \centering
    \caption{Log of imaging and spectroscopic VLT ground-based observations of the GRB\,211227A field.}
    \begin{tabular}{l|l|l|l|l|l}
    Date: $t_{\text{start}}$ (UT) & $t-t_0$ [d] & Instrument & $t_{\text{exp}}$ [$\#$ x    $s$] & Seeing [$''$] & Filter\\
    \noalign{\smallskip}
    \hline
    \hline
    \noalign{\smallskip}
    2021-12-28 04:35:34.333 & 0.211 & X-Shooter & 3 x 30 & 0.70 & $r$\\
    2021-12-28 04:38:09.773 & 0.213 & X-Shooter & 3 x 60 & 0.61 & $z$\\
    2021-12-28 04:42:16.433 & 0.216 & X-Shooter & 3 x 40 & 0.76 & $g$\\
    2021-12-29 03:51:11.544 & 1.193& FORS2 & 8 x 240 & 0.93 & $I$\\
    2021-12-29 05:08:40.348 & 1.235 & X-Shooter & 3 x 60 & 0.89 & $r$\\
    2021-12-29 05:12:45.808 & 1.239 & X-Shooter & 6 x 60 & 0.73 & $z$\\
    \noalign{\smallskip}
    \hline
    \noalign{\smallskip}
    2021-12-27 04:45:24.139 & 0.233 & X-Shooter & 4 x 600 & - & -\\ 
    \end{tabular}
    \label{log_tab_211227}
\end{table*}

\subsection{\textit{Swift}/XRT}\label{sec:211227_XRT}

\begin{figure}
    \centering
    \includegraphics[width=9cm]{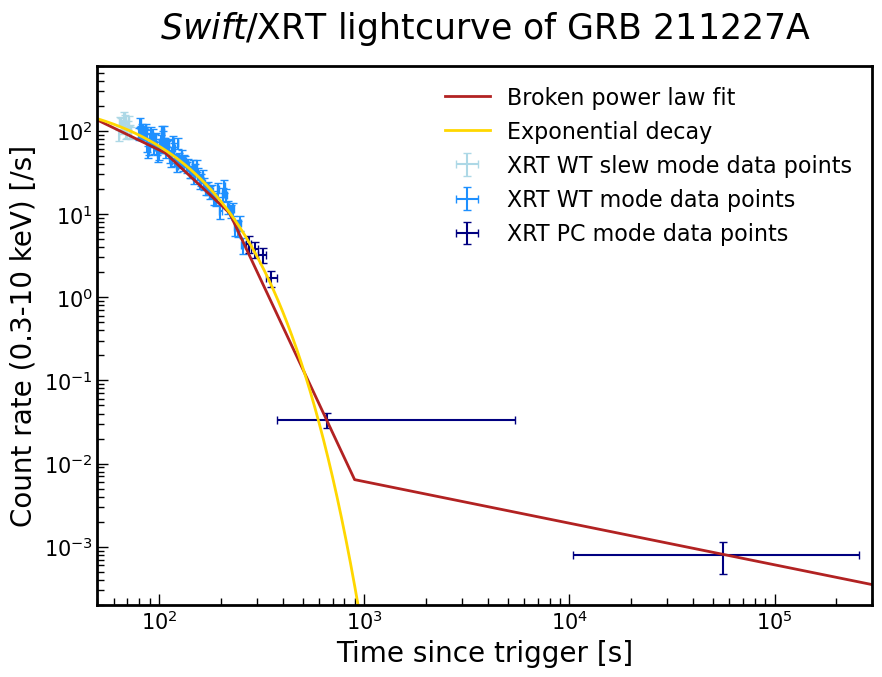}
    \caption{Light curve broken-power law fit (red solid line) for the XRT observations of GRB\,211227A. The best fit shows three breaks at T$_0$+106.4 s, T$_0$+219.8 s, and T$_0$+895 s. The yellow line represents a fit with an exponential profile of the early X-ray light curve.}
    \label{211227_XRTlc}
\end{figure}

\begin{figure*}
    \centering
    \begin{subfigure}[b]{.49\textwidth}
        \includegraphics[width=1.\textwidth]{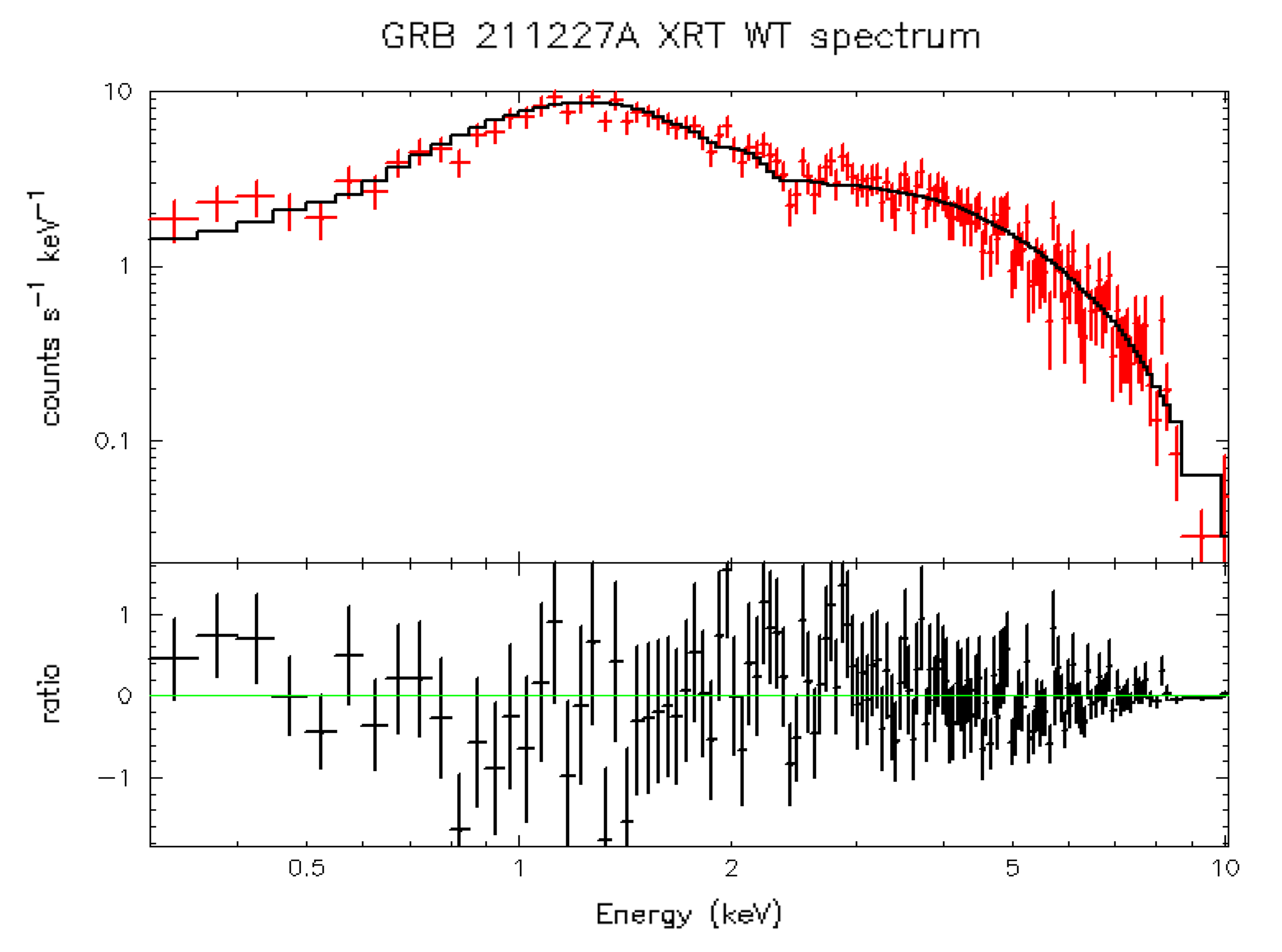}
    \end{subfigure}
    \hfill
    \begin{subfigure}[b]{.49\textwidth}
        \includegraphics[width=1.\textwidth]{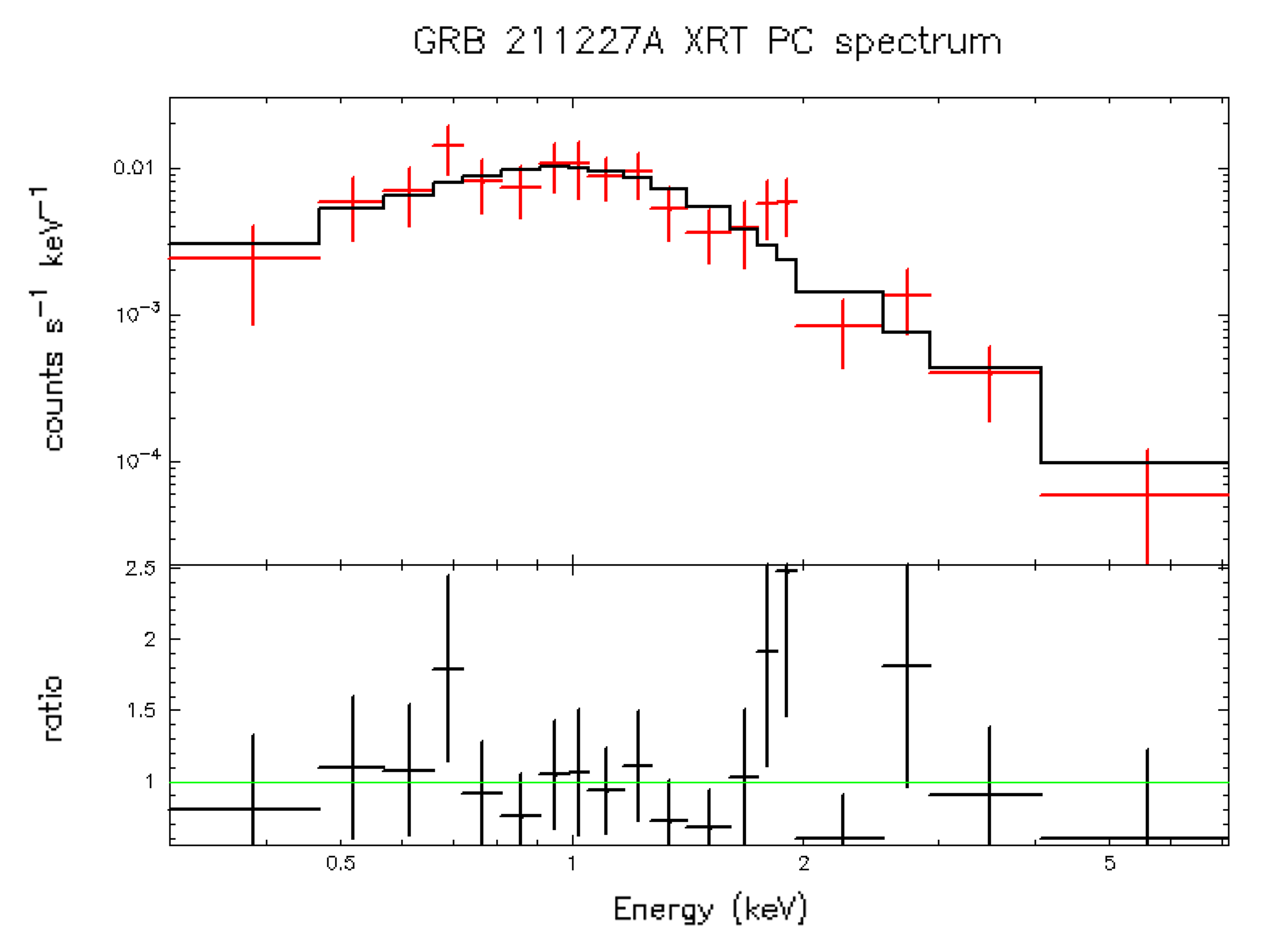}
    \end{subfigure}
    \caption{\textit{Swift}/XRT WT (left panel) and PC (right panel) mode spectra for GRB\,211227A. The best fit is shown in black (top panel). In the bottom panel the ratio between the data and the folded model is shown.}
    \label{211227_XRTspec}
\end{figure*}

We retrieved the count-rate \textit{Swift}/XRT light curve of GRB\,211227A from the GRB XRT light curve repository\footnote{\url{https://www.swift.ac.uk/xrt_curves/01091101/}} \citep{Evans07, Evans09}. The light curve comprises 8.1 ks of data obtained between T$_0$+63 s and T$_0$+57.9 ks, 182 s of which in Windowed Timing (WT) mode, while the remainder in PC mode. We found that the light curve can be fitted with a broken power-law with three breaks (see Fig. \ref{211227_XRTlc}). At early times, the curve decays with indices $\alpha_1=1.19^{+0.25}_{-0.20}$, $\alpha_2=2.30^{+0.23}_{-0.18}$ (with a first break at T$_0$+106.4 s), then, from T$_0$+219.8 s to T$_0$+895, it follows an extremely steep ($\alpha_3=5.26^{+0.35}_{-0.28}$) decay, before ending with a shallower evolution, $\alpha_4=0.5 \pm 0.5$, suggesting a possible plateau (all errors are at 90\% c.l.). However, the lack of additional data at later times prevented us from confirming this flattening. We also fit the light curve (excluding the last point) with an exponential decay, obtaining a good match: $\chi^2_r =$ 126.4/96. This early exponential profile in the X-rays is consistent with previous extended emission GRBs \citep{Gomp} and with GRB\,211211A \citep{Gompertz_211211A}.\\
We retrieved spectral data from the \textit{Swift}/XRT GRB spectrum repository\footnote{\url{https://www.swift.ac.uk/xrt_spectra/}}, and we analysed the spectra from both the WT mode and the PC mode data (from T$_0$+272.6 s). We fitted them with an absorbed power-law model within the \texttt{XSpec} package \citep{XSPEC}, keeping the Galactic contribution $N_H$ and the redshift fixed at 2.4 $\times$ 10$^{20}$ cm$^{-2}$ \citep{Willingaleetal13} and 0.228, respectively. For the WT mode data, we obtained a photon spectral index $\Gamma=1.07^{+0.07}_{-0.06}$ and a best-fitting intrinsic column density $N_H=2.7^{+0.6}_{-0.6}$ $\times$ 10$^{21}$ cm$^{-2}$ (errors at 90\% c.l.), in excess with respect to the Galactic value. The resulting fit statistics is C-stat/dof = 630.97/697. The fit of the PC mode spectrum returned $\Gamma=2.88^{+0.59}_{-0.51}$, and a best-fitting intrinsic column density $N_H=3.8^{+2.3}_{-1.8}$ $\times$ 10$^{21}$ cm$^{-2}$ (errors at 90\% c.l.), in excess with respect to the Galactic value. The resulting fit statistics is C-stat/dof = 64.3/90. The two spectra, along with their fit, are shown in Fig. \ref{211227_XRTspec}.

\subsection{Optical}

%----------------------------------------------------- FINDING CHARTS
\begin{figure*}
    \centering
    \begin{subfigure}[b]{.49\textwidth}
        \includegraphics[width=1.\textwidth]{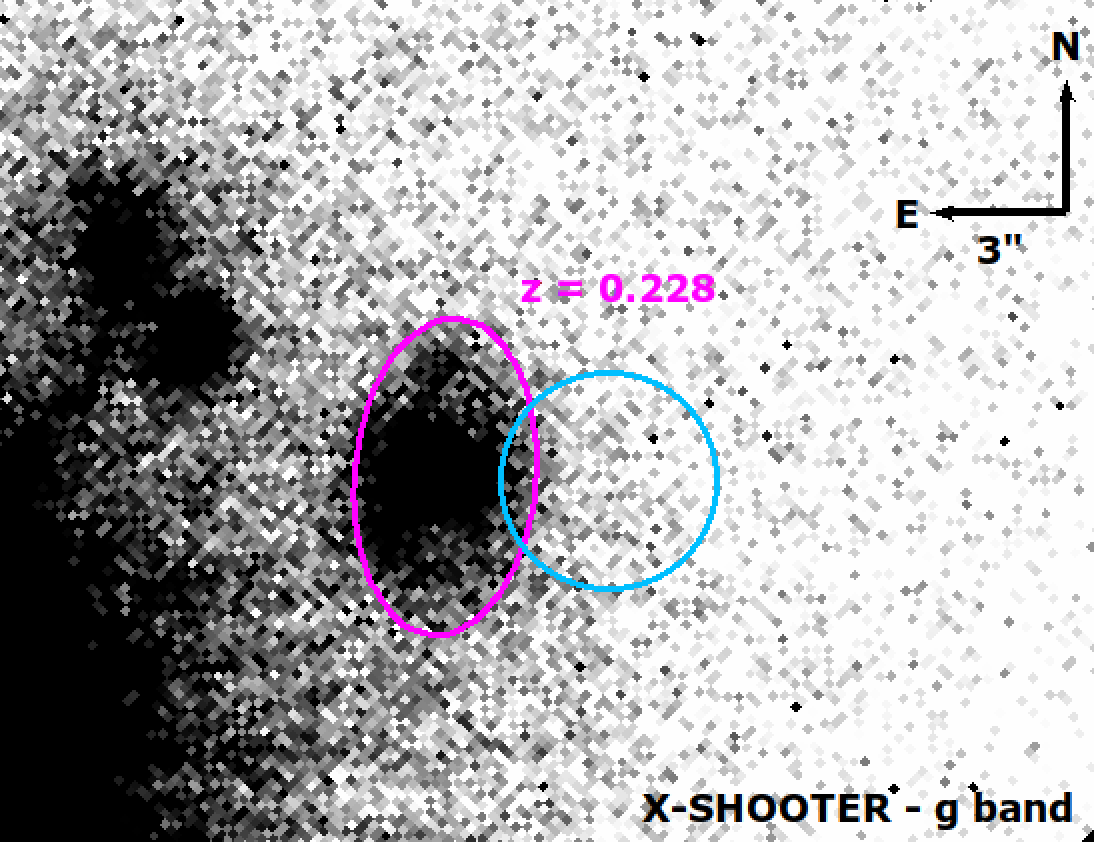}
    \end{subfigure}
    \hfill
    \begin{subfigure}[b]{.49\textwidth}
        \includegraphics[width=1.\textwidth]{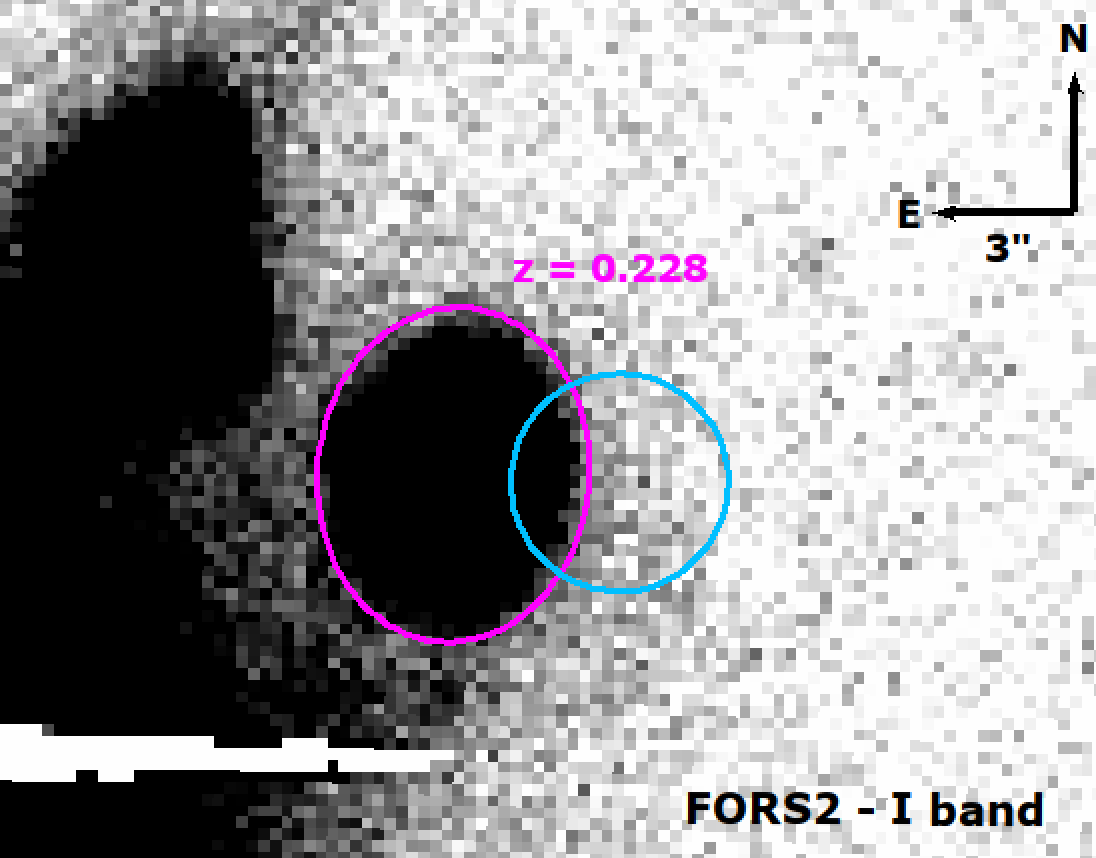}
    \end{subfigure}
    \caption{Finding chart for X-Shooter $g$-band (left panel) and the FORS2 $I$-band (right panel) observation of GRB\,211227A taken on 2021 Dec 28. Error circle (at 90\% c.l.) from \textit{Swift}/XRT observation is reported in blue. The magenta ellipse highlights the candidate host galaxy at $z$=0.228. No credible afterglow candidate was found inside the XRT circle.}
    \label{211227A_fc}
\end{figure*}

\begin{figure*}
    \centering
    \begin{subfigure}[b]{0.317\textwidth}
        \includegraphics[width=1.\textwidth]{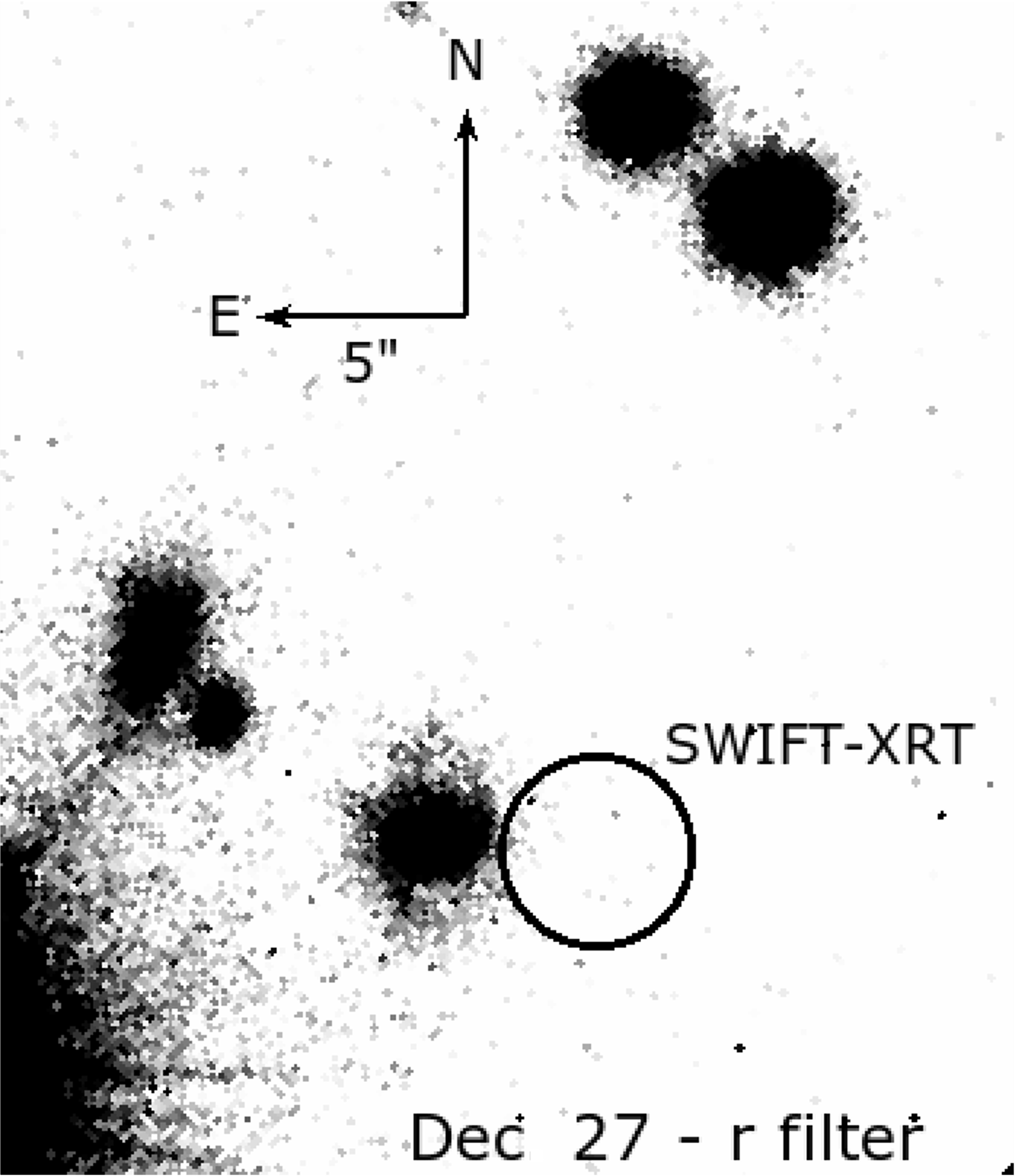}
    \end{subfigure}
    \hfill
    \begin{subfigure}[b]{0.317\textwidth}
        \includegraphics[width=1.\textwidth]{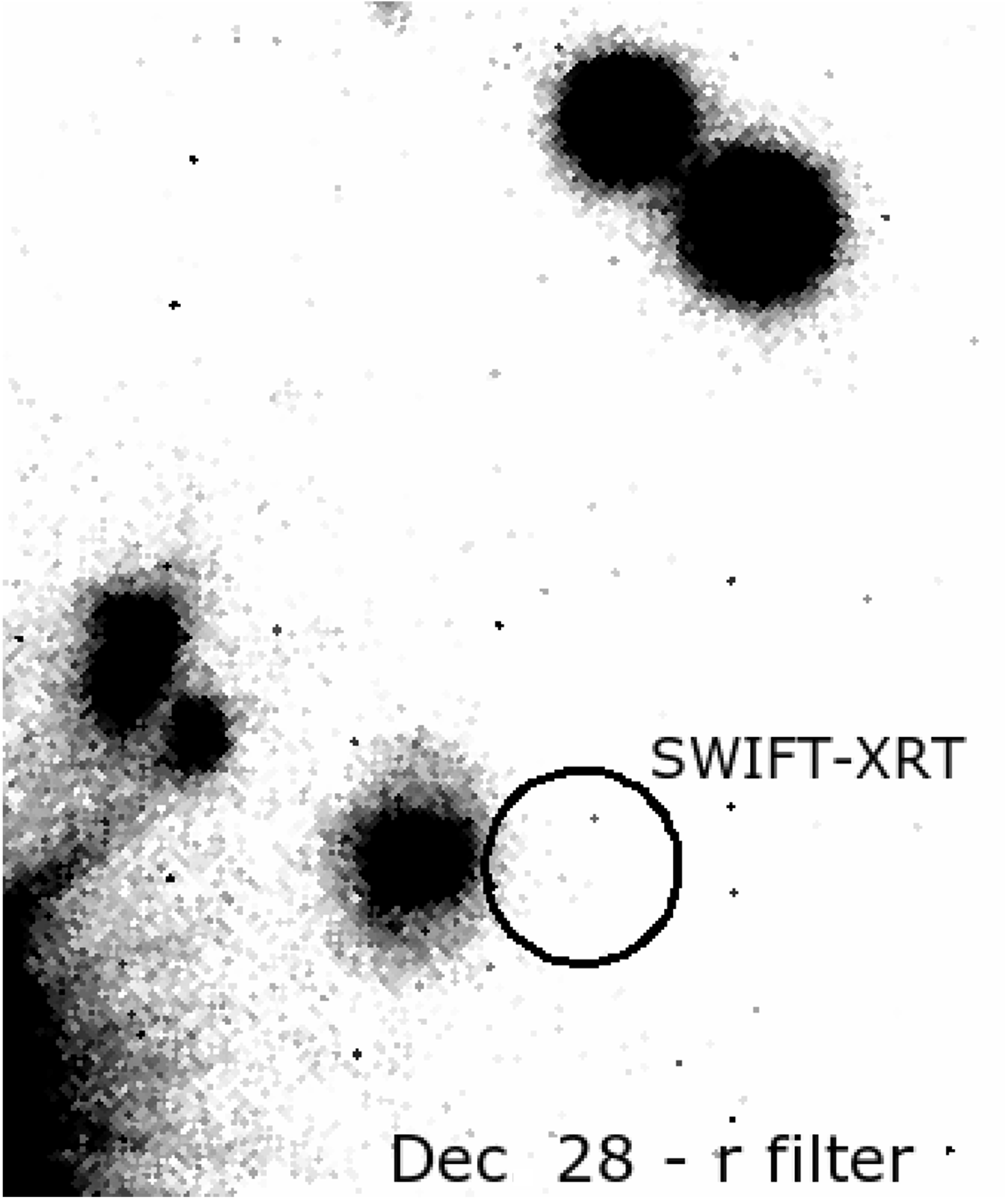}
    \end{subfigure}
    \hfill
    \begin{subfigure}[b]{0.317\textwidth}
        \includegraphics[width=1.\textwidth]{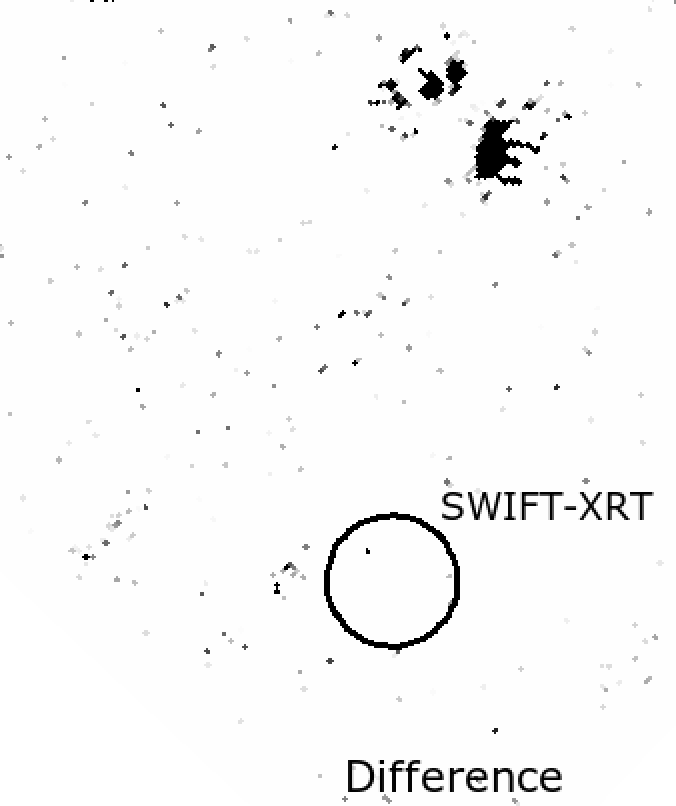}
    \end{subfigure}
    \begin{subfigure}[b]{0.317\textwidth}
        \includegraphics[width=1.\textwidth]{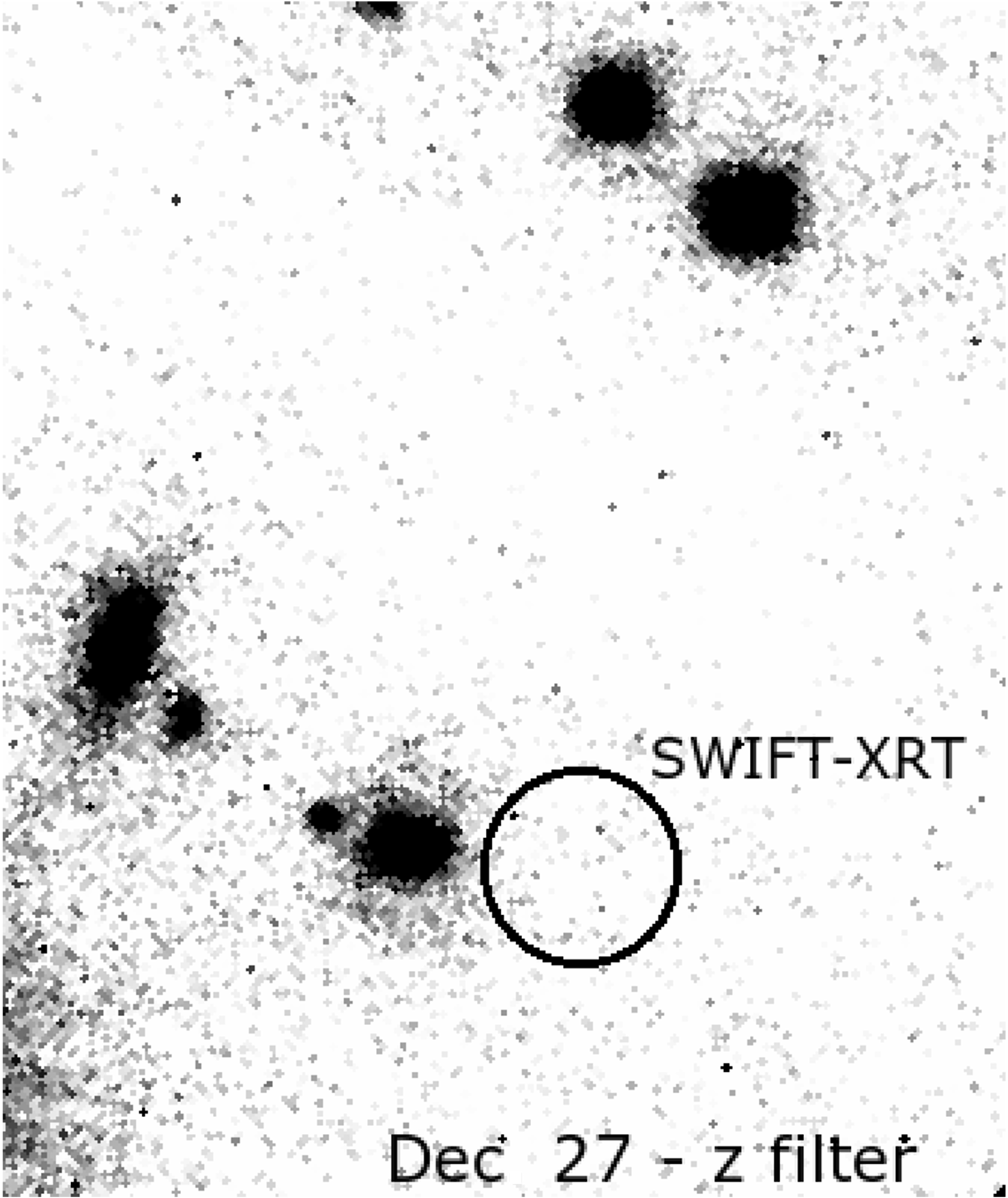}
    \end{subfigure}
    \hfill
    \begin{subfigure}[b]{0.317\textwidth}
        \includegraphics[width=1.\textwidth]{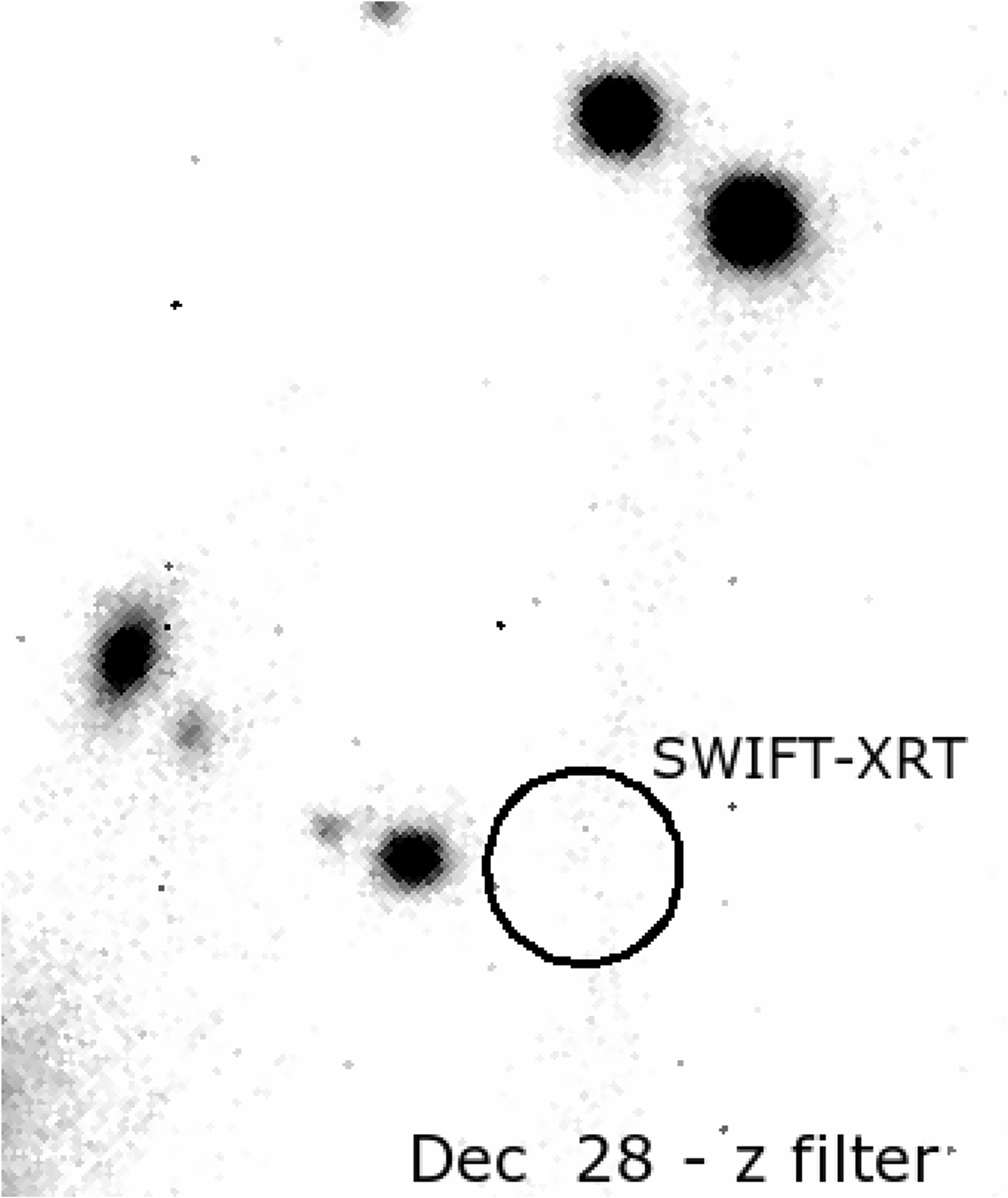}
    \end{subfigure}
    \hfill
    \begin{subfigure}[b]{0.317\textwidth}
        \includegraphics[width=1.\textwidth]{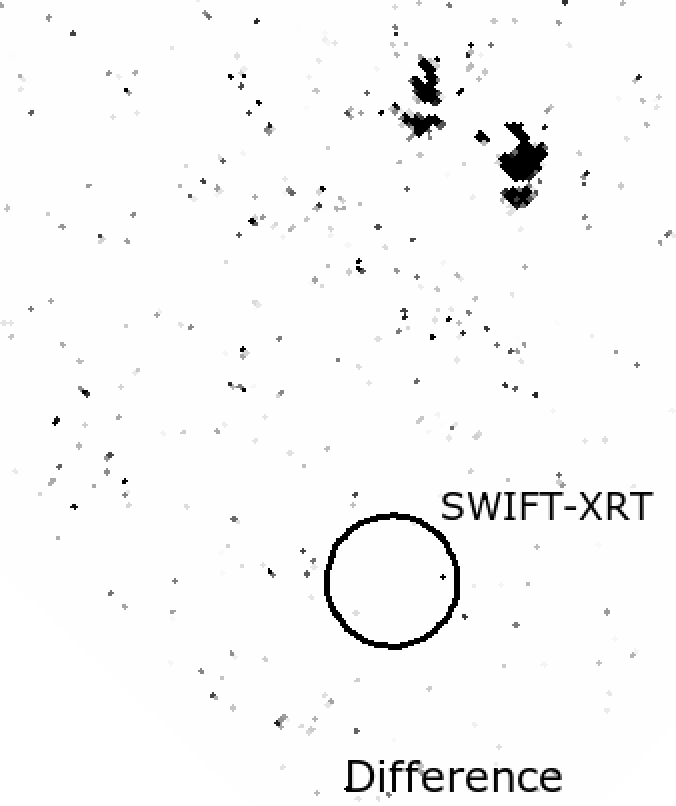}
    \end{subfigure}
    \caption{X-Shooter images of the field of GRB\,211227A.\\
    \textit{Top panels}. $r$-band images of the two epochs (Dec 27 and Dec 28) along with their subtraction. The circle represents the XRT 90\% error region.\\ \textit{Bottom panels}. Same as above for the $z$-band observations.}
    \label{211227Acharts}
\end{figure*}

We observed the field of GRB\,211227A with ESO-VLT X-Shooter and FORS2 instruments, obtaining both imaging and spectroscopic data. The complete log of the observations is reported in Table \ref{log_tab_211227}, while finding charts for the observations with the two instruments are shown in Fig. \ref{211227A_fc} for X-Shooter and FORS2 observations. We had clear night of observations, with the seeing ranging from 0.6$''$ to 0.9$''$.

\subsubsection{VLT/X-Shooter}\label{sec:XS}
\begin{figure*}
    \centering
    \includegraphics[width=16cm]{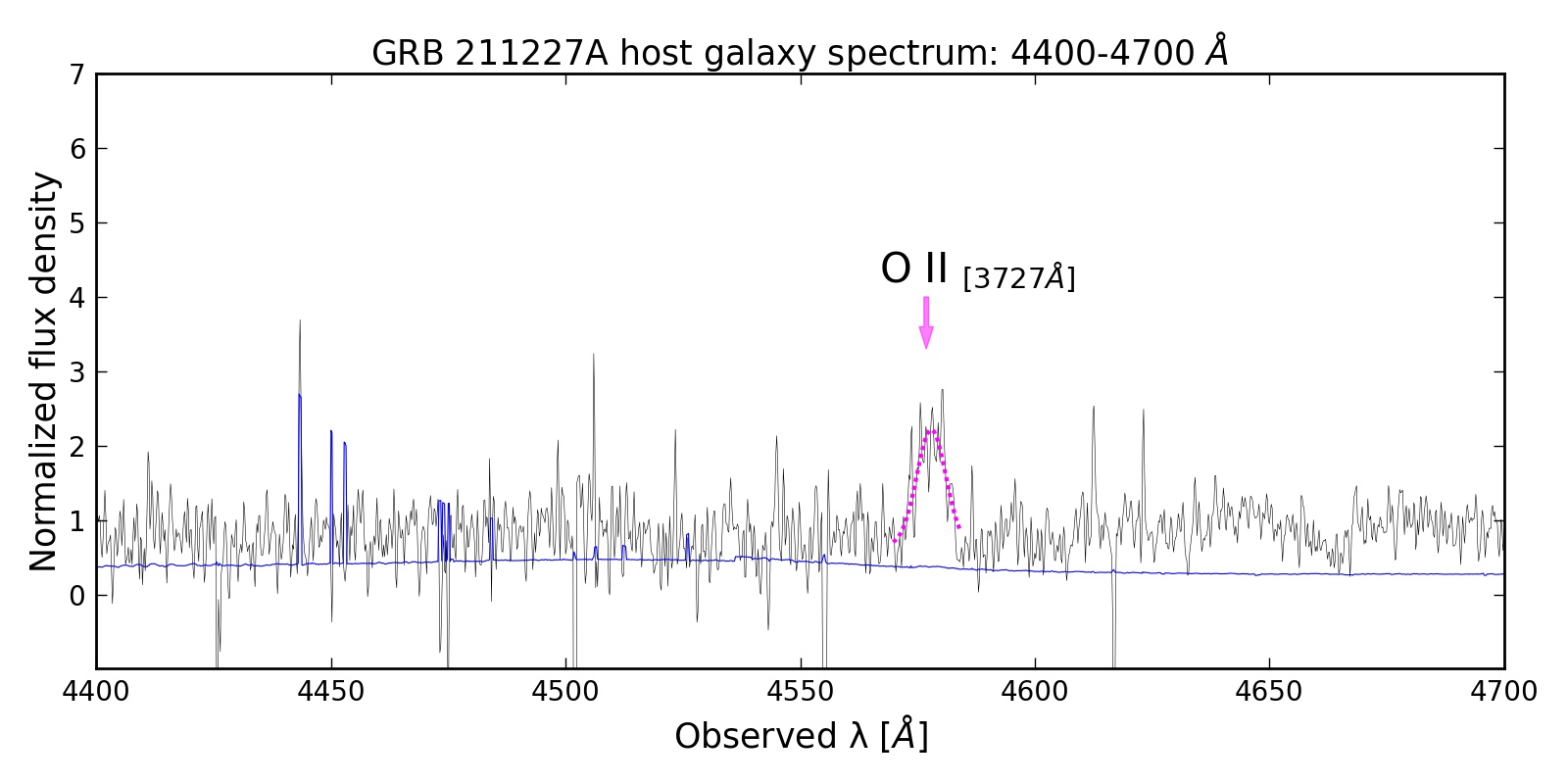}
    \includegraphics[width=16cm]{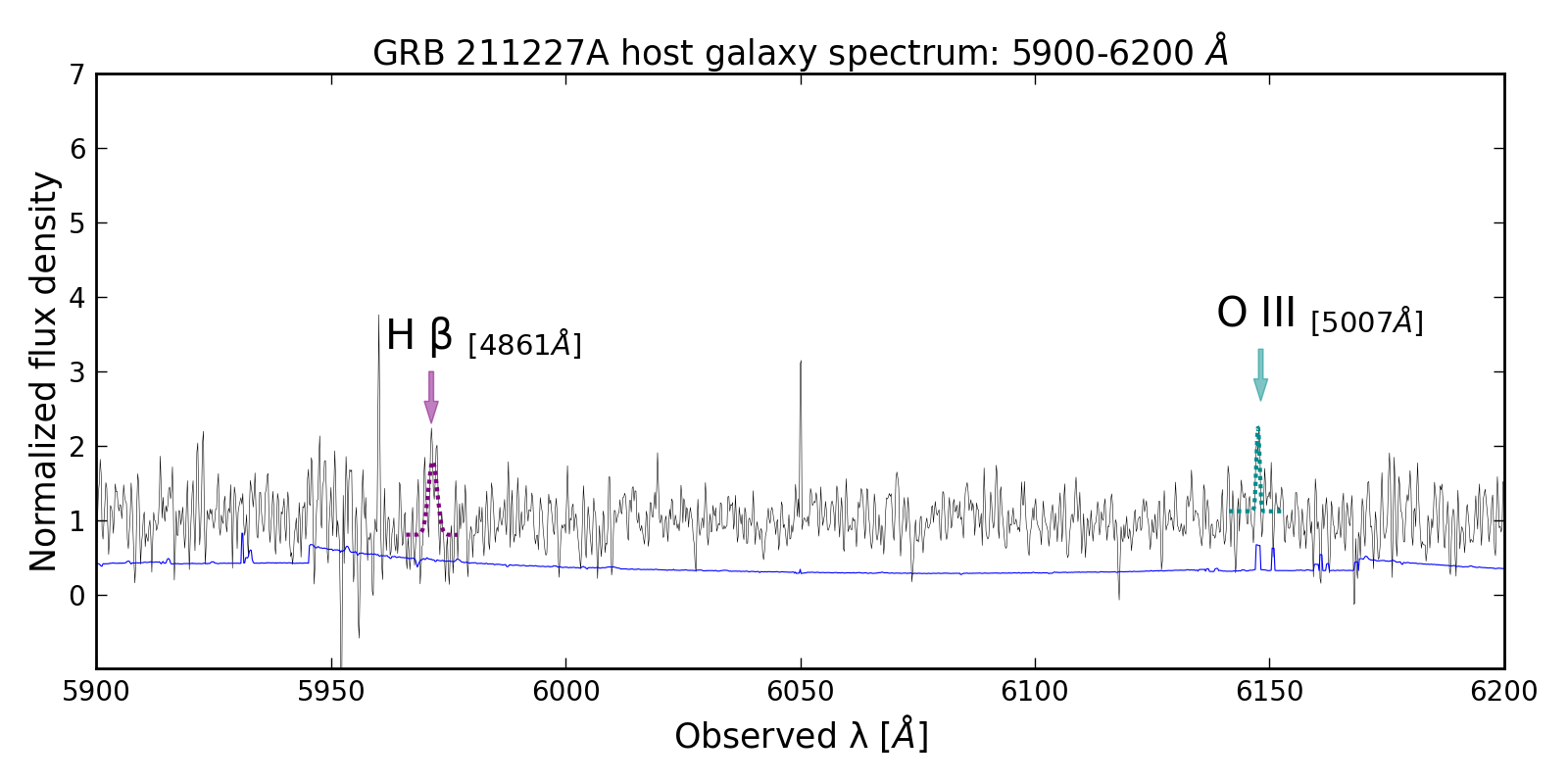}
    \includegraphics[width=16cm]{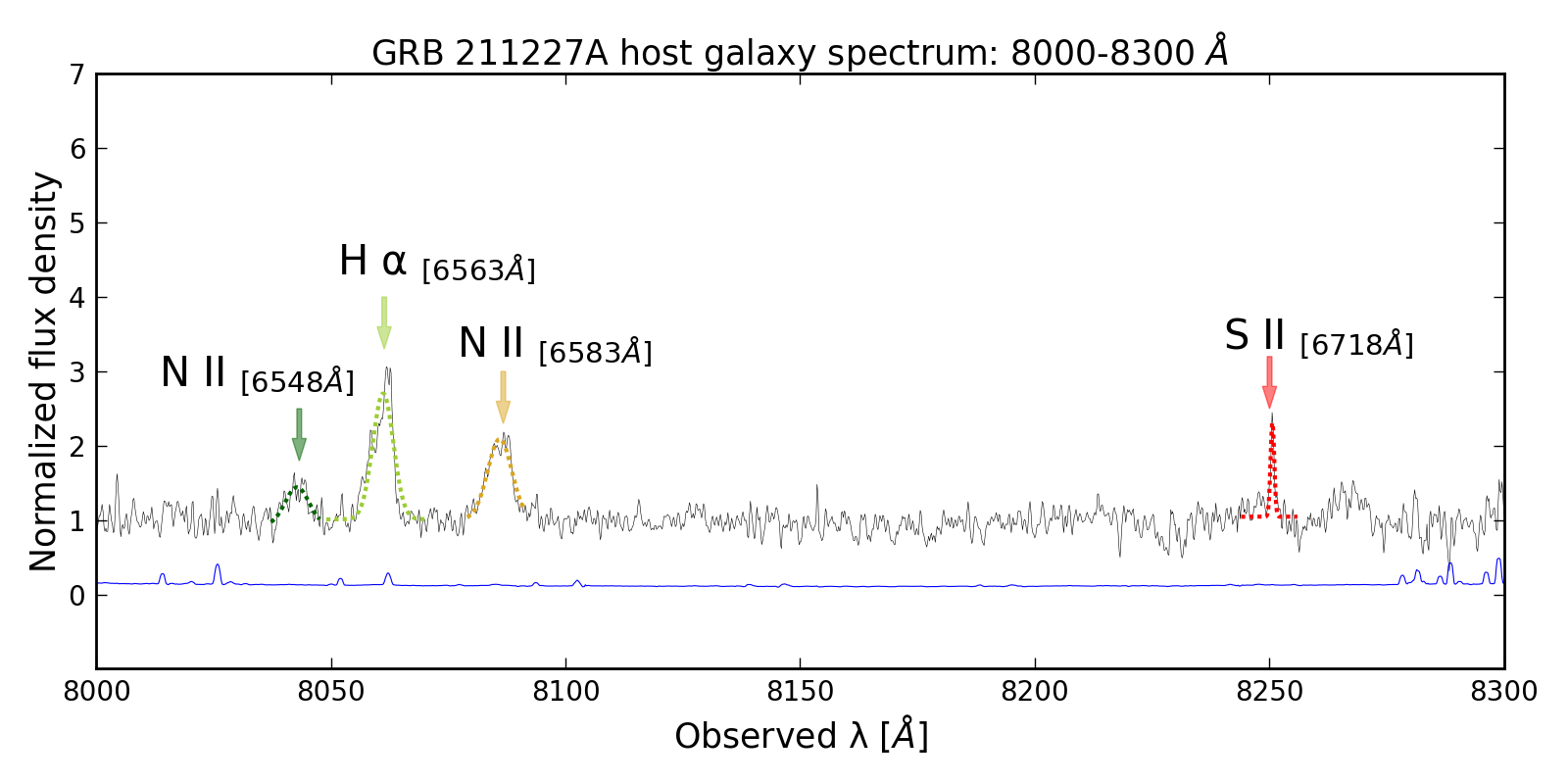}
    \caption{Slices of the spectrum obtained with VLT/X-Shooter for the host galaxy of GRB\,211227A. The emission lines that we identified are highlighted with different colours along with their best Gaussian fit. The error spectrum is reported in blue.}
    \label{Sp211227A}
\end{figure*}

X-Shooter observations were performed $\sim$ 0.2 ($r$, $z$ band) and $\sim$ 1.2 ($g$, $r$, $z$ band) days after T$_0$. We reduced raw acquisition images following the standard procedure already described for GRB\,211106A, with astrometric solution and photometric calibration both performed against the PanSTARRS DR1 catalogue \citep{PanstarrsDR1}. Aperture and PSF photometry were carried out with the DAOPHOT package. Furthermore, we performed image subtraction with \texttt{HOTPANTS} between the two epochs obtained with the X-Shooter acquisition camera with the $r$ and $z$ filters. For these two subtractions we computed an average FWHM of 3.85$''$ and 4.66$''$ for epoch 1 and 2 in the $r$ filter, and 3.01$''$ and 3.90$''$ for epoch 1 and 2 in the $z$ filter. From the analysis of the X-Shooter and FORS2 images we found no evidence for a GRB optical counterpart (Fig. \ref{211227Acharts}). We report the 3$\sigma$ magnitude limits estimated through PSF photometry for both X-Shooter and FORS2 observations in Table \ref{211227A_maglims}.
\begin{table}
    \centering
    \caption{GRB\,211227A magnitude limits.}
    \begin{tabular}{ccc}
    \hline
    \noalign{\smallskip}
    $t-t_0$ [d]& Mag. limit& Instrument / Filter\\
    \noalign{\smallskip}
    \hline
    \hline
    \noalign{\smallskip}
    0.211& 25.4& X-Shooter / $r$\\
    0.213& 24.8& X-Shooter / $z$\\
    0.216& 25.4& X-Shooter / $g$\\
    1.193& 25.8& FORS2 / $I$\\
    1.235& 25.8& X-Shooter / $r$\\
    1.239& 24.7& X-Shooter / $z$\\
    \noalign{\smallskip}
    \hline
    \end{tabular}
    \tablefoot{$3\sigma$ magnitude limits obtained for GRB211227A from X-Shooter acquisition images and FORS2 observation, as described in the text.}
    \label{211227A_maglims}
\end{table}
On Dec 26 we also carried out spectroscopic observations of the host galaxy of GRB\,211227A. We obtained 4 exposures of 600 s each, covering a wavelength range of 3000-24800$\AA$ (3000-5600, 5500-10200, and 10200-24800 $\AA$ for the UVB, VIS, and NIR arm, respectively). The slit width is 1$''$ for the UVB arm and 0.9$''$ for the others, giving a limit on the delivered resolving power of the spectra based on the tabulated values of the resolutions, which are 5400, 8900, and 5600 for the
UVB, VIS, and NIR-arm, respectively\footnote{\url{https://www.eso.org/sci/facilities/paranal/instruments/xshooter/inst.html}}.
Data reduction was carried out with the ESO-Reflex pipeline \citep[version 2.11.5,][]{REFLEX} for X-Shooter \citep{Xshoo_pipeline}, then we analysed the reduced spectrum to investigate the host galaxy properties (see Fig. \ref{Sp211227A} and \S.\ref{sec:211227_HG}). In Table \ref{lines_z} we show the identified features of the spectrum along with the redshift computed from the line centroids of their best Gaussian fit. We also computed their corresponding rest frame luminosity given a luminosity distance for the galaxy of $D_L=$ 1170.6 Mpc at the average redshift of $z=0.228$.

%----------------------------------------------------- FEATURES AND REDSHIFT
\begin{table}[]
    \centering
    \caption{Emission lines luminosity from GRB\,211227A host galaxy spectrum.}
    \begin{tabular}{c|c|c|c}
    \hline
    \noalign{\smallskip}
    $\lambda_{\text{obs}} [\AA]$ & Feature & $z$ & Luminosity [$10 ^{40}$erg/s]\\
    \noalign{\smallskip}
    \hline
    \hline
    \noalign{\smallskip}
    4577.95 & [O II]$\lambda$3727 & 0.2283 & 13.585$\pm$1.190\\
    5971.58 & H$_\beta$ & 0.2281 & 2.768$\pm$0.592\\
    6147.48 & [O III]$\lambda$5007 & 0.2278 & 1.153$\pm$0.289\\
    8042.68 & [N II]$\lambda$6548 & 0.2283 & 2.970$\pm$0.384\\
    8061.01 & H$_\alpha$ & 0.2280 & 8.415$\pm$0.405\\
    8085.86 & [N II]$\lambda$6583 & 0.2279 & 6.216$\pm$0.287\\
    8250.57 & [S II]$\lambda$6718 & 0.2281 & 0.984$\pm$0.136\\
    \noalign{\smallskip}
    \noalign{\smallskip}
    \hline
    \end{tabular}
    \tablefoot{List of  emission features identified in the spectrum of GRB\,211227A host galaxy with their respective luminosity after being corrected for slit losses and host galaxy and Milky Way extinction (see \S \ref{sec:211227_HG} for details).}
    \label{lines_z}
\end{table}

\subsubsection{VLT/FORS2}
We analysed the field of the burst for the single $I$ band FORS2 epoch, taken $\sim$ 1.2 days after the trigger. After standard data reduction, photometric calibration was achieved with Stetson field stars and astrometry was performed against the PanSTARRS DR1 catalogue. We carried out aperture and PSF photometry with the DAOPHOT package, but we did not find any candidate afterglow within the XRT error circle. We computed a 3$\sigma$ magnitude limit for the observation of 25.8 mag.

\section{Discussion}\label{discussion}

\subsection{Host galaxy identification for GRB 211106A}\label{sec:211106_HG}
\begin{figure}[]
    \centering
    \includegraphics[width=9cm]{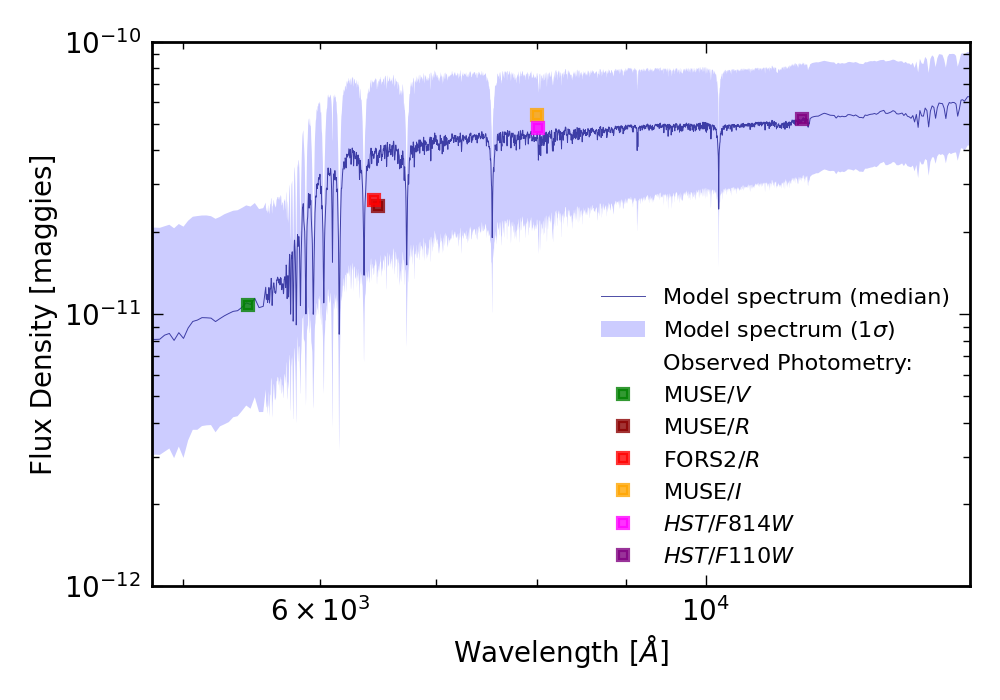}
    \caption{Best fit for GRB\,211106A host galaxy SED from the median values of the posterior distribution parameters. Within \texttt{Prospector}, \textit{maggies} are a convenient flux density ($f$) unit defined from the magnitude as $f=10^{-0.4m_{AB}}$.}
    \label{sed21}
\end{figure}

The association of GRB\,211106A with a host galaxy is particularly complex: the field where the burst appeared shows two sources close to or inside the XRT error circle, neither of which can be safely associated with the X-ray afterglow. We first analysed the one located at the coordinates (J2000) R.A. = 22:54:20.29, Dec. = $-$53:13:53.6 and at a redshift $z$ = 0.097, whose corresponding luminosity distance is $D_L\sim460$ Mpc. We computed the probability of chance association ($P_{cc}$) of this source with the best position of the afterglow - the mm-band one (see \S \ref{sec:211106_aftgl}) - following \cite{BLOOM02}. The projected separation of the two sources is 3.76$''$, and the source has a magnitude $R \sim$ 21.5 mag \citep{MUSE_GCN}. In order to obtain the effective radius of the galaxy for a better computation of $P_{cc}$, we performed surface photometry with the \texttt{ellpro} package under the Starlink software \citep{GAIA}, then we fitted the obtained profile with a Sérsic model:
    \begin{eqnarray}
    \centering
    \Sigma(r) = I_0\exp\{-k_n[(r/r_e)^{1/n}-1]\},
    \label{sersic}
    \end{eqnarray}
where \textit{n} is the concentration parameter (\textit{n} = 1 is equivalent to an exponential disk profile, while \textit{n} = 4 is the de Vaucouleurs profile typical of elliptical galaxies), $k_n \sim$ 2\textit{n} $-$ 1/3 + 4/405\textit{n} + 46/25515$n^2$ \citep{CIOTTI&BERTIN99} is a constant that depends on \textit{n}, $r_e$ is the effective radius, and $I_0$ is the effective surface brightness in flux units. The three parameters $n,r_e,I_0$ were left free to vary. The fit returned a value for the effective radius of 1.62$''$, corresponding to 3 kpc at $z$ = 0.097 and leading to a normalized projected offset of the host galaxy from the afterglow position of 2.32. In the end, we obtained a probability of chance association $P_{cc} = $ 5.4\%: no firm conclusion can be drawn about GRB\,211106A host galaxy from this result.\\
Hence, we investigated the source detected in our FORS2 images. Given the compatibility of their positions and the lack of significant fading among the three FORS2/$R$ band observations and the \textit{HST} ones (\S \ref{sec:211106_aftgl}), we can safely assume that they are the same source. By adopting $R$ = 26.51 mag (epoch 3 result, see \S \ref{211106A/FORS}) and an angular offset of 0.113$''$ from the best afterglow position, we obtained $P_{cc}$ $\sim$ 0.1$\%$. Unfortunately, the source was not resolved enough in our images in order to perform surface photometry, therefore we could not compute the projected offset in kpc and its normalized value. 
In light of the values obtained for the chance association probability, the best host galaxy candidate is the source detected by \textit{HST} and FORS2, as also proposed by \cite{Laskar22}.
\begin{comment}
Given the faintness and the small angular separation of the source with respect to the bright extended galaxy, we considered at first a small satellite galaxy nature for this object. Therefore, we computed the absolute magnitude in the $R$ band from $z$=0.097: $M_R = -$11.87. We compared this value to typical optical magnitudes of a known globular cluster and a dwarf galaxy: $\omega$-Centauri and the Small Magellanic Cloud (SMC). We retrieved their absolute magnitudes in the $V$ band $M_V = -$8.45 (from \textit{CDS}\footnote{\url{https://cds.u-strasbg.fr/}}) and $-$17.02 (from \textit{NED}\footnote{\url{http://ned.ipac.caltech.edu/}}), respectively, hence the source, if at $z$ = 0.097, would be too bright to be a globular cluster and too faint to be a dwarf galaxy.
\end{comment}
\begin{table}[]
    \centering
    \caption{GRB\,211106A host galaxy magnitudes.}
    \begin{tabular}{c|c|c|c|c}
    \hline
    \noalign{\smallskip}
    Filter & $\lambda_{0}$ [$\AA$] & FWHM [$\AA$] & mag & $\sigma_{mag}$ \\
    \noalign{\smallskip}
    \hline
    \hline
    \noalign{\smallskip}
    %$u$ & 3556.5 & 582.3 & 22.98 & 0.97 \\
    %$g$ & 4702.5 & 1262.7 & 20.57 & 0.04\\
    %$r$ & 6175.6 & 1149.5 & 19.40 & 0.03 \\
    %$i$ & 7490 & 1239 & 18.85 & 0.03 \\
    %$z$ & 8946.7  & 994.4 & 18.80 & 0.09 \\
    %$J$ & 12520 & 1720 & 18.30 & - \\
    %$K$ & 21470 & 3090 & 16.91 & - \\
    MUSE/$V$ & 5439.27 & 852.44 & 27.42 & 0.08 \\
    MUSE/$R$ & 6436.47 & 1567.06 & 26.45 & 0.10\\
    FORS2/$R$ & 6521.67 & 1645.55 & 26.51 & 0.23 \\
    MUSE/$I$ & 8054.53 & 1543.11 & 25.67 & 0.26 \\
    $HST$/$F814W$ & 8102.91  & 2098.15 & 25.79 & 0.07 \\
    $HST$/$F110W$ & 11696.34 & 4637.84 & 25.71 & 0.02 \\
    \noalign{\smallskip}
    \hline
    \end{tabular}
    \tablefoot{Observed magnitudes for the host galaxy of GRB\,211106A.}
    \label{sedobs1}
\end{table}\\
To further investigate the proposed host galaxy, we built a spectral energy distribution (SED) aimed at getting a photometric redshift estimation. For this purpose, we considered \textit{HST} magnitudes obtained with the ${F814W}$, ${F110W}$ filters, the VLT/FORS2 ${R}$-band magnitude, and the VLT/MUSE corresponding magnitudes in the $V$, $R$, $I$ filters (see Table \ref{sedobs1}).\\
To perform the SED fit we followed a similar approach to the one described in the recent work by \cite{Nugentetal22}. We used the Python package for inference of stellar population \texttt{Prospector} \citep{proben}. Within this package, we used \texttt{emcee} affine-invariant ensemble sampler for Monte Carlo Markov-Chain \citep{emcee} in order to sample the posterior distributions for the parameters in our analysis, while the spectral templates were built through the \textit{Flexible Stellar Population Synthesis} (FSPS) package (\citealt{conr1}; \citealt{conr2}) by making use of \texttt{MIST} \citep{MIST} and \texttt{MILES} \citep{MILES} stellar population synthesis libraries.\\
We performed the fit with a Chabrier initial mass function (IMF, \citealt{Chabrier03}), a fixed dust emission contribution from \cite{DeL}, and a parametric \textit{delay-$\tau$} star formation history (SFH). We included the intrinsic host attenuation as a SMC-like extinction in terms of optical depth in the $V$ band $\tau_{V}$, from which we derived $A_{V}$ from the following relation: $A_{V}\simeq1.086\tau_{V}$.\\
The free parameters considered in the model were redshift, age, stellar mass, the fraction of surviving mass at the age of the host, the metallicity of the galaxy, and the  SFH \textit{e}-folding factor. The best fit, along with the host galaxy observed magnitudes, are shown in Fig. \ref{sed21}, and a list of the results for the parameter space is reported in Table \ref{sedpar21}.
\begin{table}[]
    \caption{GRB\,211106A host galaxy SED fitting results.}
    \centering
    \begin{tabular}{c|c}
    \hline
    \noalign{\smallskip}
    Parameter & posterior median value \\
    \noalign{\smallskip}
    \hline
    \hline
    \noalign{\smallskip}
    %stellar mass $M_*$ [10$^{10}$ \(M_\odot\)] & 3.56$_{-0.86}^{+1.02}$\\
    %extinction $A_V$ [mag] & 0.38$_{-0.20}^{+0.25}$\\
    %metallicity $Z$ [\(Z_\odot\)] & 0.85$_{-0.33}^{+0.47}$\\
    %age $t_{age}$ [Gyrs] & 4.42$_{-1.78}^{+3.87}$\\
    %folding factor $\tau$ [Gyrs] & 0.65$_{-0.43}^{+0.76}$\\
    redshift $z$  & 0.64$_{-0.05}^{+0.05}$\\
    stellar mass $M_*$ [10$^{8}$ \(M_\odot\)] & 1.95$_{-0.21}^{+0.24}$\\
    extinction $A_V$ [mag] & 0.19$_{-0.13}^{+0.23}$\\
    metallicity $log(Z/$\(Z_\odot\)) & 0.16$_{-0.50}^{+0.48}$\\
    age $t_{age}$ [Gyrs] & 0.56$_{-0.21}^{+0.27}$\\
    folding factor $\tau$ [Gyrs] & 0.04$_{-0.03}^{+0.05}$\\
    survived mass fraction $m_{frac}$ & 0.73$_{-0.02}^{+0.02}$\\
    \noalign{\smallskip}
    \hline
    \end{tabular}
    \tablefoot{Median values (50\% quantile) of the posterior distribution for the parameter set described in the text. Quoted errors represent the 15.9\% and 84.1\% quantiles, which on the marginalized posterior distribution over each parameter represents the 1$\sigma$ error.}
    \label{sedpar21}
\end{table}\\
We compared the results of our SED analysis with common features of a wide range of SGRB hosts discussed by \cite{Nugentetal22}. We found that GRB\,211106A host galaxy does not show unusual features with respect to the median values of typical SGRB hosts and their 68\% credible intervals regarding age, metallicity, and extinction. However, we noted that the inferred mass is about one order of magnitude lower than the considered median value, though consistent at $2\sigma$ level. Most likely, this can be justified by the bias to detect more easily brighter, usually more massive, galaxies, while the magnitudes from FORS2 and \textit{HST} observations are very close to the instrumental limits. Furthermore, we observed that the $A_V$ we obtained is significantly lower than the ones reported in \cite{Laskar22} from its best-fit synchrotron model to the afterglow observations (namely in the 4.65$-$5.62 mag range). This discrepancy could be due to the burst occurring in a particularly dense host environment and thus not representative for the host galaxy global properties.\\  
The SED fit enabled us to infer a redshift $z = 0.64_{-0.05}^{+0.05}$ (consistent with the redshift range estimated by \citealt{Laskar22}), which will be used for discussion in the following analysis.

\subsection{GRB\,211227A host galaxy analysis}\label{sec:211227_HG}

%----------------------------------------------------- METALLICITY DIAGNOSTIC
\begin{table}[]
    \centering
    \caption{GRB\,211227A host galaxy metallicity indicators.}
    \begin{tabular}{l|c|c}
    \hline
    \noalign{\smallskip}
    Diagnostic & Value & 12 + log(O/H) \\
    \noalign{\smallskip}
    \hline
    \hline
    \noalign{\smallskip}
    $F$[N II$\lambda$6583]/$F$[H$_\alpha$] & 0.74$\pm$0.03 & >9.25\\
    $F$[O III$\lambda$5007]/$F$[N II$\lambda$6583] & 0.19$\pm$0.05 & [9.15-9.25]\\
    $F$[N II$\lambda$6583]/$F$[O II$\lambda$3727] & 0.46$\pm$0.02 & [8.95-9.05]\\
    $F$[N II$\lambda$6583]/$F$[S II$\lambda$6718] & 6.32$\pm$0.29 & >9.25\\
    $F$[O III$\lambda$5007]/$F$[OII$\lambda$3727] & 0.09$\pm$0.02& >9.25\\
    $F$[O III$\lambda$5007]/$F$[H$_\beta$] & 0.42$\pm$0.11& [8.95-9.05]\\
    \noalign{\smallskip}
    \hline
    \end{tabular}
    \tablefoot{Value ranges obtained for some oxygen abundance indicators derived from emission lines fluxes.}
    \label{metallicity}
\end{table}

\begin{figure} 
\centering 
\includegraphics[width=9cm]{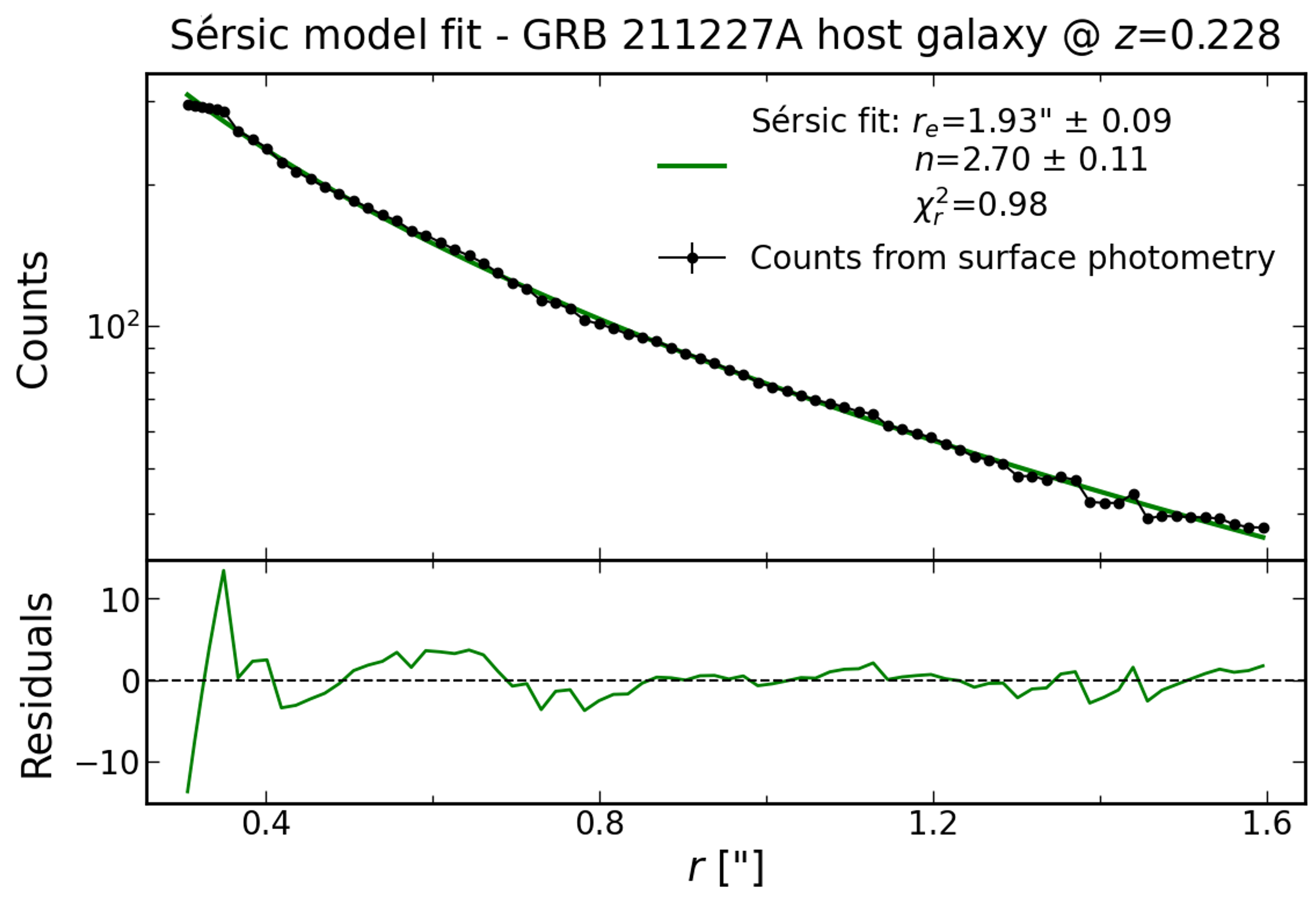} \caption{Fit of the profile of the GRB\,211227A candidate host galaxy at $z$ = 0.228. The effective radius is 1.93$''$, corresponding to 7.26 kpc at that distance.} 
\label{sersic-fit} 
\end{figure}
\begin{figure}[]
    \centering
    \includegraphics[width=9cm]{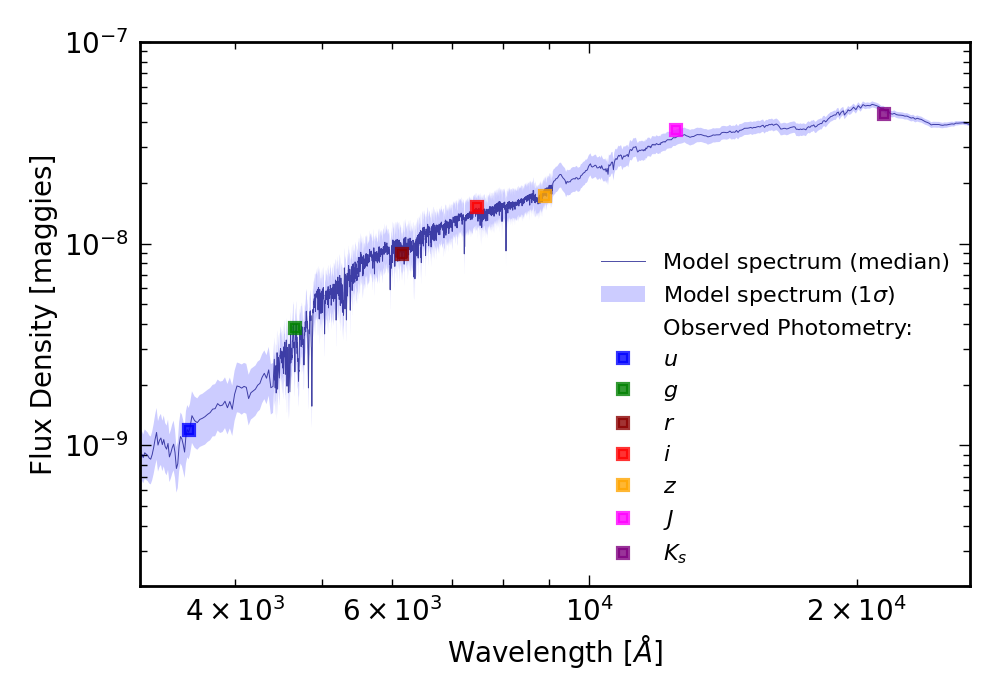}
    \caption{Best fit for GRB\,211227A host galaxy SED from the median values of the posterior distribution parameters.}
    \label{sed}
\end{figure}

\begin{figure}[]
    \centering
    \includegraphics[width=9cm]{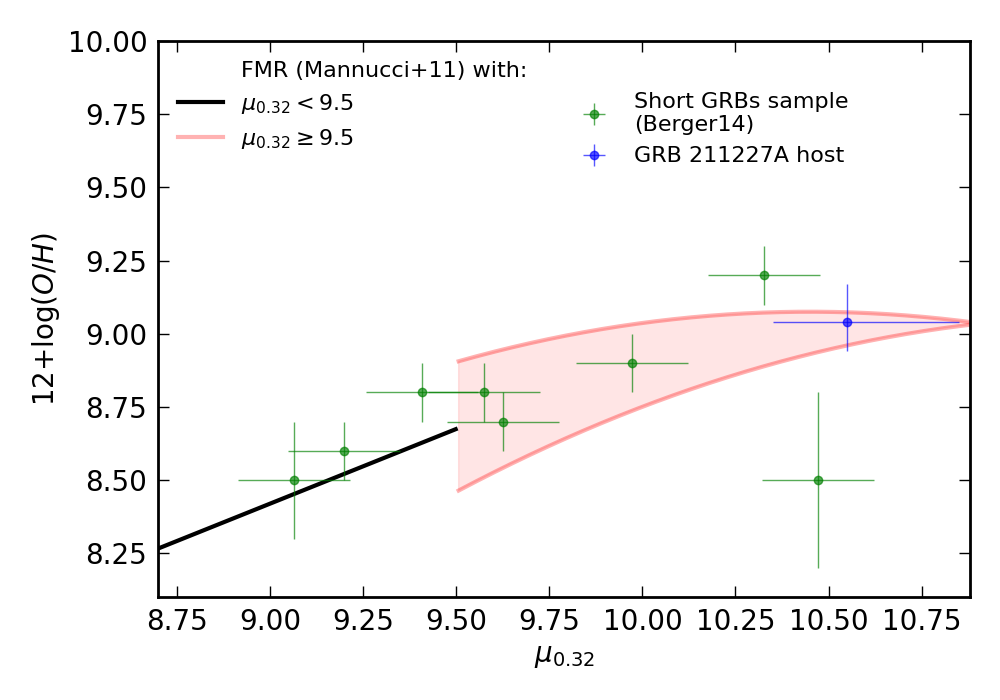}
    \caption{FMR by \cite{Man11} and SGRB hosts with determined SFR, mass, and oxygen abundance from \cite{Berger14}. GRB\,211227A host is reported for the value of SFR computed from the SED analysis, together with the FMR for a range of SFR consistent with the 1$\sigma$ uncertanties on the same SFR value.}
    \label{fmr}
\end{figure}

 
The association of GRB\,211227A with the host galaxy is less complex than for GRB\,211106A. As shown in \S \ref{GRB211227Aaft}, the XRT error circle lies 3.67$''$ away from a $z$ = 0.228 galaxy with $r$-band magnitude 19.4. A fit with a Sérsic model (see Fig. \ref{sersic-fit}) yielded an effective radius $r_{\text{eff}}=1.93''$, corresponding to 7.26 kpc at $z$ = 0.228, from which we obtained a normalized offset of 1.90. The corresponding chance association probability was thus $P_{cc}=$ 1.3\%, suggesting that the source we analysed is probably the host galaxy of GRB\,211227A. From the analysis of spectroscopic observation of this source, we identified several emission lines, all at a common redshift $z$ = 0.228 (see Table \ref{lines_z}), in agreement with \cite{GRB211227A_GCN_XSHOOTER}. We also detected, at the same redshift, the absorption lines corresponding to Ca II $\lambda$3933/3968 and to the Na I D doublet ($\lambda$5890/5896).\\
We focused our analysis on the detected emission lines, for which we computed the fluxes and corresponding luminosities presented in Table \ref{lines_z} in order to characterise the host galaxy. First, by matching photometric observations to our spectral data, we inferred a mean slit-loss correction factor of 2.91. We have also taken into account the extinction, both in the Milky Way and in the host galaxy. For the former, we obtained $A_V = R_V\times E(B-V)=0.06$ mag, where we considered the typical value for our galaxy of $R_V=3.10$ and $E(B-V)$ = 0.02 \citep{Schelgeletal98}. Then, following \cite{Pei}, we derived the value from the Milky Way extinction curve corresponding to the central wavelength of each detected emission lines. We estimated the attenuation in the host galaxy from the flux ratio of H$_{\alpha}$ and H$_{\beta}$ emission line fluxes, the ratio of which is known as the Balmer decrement. The intrinsic, dust corrected ratio is $r_{int} = 2.86$, assuming a case B recombination, in the low-density limit and electron temperature of 10$^4$ K \citep{Osterbrock89}. We observed a Balmer decrement of $r_{obs} = 4.18 \pm 0.20$, from which we derived:
\begin{eqnarray}
A_V = \frac{-2.5\log{(r_{obs}/r_{int}})}{A_{H_{\alpha}}-A_{H_{\beta}}} = 1.06 \pm 0.13 \text{ mag},
\label{eqA}
\end{eqnarray}
where the attenuation at the wavelength of $H_{\alpha}$ and $H_{\beta}$ is A$_{H_{\alpha}}=0.79$ and A$_{H_{\beta}}=1.18$. We obtained these values using a SMC-type extinction curve \citep[from][]{Pei} with $R_V=2.93$.\\
In order to examine the galaxy nature, we computed its position in the Baldwin-Phillips-Terlevich (BPT, \citealt{BPT}) diagram by means of two emission lines ratios: [NII]$\lambda$6583/H$_\alpha$ and [OIII]$\lambda$5007/H$_\beta$. Taking into account the diagram for the Sloan Digital Sky Survey galaxies reported in \cite{Kew13}, we found that GRB\,211227A putative host galaxy lies in a \textit{mixing sequence} zone, slightly outside the \textit{Star-forming abundance sequence} region. Moreover, the galaxy is placed below the clear bifurcation between star forming galaxies and AGNs, therefore it is difficult to distinguish how much contribution to the emission lines is given by the gas ionized by hot stars or non-thermal sources such as AGN and/or shocks (for a detailed discussion on this topic, see \citealt{Kew13}). Nevertheless, keeping this caveat in mind, we investigated the star formation rate (SFR) and metallicity of the galaxy, assuming the star forming component to be the leading ionizing power actor over a possible AGN one.\\ 
We measured for the [O II] emission line a luminosity of $(1.36\pm0.12)\times$10$^{41}$ erg/s, corresponding to a SFR of $1.90\pm0.57$ $M_{\odot}$ yr$^{-1}$ \citep[derived following][]{Kennicutt98}; we also measured a luminosity for the H$_\alpha$ emission line of $(8.42\pm0.41)\times$10$^{40}$ erg/s, which translated into a SFR of $0.67\pm0.03$ $M_{\odot}$ yr$^{-1}$.
Considering a rest frame $B$ band absolute magnitude $M_B=-20.36$, extracted from the host spectrum, corresponding to a luminosity $L_B\simeq0.55L_*$ (assuming the absolute $B$ band magnitude of field galaxies to be $M_*=-21$ as done by \citealt{D'Avanzo09}),
the specific SFR turned out to be $3.45\pm0.88$ and $1.22\pm0.05$ $M_{\odot}$ yr$^{-1}L_*^{-1}$ for the [O II] and the H$_\alpha$ emission line, respectively.\\
In order to estimate the metallicity, following \cite{Nagaoetal06}, we computed the ratio of the emission lines indicated in Table \ref{lines_z} and derived the corresponding  ranges of oxygen abundance. The results are presented in Table \ref{metallicity}. The discrepancy of the values obtained could be due to the non-negligible contribution to the lines from AGN and/or shocks, especially regarding the high [N II]$\lambda$6583 flux, while another explanation is the different absolute scale differences between individual strong-line diagnostics \citep{Kewley08}. Nevertheless, we noticed that the indicators tend to favor a super-solar metallicity (taking 12+log(O/H)=8.69 as solar metallicity reference, from \citealt{Asplundetal09}).
\begin{table}[]
    \centering
    \caption{GRB\,211227A host galaxy magnitudes.}
    \begin{tabular}{c|c|c|c|c}
    \hline
    \noalign{\smallskip}
    Filter & $\lambda_{0}$ [$\AA$] & FWHM [$\AA$] & mag & $\sigma_{mag}$ \\
    \noalign{\smallskip}
    \hline
    \hline
    \noalign{\smallskip}
    %$u$ & 3556.5 & 582.3 & 22.98 & 0.97 \\
    %$g$ & 4702.5 & 1262.7 & 20.57 & 0.04\\
    %$r$ & 6175.6 & 1149.5 & 19.40 & 0.03 \\
    %$i$ & 7490 & 1239 & 18.85 & 0.03 \\
    %$z$ & 8946.7  & 994.4 & 18.80 & 0.09 \\
    %$J$ & 12520 & 1720 & 18.30 & - \\
    %$K$ & 21470 & 3090 & 16.91 & - \\

    $u$ & 3821.47 & 600.87 & 22.314 & 0.015 \\
    $g$ & 4874.36 & 1452.65 & 21.050 & 0.003\\
    $r$ & 6268.45 & 1241.21 & 20.123 & 0.003 \\
    $i$ & 7619.31 & 1588.68 & 19.549 & 0.002 \\
    $z$ & 8661.19  & 921.69 & 19.415 & 0.004 \\
    $J$ & 12520 & 1720 & 18.596 & 0.145 \\
    $K$ & 21470 & 3090 & 18.388 & 0.170 \\
    \noalign{\smallskip}
    \hline
    \end{tabular}
    \tablefoot{Catalogue magnitudes for the host galaxy of GRB\,211227A.}
    \label{sedobs}
\end{table}\\
To further investigate the nature of the host galaxy, we performed a SED fit: we collected $u$, $g$, $r$, $i$ and $z$ magnitudes from the CFHTLS survey \citep{CFHTLS}, and $J$, $K$ infrared magnitudes from the VISTA VHS DR5 \citep{VISTA_VHS} catalogue (see Table \ref{sedobs}). To perform the SED fit we followed the same procedure, and used the same spectral templates, described in \S \ref{sec:211106_HG}. However, differently from the SED fit for GRB\,211106A host, we fixed the redshift to the value derived from the spectrum: $z$=0.228. The best fit, along with the host galaxy catalogue magnitudes, are shown in Fig. \ref{sed}, and a list of the results for the parameter space is reported in Table \ref{sedpar}.
\begin{table}[]
    \caption{GRB\,211227A host galaxy SED fitting results.}
    \centering
    \begin{tabular}{c|c}
    \hline
    \noalign{\smallskip}
    Parameter & posterior median value \\
    \noalign{\smallskip}
    \hline
    \hline
    \noalign{\smallskip}
    %stellar mass $M_*$ [10$^{10}$ \(M_\odot\)] & 3.56$_{-0.86}^{+1.02}$\\
    %extinction $A_V$ [mag] & 0.38$_{-0.20}^{+0.25}$\\
    %metallicity $Z$ [\(Z_\odot\)] & 0.85$_{-0.33}^{+0.47}$\\
    %age $t_{age}$ [Gyrs] & 4.42$_{-1.78}^{+3.87}$\\
    %folding factor $\tau$ [Gyrs] & 0.65$_{-0.43}^{+0.76}$\\
    stellar mass $M_*$ [10$^{10}$ \(M_\odot\)] & 3.42$_{-0.22}^{+0.28}$\\
    extinction $A_V$ [mag] & 0.19$_{-0.11}^{+0.10}$\\
    metallicity $log(Z/$\(Z_\odot\)) & 0.35$_{-0.10}^{+0.13}$\\
    age $t_{age}$ [Gyrs] & 5.64$_{-0.68}^{+0.86}$\\
    folding factor $\tau$ [Gyrs] & 0.95$_{-0.12}^{+0.15}$\\
    survived mass fraction $m_{frac}$ & 0.64$_{-0.01}^{+0.01}$\\
    \noalign{\smallskip}
    \hline
    \end{tabular}
    \tablefoot{Median values of the posterior distribution for the parameter set described in the text with $1\sigma$ errors.}
    \label{sedpar}
\end{table}\\
Regarding the metallicity, we computed an oxygen abundance 12+log(O/H)=9.04$_{-0.10}^{+0.13}$, fairly super-solar, to some extent in agreement with what is suggested from the metallicity ranges in Table \ref{metallicity}.
\begin{comment}
the extinction in the $V$ band is consistent with the value computed from Eq. \ref{eqA} at 1$\sigma$ level.     
...    
However, it is in agreement at $1\sigma$ level with the mean value of the three indicators shown in Table \ref{metallicity}: 9.09$^{+0.7}_{-0.4}$.
\end{comment}
To frame out the stellar mass of the galaxy, we took into account the relation between stellar mass and absolute $K$ band magnitude proposed in \cite{Savaglio+09} for a sample of GRB host galaxies. For the host of GRB\,211227A we computed an absolute magnitude $M_{K}=-21.95\pm0.17$, corresponding to a stellar mass of $M_*^\text{Sav}=(1.18\pm0.22)\times$10$^{10}$\(M_\odot\). The median of the posterior distribution on the stellar mass $M_*$ is slightly higher than the value inferred from the $K$ band absolute magnitude and consistent at 2$\sigma$ level with the dispersion of the sample, pointing towards the goodness of our result.\\
Besides, we were able to estimate the SFR from the \textit{delay-$\tau$} SFH, the age of the host, and the total mass formed at $t_{age}$: $M_{tot}=M_*/m_{frac}$. From the median posterior values we derived:
\begin{eqnarray}
\text{SFR}(t_{age})=\dfrac{M_{tot}}{\int_0^{t_{age}} te^{-t/\tau}dt}t_{age}e^{-t_{age}/\tau} = 0.84^{+2.17}_{-0.69} \mathrm{M_\odot\,yr^{-1}},
\label{eqSFR}
\end{eqnarray}
which, given a mean conversion factor of $\sim1.7$ in order to account for the different IMF assumed by the relations in \cite{Kennicutt98} (i.e., Salpeter IMF, \citealt{Salp}) with respect to the Chabrier IMF from the SED analysis, is consistent with the values estimated from the [O\,II] and H$_\alpha$ line luminosities due to the wide 1$\sigma$ uncertanties derived from $t_{age}$ and $\tau$, and consequently, making it difficult to discriminate the contribution of AGN and/or shocks on the [O\,II] and H$_\alpha$ lines.\\ 
We considered the results of our SED analysis as a reference to frame out GRB\,211227A host galaxy properties with respect to the sample by \cite{Nugentetal22}. We found that GRB\,211227A host galaxy is in good agreement at 1$\sigma$ c.l. with median values of typical SGRB hosts regarding stellar mass, SFR, and extinction, and 2$\sigma$ consistent with respect to usual metallicities. It is interesting to note that the host appears to be more massive than the average lying in the upper tail of the sample distribution. This fact, coupled to a SFR ranging from moderate to substantial values, suggests a transitioning nature for the host, with a mass typical of quiescent galaxies but a SFR more consistent with star-forming hosts.\\
Since GRB\,211227A shows an extended emission, we also compared its host galaxy properties to the typical values for LGRB hosts, which turn out to have a lower metallicity (median 12+log(O/H) $\simeq$ 8.3, \citealt{Staneketal06,Modjazetal08,Levesqueetal10, Verganietal15_Long1, Japeljetal16_Long2, Palmerioetal19_Long3}), lower masses (median M$_*^{\text{LGRB}}\simeq$ 1.58$\times$10$^{9}$ M$_\odot$, \citealt{Savaglio+09, z_080123, Verganietal15_Long1, Japeljetal16_Long2, Palmerioetal19_Long3}), and they are younger and star-forming \citep{BLOOM02, Wain, Verganietal15_Long1, Japeljetal16_Long2, Palmerioetal19_Long3}. This suggests that a LGRB host nature is disfavoured for the galaxy. As a final gauge, we explored the fundamental metallicity relation (FMR, \citealt{Mannuccietal10}; \citealt{Man11}), tested for the first time on SGRBs hosts by \cite{Berger14}, between the oxygen abundance and the quantity $\mu_{0.32}$=log($M_{*}$)-\,0.32$\times$log(SFR). We made use of metallicity, mass, and SFR values derived from the SED fit, leading to $10.55^{+0.29}_{-0.20}$. 
A plot of our results is shown in Fig. \ref{fmr}. The value is consistent with the FMR, supporting, despite the small sample and the possible lines contamination, the picture of SGRB hosts following the relation for field galaxies along a wide range of stellar mass and SFR values, and constraining it in a region of poor sampling.

\subsection{Afterglow emission}\label{sec:AG_emission}
After the analysis of the host galaxies associated with the two bursts, we investigated their optical afterglow emission with respect to their X-ray emission. We computed the X-ray to optical slope and we compared them with the population of Long GRBs analysed in the same way by \citet{Melandrietal12}. To do this, we estimated the flux both in the X-rays and in the optical band ($R$ filter) at $t_{11}$ = 11 h in order to compute the spectral index $\beta_{OX}$. Moreover, we computed the X-ray spectral index $\beta_X$ from the late-time spectrum fit, defined as $\beta_X = \Gamma_X$ - 1, where $\Gamma_X$ is the photon index obtained from the X-ray spectral fit.

\subsubsection{GRB\,211106A}
Starting from the X-ray light curve fit for GRB\,211106A (\S \ref{sec:211106A_XRT}), we estimated the X-ray flux at $t_{11}$ = 11 h, that is $F_{X,\text{11 hr}}$ = 0.05 $\mu$Jy, while for the optical afterglow we only had an upper limit of $R$ = 26.7 mag at $t-t_0$ = 2.82 d. We assumed that, at such a late time, the optical afterglow follows a similar decay with respect to the X-ray afterglow, therefore we estimated the optical limit at $t_{11}$ assuming a decay index $\alpha_X = 1.07$, and we obtained $F_{R,\text{11 hr}} < 0.53$ $\mu$Jy. The resulting limit for the X-ray to optical slope turns out to be $\beta_{OX} \leq$ 0.3. Moreover, we computed the value for $\beta_X$ starting from the joint spectral fit of \textit{Swift}/XRT, \textit{Chandra} and \textit{XMM-Newton} data performed by \citet{Laskar22}, giving $\Gamma_X = 1.9 \pm 0.3$. Thus, $\beta_X = 0.9$ and $\beta_{OX} - \beta_X \leq -0.6$. According to the definition of a ``dark'' GRB \citep{Jakobssonetal04, VanderHorstetal09, Melandrietal12}, $\beta_{OX} - \beta_X \leq - 0.5$, we can classify GRB\,211106A as a dark burst. The darkness of a burst was found to be related to high redshift or large dust column density for Long GRBs \citep{Melandrietal12}, while for Short GRBs it may often be a consequence of the intrinsic optical faintness of the afterglow due to a low density circumburst medium \citep[see][]{Asquinietal19}. GRB\,211106A shows a relatively high $N_\text{H,int}$ \citep{Laskar22}, therefore intrinsic absorption can explain its darkness (for a recent comprehensive work on short GRBs darkness, see \citealt{Gobat+23}).

\subsubsection{GRB\,211227A}
We repeated the same procedure for GRB\,211227A, and we estimated the X-ray flux $F_{X,\text{11 hr}}$ = 0.005 $\mu$Jy, while for the optical afterglow we considered again the closest limit, that is $r$ = 25.4 mag, obtained at $t-t_0$ = 0.211 d ($\sim$5 hours after $t_0$). In this case we fitted the X-ray light curve with a broken power law, and since both our limit and $t_{11}$ lie on the same branch of the broken power law (the last one), we estimated the optical limit at 11 hr assuming a decay index $\alpha_X = 0.5$: $F_{r,\text{11 hr}} < 0.17 \mu$Jy. The corresponding limit for the X-ray to optical slope turns out to be $\beta_{OX} \leq$ 0.5. We retrieved the X-ray spectral index from the \textit{Swift}/XRT spectrum repository, and we obtained $\beta_X = 2.0$, from which we derived $\beta_{OX} - \beta_X \leq -1.5$. Also in this case, GRB\,211227A fulfills the condition for a ``dark'' GRB: $\beta_{OX} - \beta_X \leq -0.5$. At variance with GRB\,211106A, this burst shows a medium $N_\text{H,int}$ ($\sim 10^{21}$ cm$^{-2}$, see \S\ref{sec:211227_XRT}), therefore it is more likely that the darkness of the burst is due to a tenuous circumburst medium.

\subsection{Kilonova limits}\label{sec:KN_lim}
In this section we look for possible evidence of kilonova emission associated with the two GRBs presented in this work. The procedure adopted here follows the one described by \citet{ROSSI20-KN}: first, we put AT2017gfo spectra taken with a $\sim$daily cadence, from 1.5 to 10.5 days after the trigger, from WISeREP repository\footnote{ \url{https://www.wiserep.org/search/spectra}}, in the rest frame.  Then, in order to compare the AT2017gfo and GRB luminosity at the same wavelength, we computed an effective rest frame filter, centered at $\lambda_{X,\text{ eff}} = \lambda_X/(1+z_{\text{GRB}})$, where $X$ is the band corresponding to the filter used for GRB observations. After integrating the AT2017gfo luminosity spectra taken at different epochs over the effective rest frame filter, we built the AT2017gfo luminosity light curves in each rest frame filter, $L_{X,\text{ eff}} (t)$. We applied this procedure to all filters adopted for our observations for both GRBs.

\subsubsection{GRB\,211106A}\label{sec:KN_lim211106A}
 
\begin{figure}[]
    \centering
    \includegraphics[width=9cm]{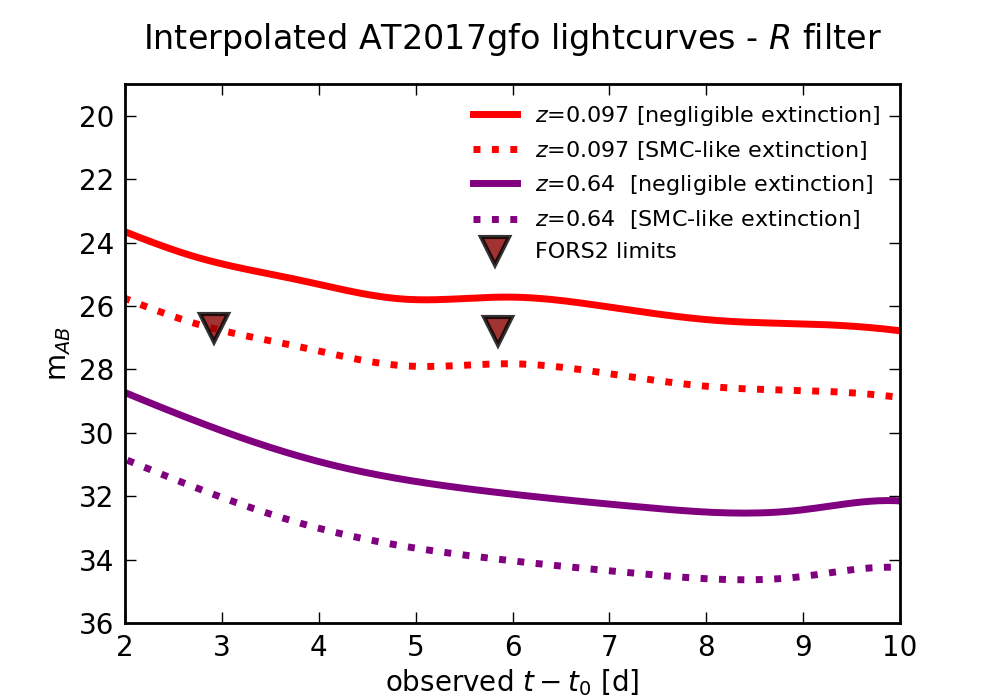}
    \caption{Interpolated $R$ filter magnitudes for AT2017gfo at the GRB\,211106A luminosity distances corresponding to $z$ = 0.097 and $z$ = 0.64. The light curves considering both no local extinction and $A_V$=2.6 \citep[from ][]{Laskar22} are shown.}
    \label{FORSint}
\end{figure}

\begin{figure}[]
    \centering
    \includegraphics[width=9cm]{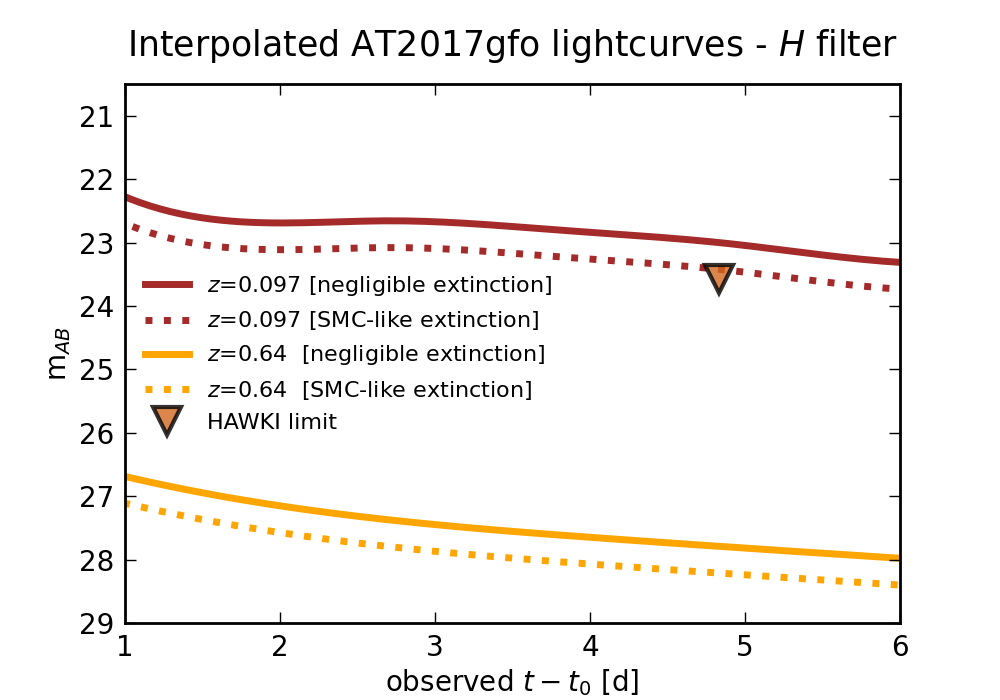}
    \caption{Same as Fig. \ref{FORSint}, but for the HAWK-I $H$-band observation of GRB\,211106A.}
    \label{HAWKint}
\end{figure}

\begin{figure*}
    \centering
    \begin{subfigure}[b]{.49\textwidth}
        \includegraphics[width=1.\textwidth]{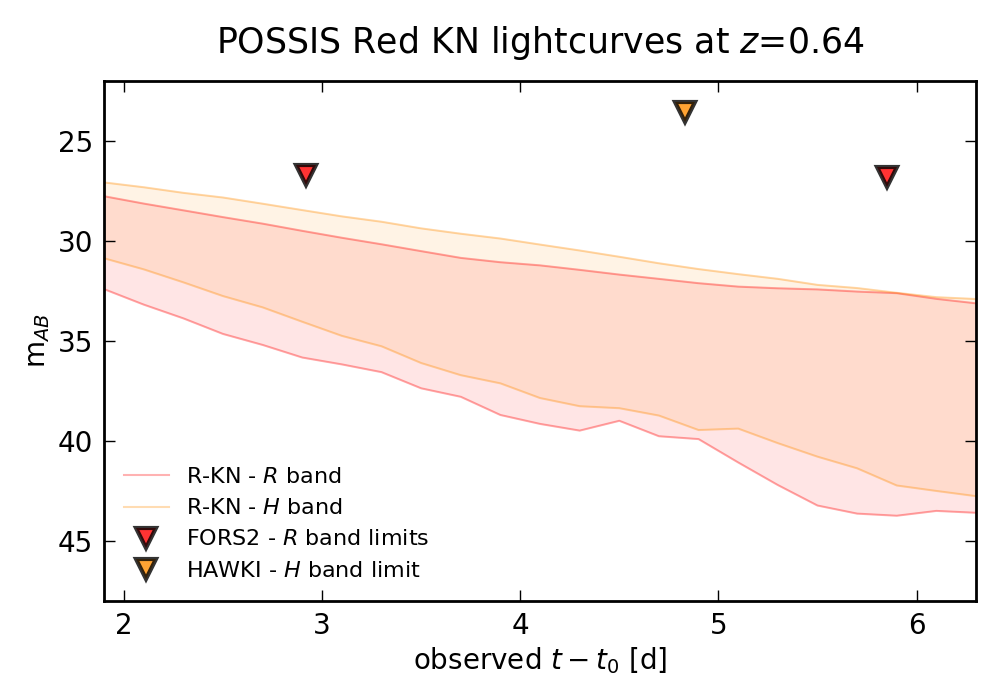}
    \end{subfigure}
    \hfill
    \begin{subfigure}[b]{.49\textwidth}
        \includegraphics[width=1.\textwidth]{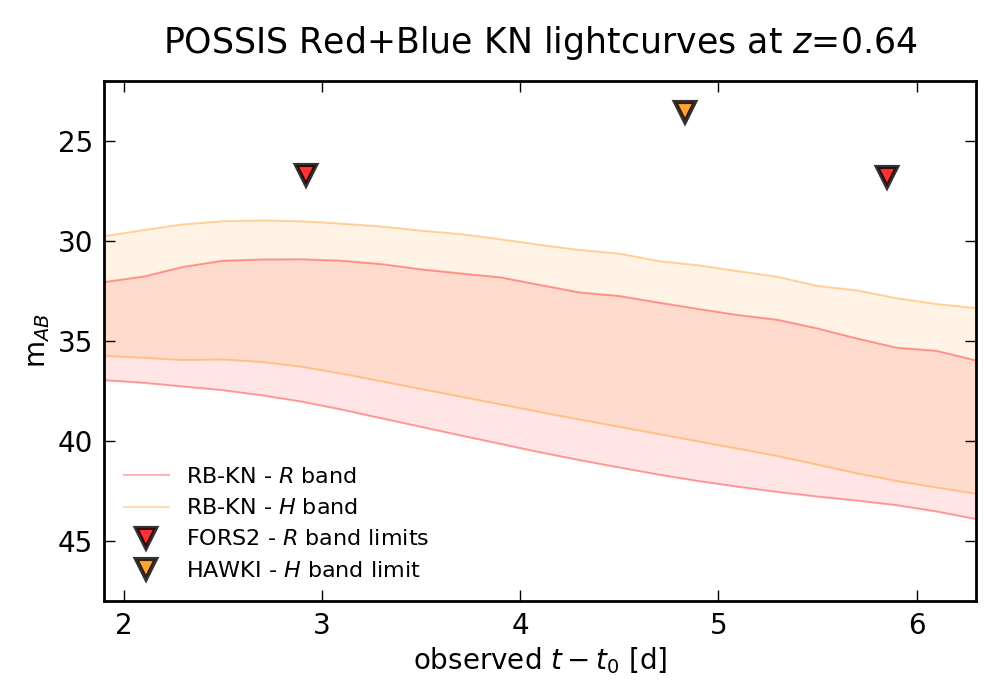}
    \end{subfigure}
    \caption{Simulated light curves of 81 "Red" kilonovae (left panel) and 140 "Red+Blue" kilonovae (right panel) configurations, in the $R$ and $H$ filters at $z=$ 0.64, as seen for an on-axis observer. All the models lie below our photometric limits for GRB\,211106A.}
    \label{bulla211106A}
\end{figure*}

We show in Fig. \ref{FORSint} and Fig. \ref{HAWKint} the light curves - built as described above - in $R$, $H$ bands, and the magnitude limits on our observations performed with FORS2 and HAWK-I instruments (see \S \ref{Obs&analysis211106A}). We investigated the kilonova detectability for $z$=0.097 - as a reference - and the photometric $z$=0.64, both assuming zero and non-negligible extinction. For the latter, we considered a coefficient $A_V=$ 2.6 mag, which was found as a lower limit from the afterglow analysis by \cite{Laskar22}, derived taking into account a SMC-like extinction model. Following \cite{Pei}, and taking into account the same model, we derived $A_R$=2.10 mag and $A_H$=0.42 mag for our filters, accounting also for the Milky Way extinction along the line of sight\footnote{ \url{https://ned.ipac.caltech.edu/extinction_calculator}}. We estimated the expected magnitude in FORS2 epochs 1 and 2 by interpolating the light curve, while the third epoch was taken far later than the last AT2017gfo observation. At $z$=0.097, the KN emission would have become fainter than the mean magnitude limit of the three epochs in $R$-band $\sim$9 d after $t_0$ in case of negligible extinction and slightly before the time of our first epoch in the absorbed case.\\
For negligible absorption, in epoch 1 and 2 an AT2017gfo-like event would have been $\sim$1-2 magnitudes brighter than our limits, thus it would have been strongly detectable. On the other hand, in the case of high extinction it would have been more challenging to identify the KN in epoch 1 given an extrapolated magnitude comparable to the magnitude limit, while it would have been more than 1 magnitude below our threshold for epoch 2.\\
By placing our AT2017gfo-like light curve template at increasing values for $z$, we found that the KN would have been undetectable even in epoch 1 for $z\geq0.22$ in the cases of zero extinction.
\\
The same analysis performed for the HAWK-I $H$ band showed that, at the time of the observation, an AT2017gfo-like KN at $z$=0.097 would have been about 0.6 and 0.2 mag brighter with respect to our limit, in the negligible and high extinction case, respectively. Here, the redshift threshold for the detection is $z $=0.13 and $z $=0.11 for the two cases.\\
Hence, the absence of a KN identification supports a GRB distance larger than $z$ =0.097. Such a conclusion is in agreement with the photometric redshift $z$=0.64 we derived for the GRB\,211106A host galaxy and with the $z\sim$0.5-1.0 redshift range estimated by \cite{Laskar22}. At such a redshift, the expected light curves would have been significantly fainter than our magnitude limit in both $H$ and $R$ filters.\\
However, the KN non-detection could be ascribed also to an event with different characteristics with respect to AT2017gfo, since KNe associated to short GRBs can actually display significant differences in their luminosity, spectral properties, and temporal evolution (see, e.g., \citealt{ROSSI20-KN}). To further investigate the constraints on KN emission provided by our optical/NIR upper limits, we made use of multidimensional models of kilonovae computed with the time-dependent 3D Monte Carlo code \texttt{POSSIS}\footnote{\url{https://github.com/mbulla/kilonova_models}}, by \cite{Bulla19}. In particular, we focused on two classes of models: those presented by \cite{Anand}, taylored for NS-BH mergers, and the ones from \cite{Diet} for NS-NS mergers. The former allowed us to compute light curves for the case of a "Red" kilonova (R-KN) with a tidal dynamical ejecta contribution of lanthanide-rich elements only, while the latter provided us the case of a "Red+Blue" kilonova (RB-KN) with also a polar lanthanide-free component. Both models include a contribution of spherical disk wind ejecta. Therefore, we could test different shapes against our photometric limits in order to constrain some of the kilonova parameters for GRB\,211106A. Specifically, we built $R$ and $H$ band light curves from models placed at the distance of our photometric redshift, $z$=0.64, as seen from an on-axis observer. All the examined light curves were a function of three fundamental parameters: the mass of the dynamical ejecta $M_{dyn}$, the mass of the disk wind ejecta $M_{dw}$, and the half-opening angle of the lanthanide-rich component $\Phi$, whose ranges are listed in Table \ref{tabulla} for the two classes of models.\\
We examined 81 different configurations of R-KN and 140 of RB-KN, reported in Fig. \ref{bulla211106A}. We found that none out of 81 models of R-KN and 140 models of RB-KN would have been detectable, even in our first FORS2 $R$-band observation.
\begin{comment}
The only 5 R-KN configurations of ejecta that would have been detectable in the zero extinction scenario are the ones with the highest mass budget, with $M_{dyn}\geq0.07$\(M_\odot\) and $M_{dw}\geq0.07$\(M_\odot\), and a total ejecta mass of at least $M_{tot}=M_{dyn}+M_{dw}\geq0.16$\(M_\odot\). Investigating the RB-KN models we found that, as a general trend, the models with the larger contribution of disk wind mass and $\Phi$ in the $30^{\circ}$-$45^{\circ}$ range would have been the closest to detection, given a minimum magnitude difference between the KN and our limit in the first $R$ epoch of $\Delta m\simeq$ 0.13, for $M_{dyn}\geq0.001$\(M_\odot\), $M_{dw}\geq0.13$\(M_\odot\) and $\Phi=30^{\circ}$. For instance, in Fig. \ref{bulla211106A} we show the comparison between our photometric limits and the light curves for the 81 tested R-KN models and the 140 RB ones, with no intrinsic extinction.
\end{comment}
The kilonova at such a luminosity distance would have been very challenging to detect, requiring a fairly large amount of ejecta mass. From our detectability thresholds we were able to draw upper limits on the ejecta masses in the negligible extinction scenario: in the case of R-KN we established $M_{dyn}<0.09$\(M_\odot\) and $M_{dw}<0.09$\(M_\odot\), while for RB-KN we found $M_{dyn}<0.02$\(M_\odot\) and $M_{dw}<0.13$\(M_\odot\). 

\begin{table}
    \caption{Ranges of the values for parameters used to compute light curves in the two subsets of the models from \texttt{POSSIS}.}
    \centering
    \begin{tabular}{cccc}
    \hline
    \noalign{\smallskip}
    Models &$M_{dyn}$ [\(M_\odot\)] & $M_{dw}$[\(M_\odot\)] & $\Phi$ \\
    \noalign{\smallskip}
    \hline
    \hline
    \noalign{\smallskip}
    R-KN& [0.01-0.09] &  [0.01-0.09] & $30^{\circ}$ \\
    RB-KN&[0.001-0.02]&  [0.01-0.13] & [$0^{\circ}$-$90^{\circ}$] \\
    \noalign{\smallskip}
    \hline
    \end{tabular}
    \label{tabulla}
\end{table}
\begin{comment}
\begin{table}
    \centering
        \caption{Ranges of the values for the parameters with which the models were generated.} 
    \begin{tabular}{ccc}
    \hline
    \noalign{\smallskip}
    parameter & R & RB  \\
    \noalign{\smallskip}
    \hline
    \hline
    \noalign{\smallskip}
    $v_{dyn} [c]$& [0.1-0.3] &  [0.08-0.3]  \\
    $v_{wind} [c]$&[0.025-0.1]&  [0.025-0.08] \\
    $\kappa_{lr} [cm^2 g^{-1}]$&1&  1  \\
    $\kappa_{lf} [cm^2 g^{-1}]$&---&  0.005  \\
    $\kappa_{wind} [cm^2 g^{-1}]$&0.1&  0.1 \\
    \noalign{\smallskip}
    \hline    
    \end{tabular}
\tablefoot{Opacities are wavelength and time dependent, values are reported for 1$\mu$m and 1.5d as described in \cite{Anand}.}

    \label{tabullafix}
\end{table} 
\end{comment}

\subsubsection{GRB\,211227A}
\begin{figure}
    \centering
    \includegraphics[width=9cm]{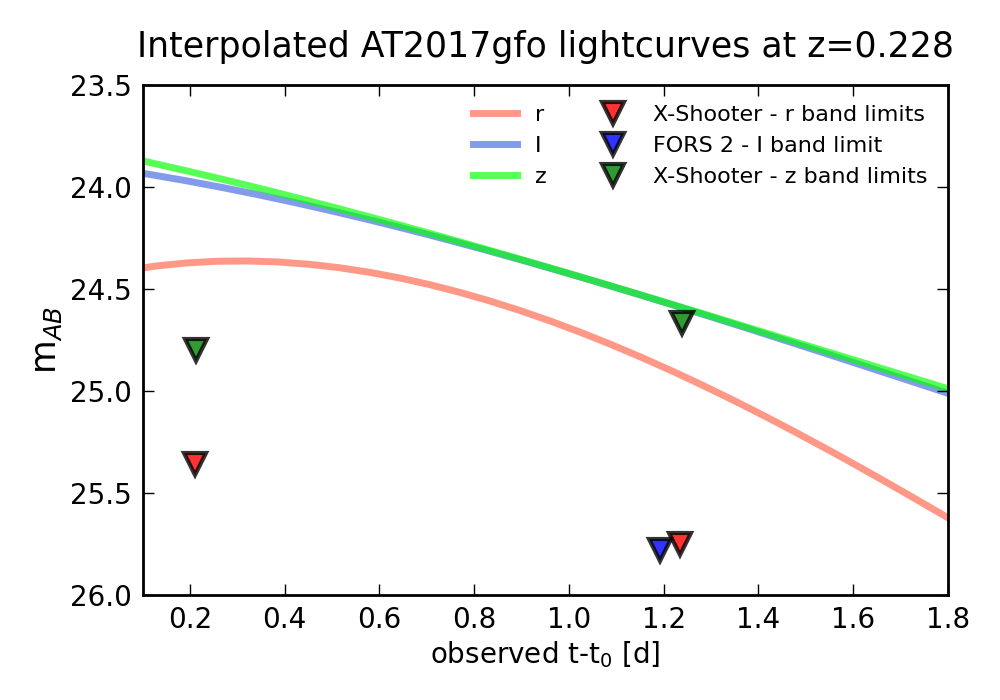}
    \caption{Interpolated magnitudes for X-Shooter $r$, $z$, and FORS2 $I$-band at the redshift of GRB\,211227A, with our observational magnitude limits marked as down-pointing triangles.}
    \label{Kilolc}
\end{figure}

\begin{figure*}
    \centering
    \begin{subfigure}[b]{.49\textwidth}
        \includegraphics[width=1.\textwidth]{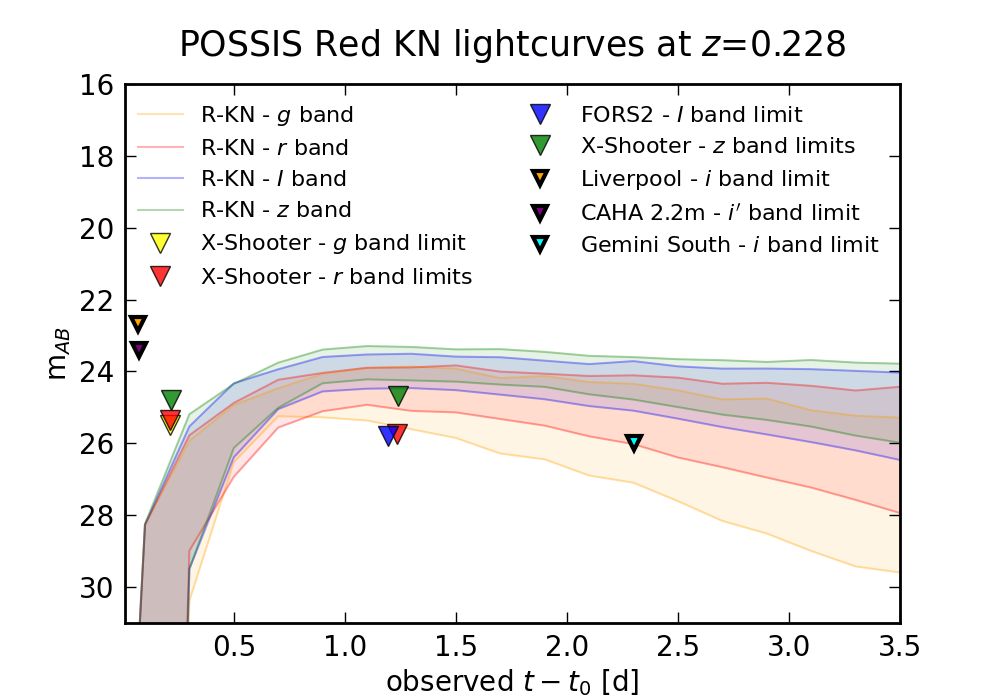}
    \end{subfigure}
    \hfill
    \begin{subfigure}[b]{.49\textwidth}
        \includegraphics[width=1.\textwidth]{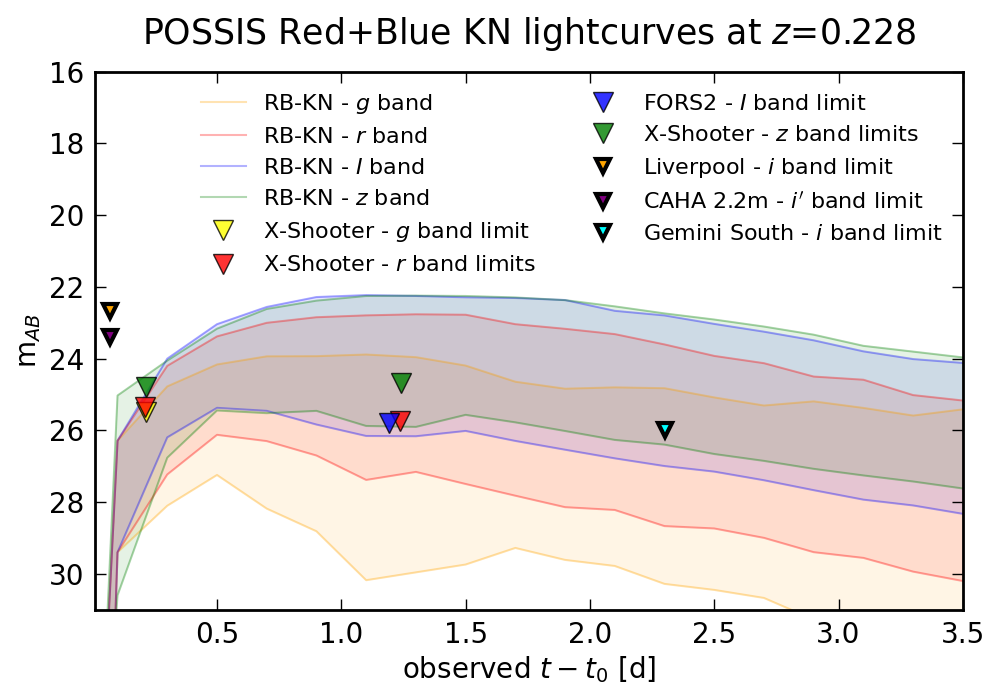}
    \end{subfigure}
    \caption{Simulated light curves of 81 "Red" kilonovae (left panel) and 140 "Red+Blue" kilonovae (right panel) configurations, in the $g$, $r$, $I$, and $z$ bands, as seen for an on-axis observer. During the second night of observation, $\sim$1.2 days after T$_0$, all the tested models would have been above our photometric limits for GRB\,211227A, in the $r$, $z$, and especially $I$ filters. As a comparison we also reported the deepest limits for the optical observations described in \S \ref{GRB211227Aaft}.}
    \label{bulla211227A}
\end{figure*}

We performed the same analysis for GRB\,211227A, without considering intrinsic optical extinction. This is because we found evidence for low $A_V$ from the SED analysis of the host galaxy (\S \ref{sec:211227_HG}). This is in agreement with the observed offset of the XRT afterglow position with respect to the host galaxy center. As shown at the beginning of \S \ref{sec:211227_HG}, the candidate host galaxy of GRB\,211227A has an effective radius of 1.9$''$ (corresponding to 7.26 kpc at $z$=0.228) and the best XRT position places the afterglow at ($3.7\pm2.4)''$ from the centre of the host galaxy, suggesting that the burst occurred outside the densest regions of its host.
In Fig. \ref{Kilolc} we report the light curve computed for the burst placed at $z$ = 0.228, along with our magnitude limits for the X-Shooter $r$ and $z$ observations and the FORS2 $I$ band observation.
\\
With the exception of the second $z$-band observation, for which the magnitude limit is comparable to the interpolated magnitude of the AT2017gfo-like event, in all the other cases the kilonova would have been $\sim$ 0.8-1.2 magnitudes above our detection thresholds. This fact, coupled with the host galaxy association, and the early times of the two observations, suggests that the main reason for the non-detection of the KN emission could be an event with different properties with respect to AT2017gfo, rather than a similar one placed at a farther distance than $z$=0.228, as for GRB\,211106A.\\
We tested models from \texttt{POSSIS}, presented in the previous section, for an on-axis observer. All the configurations are reported in Fig. \ref{bulla211227A} for R-KN and RB-KN. We found that a R-KN would have been detectable in all 81 models during the second $r$ and $z$ epochs, and the single $I$ observation. Especially due to the combination of deep limit and bright counterpart in the $I$ band, even the faintest R-KN would have been $\sim1.6$ magnitudes above the threshold of the FORS2 epoch. For the RB-KN we found that 137 out of 140 configurations would have been detectable at least in the $I$ filter, which also in this case represents the most stringent constraint.\\
We found that the three configurations below our limits would have been the ones with the lowest dynamical mass budget $M_{dyn}=0.001$\(M_\odot\), half opening angle $\Phi$=75$^{\circ}$, and with three different disk wind mass contribution $M_{dw}=0.07, 0.09, 0.13$\(M_\odot\). The angular dependence and the large mass range from the disk wind may suggest that, at the time of our second epoch of observations, the dominant mass contribution was given by the dynamical component. Therefore, we were able to draw upper limits for the dynamical ejecta mass alone. In the case of R-KN we established $M_{dyn}<0.01$\(M_\odot\) while for RB-KN we found $M_{dyn}<0.002$\(M_\odot\), with a preference for large half-opening angle. Given the similarity between GRB\,211211A and GRB\,211227A, it is worth noticing how the former mass threshold is only marginally consistent with what was derived in the KN analysis by \cite{Rastinejad_211211A} in association to GRB\,211211A, while the latter mass limit is not. These differences further underlines how KNe can present a wide range of properties and are far from being standard candles.\\
Nevertheless, in the R-KN scenario such low ejecta masses are still viable from BH-NS mergers. For instance, \cite{METZ} reports 10$^{-4}$\(M_\odot\) as the lower end of typical dynamical ejecta. Even the suppression of KN emission could be possible due to the absence of ejecta when the BH swallows the NS: this is feasible assuming large NS-BH mass ratio, low spinning BH, and stiffer NS equation of state, as discussed in the analysis of the peculiar GW event GW190814 by \cite{Ackleyetal20}. In the RB-KN scenario the situation would be more puzzling: as in the previous case, lower masses are possible for the dynamical component, however, the extended emission of the burst might suggest a magnetar nature for the central engine (\citealt{Gomp}; \citealt{Yang}), that would ensure a further magnetar wind contribution on the ejecta material of the order of $\sim$10$^{-2}$\(M_\odot\) \citep{METZ}, too high to justify the non-detection.

\subsection{Supernova limits}\label{sec:SNdet}
As we did for the kilonova detectability discussion presented in the previous section, we look for possible presence of supernova emission associated with the two bursts. For GRB\,211106A, this is done in order to investigate and possibly rule out a collapsar origin associated with a short event (as in the case of GRB\,200826A, \citealt{Rossietal22}), while for GRB\,211227A because of the long duration of its prompt emission. In the following discussion we consider SN1998bw  (\citealt{Galama98}; \citealt{Mc99}; \citealt{Patat01}; \citealt{Soller02}; \citealt{Cloc11}) and SN2006aj (\citealt{Pian06}; \citealt{Mira06}; \citealt{Ferr06}; \citealt{Soller06}) as references to investigate the supernova emission against our photometric limits. For those few cases where data were not available at the time of our observations, we made use of a back extrapolation from the retrieved lightcurves. 

\subsubsection{GRB\,211106A}
\begin{figure}
    \centering
    \includegraphics[width=9cm]{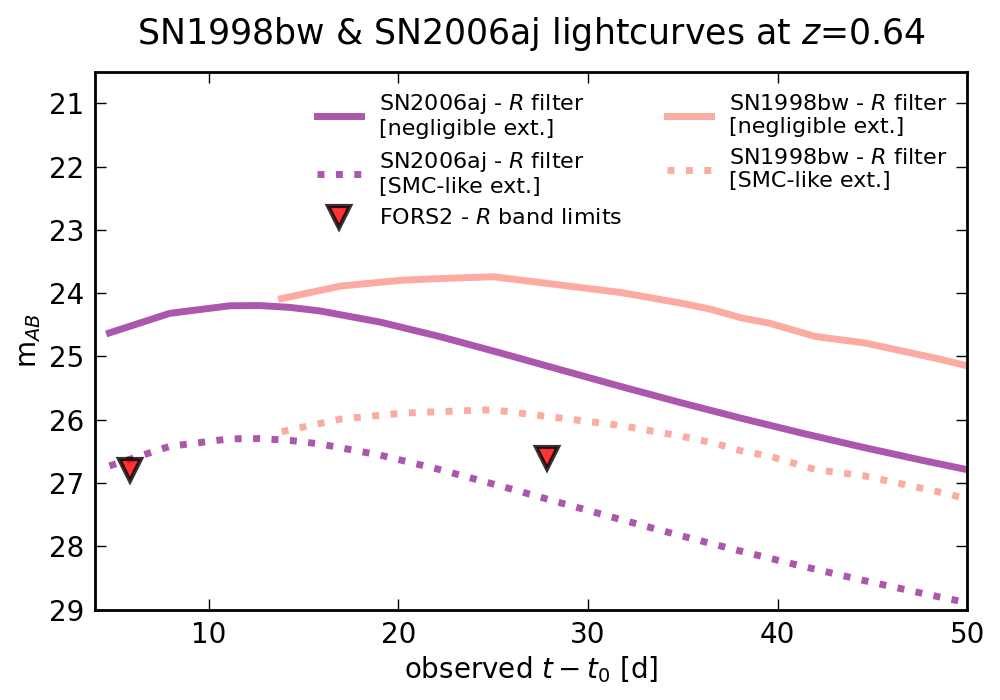}
    \caption{$R$-band lightcurves of SN1998bw and SN2006aj placed at the photometric redshift of GRB\,211106A. The light curves considering both no extinction and $A_V$=2.6 \citep[from][]{Laskar22} are shown.}
    \label{SN211106A}
\end{figure}
We show in Fig. \ref{SN211106A} the light curves of the two supernovae mentioned above at $z$=0.64, along with $R$ magnitude limits computed for the second and third FORS2 epochs, 5.9 and 27.9 days after T$_0$, respectively. We observed that, in case of negligible absorption, the two supernovae would have been significantly brighter than our detection thresholds: SN1998bw would have been $\sim$2.6 mag above our third epoch limit, SN2006aj $\sim$2.2 mag above the second one.
Including the intrinsic extinction $A_R$=2.1 mag, as we did in \S \ref{sec:KN_lim211106A}, SN1998bw would have been still confidently detectable in the third observation, while a SN2006aj-like event would have been more challenging to reveal, even if $\sim$0.1 mag more luminous than the second epoch limit.

\subsubsection{GRB\,211227A}
\begin{figure}
    \centering
    \includegraphics[width=9cm]{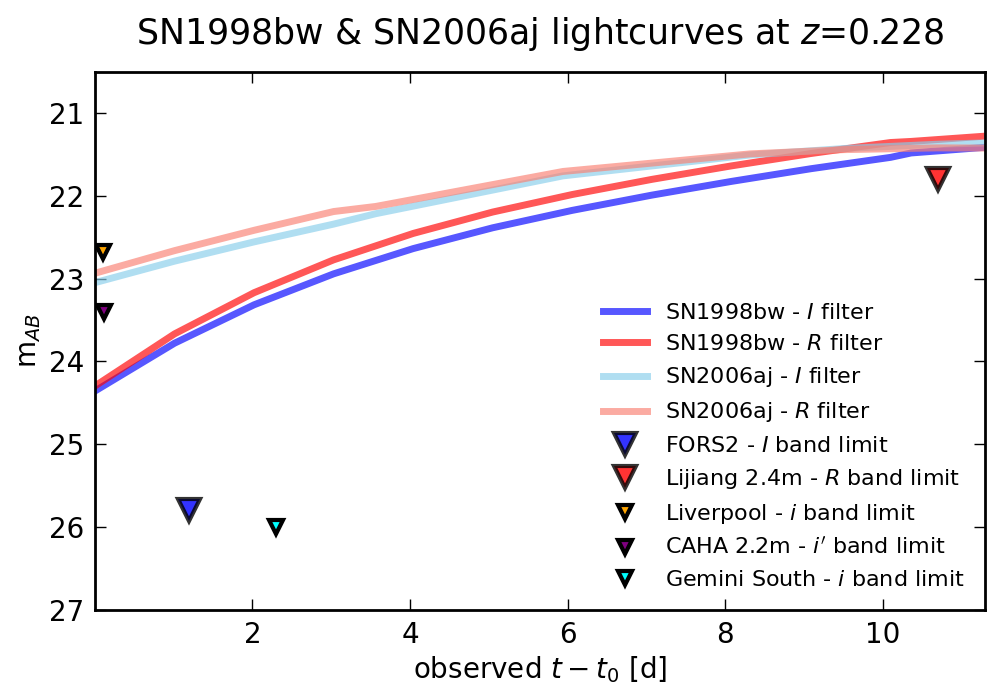}
    \caption{$R$ and $I$ band lightcurves of SN1998bw and SN2006aj placed at the redshift of GRB\,211227A. Along with our FORS2 $I$-band limit the magnitude limit for the $R$-band observation with the Lijiang 2.4m telescope from \cite{Luetal22} is reported. As a comparison we also outlined the deepest limits for the optical observations described in \S \ref{GRB211227Aaft}.}
    \label{SN211227A}
\end{figure}
We also tested the two supernova lightcurves, placed at $z$=0.228, against our photometric limits, to investigate a possible collapsar nature for GRB\,211227A progenitor. \cite{Luetal22} showed that a SN1998bw-like event at $z$=0.228 would have been detected in the $R$-band at $\sim$10.7d after T$_0$, in case of negligible extinction along the line of sight. In order to examine this conclusion we compared SN1998bw and SN2006aj-like $I$ band lightcurves, at $z$=0.228, against our deepest photometric limit, obtained with FORS2 in the $I$ band at $\sim$1.2d after T$_0$.\\
The lightcurves for $I$ and $R$ filter of SN1998bw and SN2006aj are shown in Fig. \ref{SN211227A}. We observed that both supernovae would have been brighter than our $I$ band limit, $\sim$2 mag for SN1998bw and $\sim$3 mag SN2006aj. This result is in agreement with \cite{Luetal22}, pointing towards a compact binary merger progenitor for GRB\,211227A.

\subsection{GRB\,211006A and GRB\,211227A in the context of short GRBs}\label{sec:SBAT4}

In order to deeper investigate the characteristics for the two GRBs discussed in this work, we compared their observed and rest frame properties with those of the short GRB sample named S-BAT4 \citep{Davanzo2014}. The sample consists of a sub-sample of the {\it Swift} short GRBs with favourable observing conditions for redshift determination from the ground, promptly re-pointed by \textit{Swift}/XRT, and which are bright (in terms of the observed peak flux\footnote{This flux-limited sample selects short GRBs having the peak photon flux $P\ge 3.5$ ph s$^{-1}$ cm$^{-2}$ using the $15-150$ keV {\it Swift}/BAT light curves binned with $\delta t = 64$ ms. This corresponds to an instrument that is $\sim 4$ times less sensitive than {\it Swift}.}) in the $15-150$ keV {\it Swift}/BAT energy band. Although relatively small (16 events up to June 2013), this sample of short GRBs is complete in flux and has the highest completeness in redshift (70\%) with respect to the short GRB samples presented in the literature to date. Following the same criteria adopted for the original sample, we extended it by adding 26 new SGRBs detected from July 2013 to December 2021, 24 of which with redshift determination (15 spectroscopic and 8 photometric redshifts, while one spectroscopic redshift and one photometric redshift were retrieved also for GRB\,090515A and GRB\,130515A of the original sample, respectively). The sample built this way consists of 42 SGRBs, 37 of them with redshift measurement, corresponding to a redshift completeness of $\sim$88\%, which reduces to 67\% if we consider only events with spectroscopic redshift measurement. 8 SGRBs with extended emission (EE) are also included, considering also GRB\,211127A, whose observed properties make it belong to the S-BAT4 sample. Given the lack of an on-board \textit{Swift}/BAT trigger, GRB\,211106A is not part of the S-BAT4. The detailed overview of the extended S-BAT4 sample will be presented in a forthcoming dedicated paper.

\subsubsection{Prompt emission correlations}\label{sec:ELcorr}

\begin{figure}[]
    \centering
    \includegraphics[width=9cm]{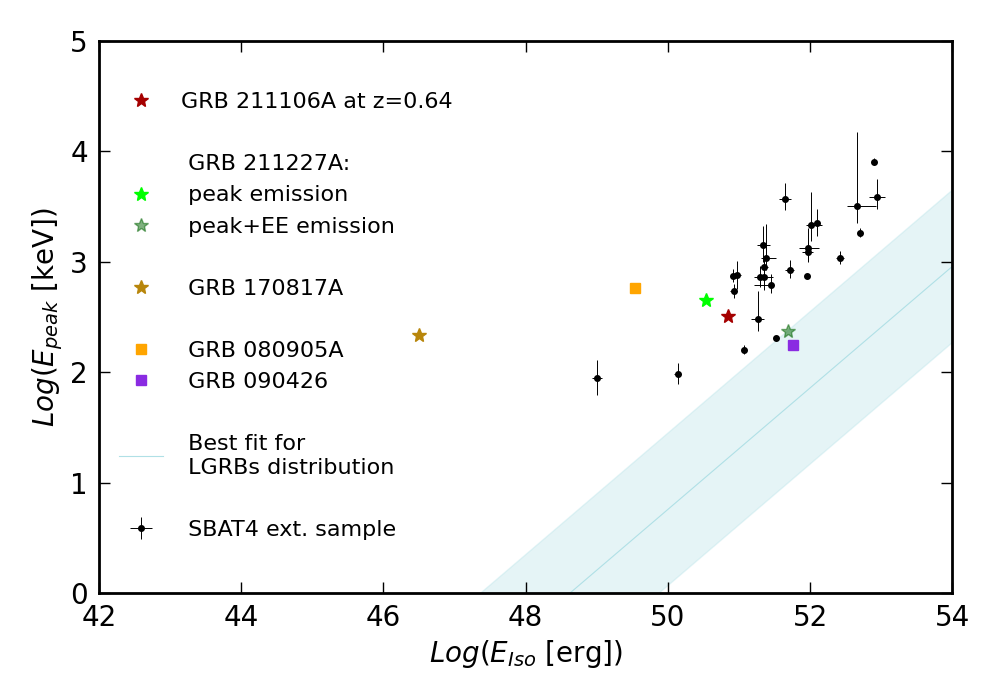}
    \centering
    \caption{Amati correlation for long GRBs. Shaded region around best fit lines highlights the 3$\sigma$ scatter region of the long GRBs distribution (values from \citealt{NAVA2012}). SGRBs are represented by black dots. GRB\,211106A prompt quantities are shown for $z$=0.64. GRB\,170817A is included for comparison \citep[values from][]{GOLDSTEIN17}.}
    \label{CORR}
\end{figure}

\begin{figure}
    \centering
    \includegraphics[width=9cm]{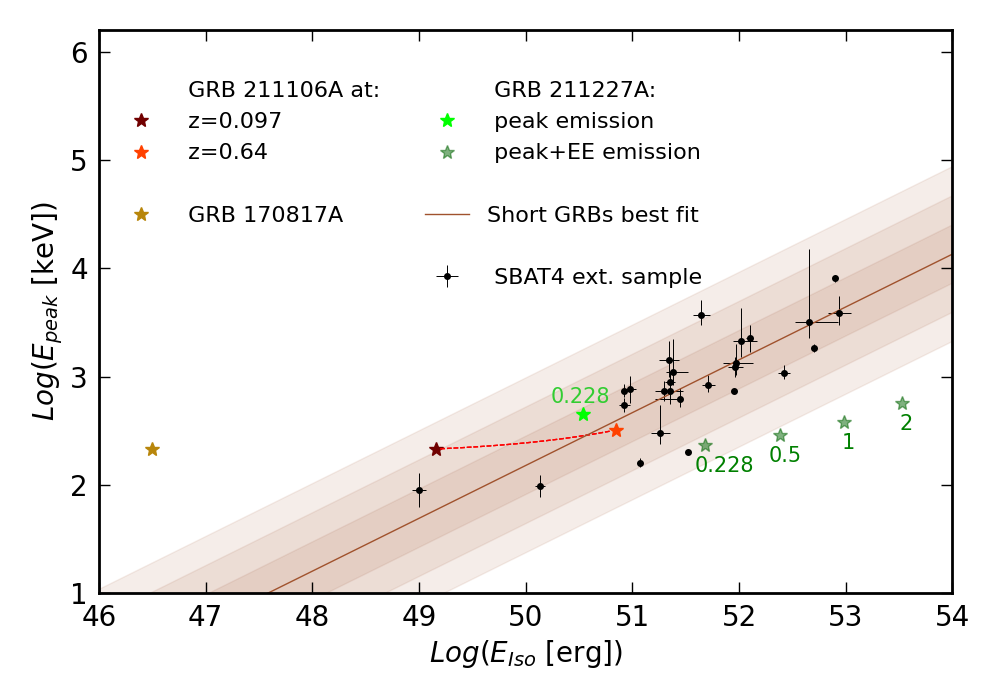}
    \caption{Amati relation for the S-BAT4 extended sample. SGRBs are represented by black dots, shaded regions represent the 1, 2, and 3 $\sigma$ scatter of the sample around the best fit. GRB\,211106A becomes more consistent with the best fit at increasing values of redshift. GRB\,211227A is in agreement for both the main and the whole emission with the correlation for the redshift of $z$=0.228. As a reference we reported also the prompt energy features for $z$ = 0.5, 1, and 2 for the whole emission phase.}
    \label{CORR2}
\end{figure}

For the GRBs of the extended S-BAT4 sample we retrieved all the needed information to characterize the prompt phase from the literature and/or GCN circulars\footnote{\url{https://gcn.gsfc.nasa.gov/gcn3_archive.html}}. For the bursts with a measured redshift we computed the isotropic equivalent energy ($E_{iso}$) - in the rest frame common energy range 1 keV - 10 MeV -  and the rest frame peak energy ($E_{peak}$).  The $E_{iso}$ was computed converting all the retrieved fluence values from the original energy band to a common (in the rest frame) band by means of the following relation:
    \begin{eqnarray}
    \centering
    F^{rf}(\text{1 keV-10 MeV})=F^{obs}(\text{$low$-$up$}) \dfrac{\int_{1\text{ keV/}(1+z)}^{10\text{ MeV/}(1+z)}EN(E)dE}{\int_{low}^{up}EN(E)dE},
    \label{flu}
    \end{eqnarray}
where $N(E)$ is either a Band function (\citealt{Band}) or a Cut-off power law for the majority of the bursts considered, $z$ is the redshift of the event, and $low$-$up$ is the energy range in which the original fluence $F^{obs}$ was measured.
For some of the new bursts added to the S-BAT4 with a redshift measurement, observed only by \textit{Swift}/BAT, we were not able to infer the rest frame energy properties as the spectrum was best fitted by a single power law.
For the original sample, we updated the prompt emission spectral quantities for GRB\,111117A, for which a spectroscopic redshift $z$ = 2.211 was determined \citep{GRB111117A_z} in place of the previous photometric one, $z$ = 1.3 reported in \cite{Davanzo2014}.
An extensive analysis on the results is beyond the scope of this work and will be discussed in future papers. 
For GRB\,211106A, assuming the photometric redshift, $z$=0.64, we computed the rest frame prompt properties of the burst: $E_{iso}=7.0^{+1.6}_{-1.4}\times$10$^{50}$ erg and $E_{peak}=321^{+128}_{-74}$ keV from the spectral parameters obtained by Konus-\textit{Wind} \citep{KONUS_GCN}. For GRB\,211227A, which belongs to the S-BAT4 sample, we likewise made use of the spectral parameters obtained by Konus-\textit{Wind} \citep{zhu22}. We considered the values reported for both the main emission, lasting $\sim$4s, and the long rebrightening phase seen up to $\sim$84s. From the analysis of the peak we computed $E_{iso}=3.4^{+0.3}_{-0.7}\times$10$^{49}$ erg and $E_{peak}=449^{+1475}_{-220}$ keV. If we also consider the extended emission, the rest frame prompt properties become: $E_{iso}=(4.8 \pm 0.4)\times$10$^{51}$ erg and $E_{peak}=236^{+55}_{-52}$ keV.\\
In Fig. \ref{CORR} we reported the S-BAT4 sample in the $E_{peak} - E_{iso}$ plane. SGRBs lie almost entirely - with the exception of GRB\,090426 - outside the 3$\sigma$ region in the $E_{peak} - E_{iso}$ correlation found for long GRBs \citep{AMATI_RELATION}, suggesting for a separate short GRB correlation with a slope similar to the long GRB relation with a different normalisation. A similar result was presented and discussed in \cite{Davanzo2014}, implying that the distribution of GRBs over the $E_{peak} - E_{iso}$ plane, which clearly shows two different regions, can be used as a discriminant for the 
 GRB progenitor type (namely compact binary merger or collapsar).\\ 
We performed a fit to the SGRBs of the extended S-BAT4 sample\footnote{We excluded GRB\,080905A and GRB\,090426, as discussed in \cite{Davanzo2014}, as well as GRBs with only upper limits available. GRB\,211227A was also excluded in order to test its properties with respect to the sample.} in the $E_{peak}-E_{iso}$ plane with the relation \(y = 10^Ax^B\), following the \textit{OLS bisector} method presented in \cite{ISOBE}. The best bisector fit provided a normalization A $=-22.2\pm3.5$ and a slope B $=0.5\pm0.1$, in agreement with the findings of \cite{Davanzo2014}. 
GRB\,211106A, if at $z$=0.64, would be consistent with the best fit of the relation found for SGRBs (Fig. \ref{CORR2}), and, in general, needs to be at $z\gtrsim0.2$ and $z\gtrsim0.1$ to be consistent at 1$\sigma$ and 2$\sigma$ level, respectively. This, together with the discussion in \S \ref{sec:211106_HG} and \S \ref{sec:KN_lim}, rules out an host galaxy at $z$=0.097, that would be consistent only at 3$\sigma$ level with the SGRBs Amati relation.
GRB\,211227A prompt energy quantities are consistent with the rest of the sample in the Amati plane, if we consider the whole burst emission at $z$=0.228, at 3$\sigma$ level.
Given the computed $P_{cc}=1.3\%$, we tested the burst also at increasing values of the redshift (as done by \citealt{GRB211227A_GCN_KONUS}) and we verified that the burst would behave as a long GRB in terms of prompt features, leaving the short Amati relation 3$\sigma$ region for $z \gtrsim$ 0.4.
However, we lack a SN detection, the host galaxy properties are consistent with typical values of SGRBs hosts (\S \ref{sec:211227_HG}), and the shape of the prompt emission lightcurve is clearly distinguished between the initial pulse and a softer long lasting emission (Fig. \ref{GRB211227_BAT}). If we analyse the burst in the context of SGRBs with extended emission, considering the prompt energy quantities of the initial pulse alone, it is in good agreement with the best short Amati relation fit at $1\sigma$ level, at $z$=0.228. This fact, coupled with the spectral analysis, reflects the progenitor scenario from compact object merger, despite the long T$_{90}$.

\begin{comment}
Focusing on GRB\,211106A, we computed the isotropic equivalent quantities and the rest frame peak energy for $z$=0.097: $E_{iso}= 1.40^{+0.33}_{-0.28}\times$10$^{49}$ erg and $E_{peak} = 215_{-49}^{+86}$ keV.
In Fig. \ref{CORR}, we have also included GRB\,170817A, for which the combined detection with the gravitational wave event GW\,170817 \citep{GW170817} has guaranteed a straightaway and robust observation of the burst. $E_{peak}$ was derived by \cite{GOLDSTEIN17} from the best fit for the initial hard spike of the burst, while \cite{SALAFIA} computed $E_{iso}$.
\end{comment}
    
\subsubsection{Rest frame X-ray afterglows}

\begin{figure}[]
\centering
\includegraphics[width=9cm]{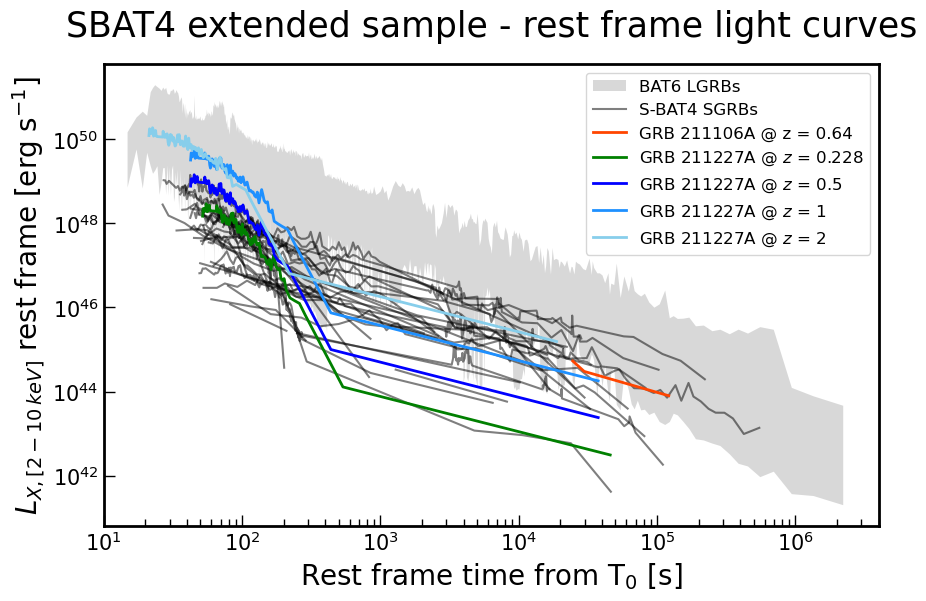}
\caption{Rest frame X-ray light curves in common energy range (2-10) keV for all SGRBs in the S-BAT4 extended sample (black lines). The same curves for GRB\,211106A and GRB\,211227A are highlighted with coloured solid lines. Long GRBs belonging to the BAT6 sample \citep{BAT6} are shown for comparison and they correspond to the grey-shaded area. The consistency of these curves with the S-BAT4 extended sample strengthens their classification as short GRBs.}
\label{LX_rf}
\end{figure}

\begin{comment}
\begin{figure}[]
\centering
\includegraphics[width=9cm]{img/LX_rf_lc_longV2.png}
\caption{}
\end{figure}   
\end{comment}

As a final gauge on GRB\,211106A distance and GRB\,211227A nature, we analyzed their X-ray afterglow emission compared to the S-BAT4 SGRBs. We computed the rest frame X-ray light curves for all SGRBs with redshift, as follows: first, we retrieved from the \textit{Swift} Burst Analyser\footnote{\url{http://www.swift.ac.uk/burst_analyser/}} \citep{Evans09} the observed 0.3–10 keV unabsorbed fluxes and the time-resolved measured photon spectral index, $\Gamma$, for each GRB of the sample. We then put the fluxes in the common rest frame (2-10 keV) energy band by means of the following relation (as done by \citealt{Gehrels08}):
    \begin{eqnarray}
    \centering
    f_{X}^{ rf}(\text{2-10 keV})=f_{X}(\text{0.3-10 keV})\dfrac{\bigg(\dfrac{10}{1+z}\bigg)^{2-\Gamma}-\bigg(\dfrac{2}{1+z}\bigg)^{2-\Gamma}}{10^{2-\Gamma}-0.3^{2-\Gamma}}.
    \label{fx11}
    \end{eqnarray}\\
We show the curves we obtained for each GRB in Fig. \ref{LX_rf}, along with the one obtained for GRB\,211106A. This is is consistent with the typical luminosity of the sample if at $z$=0.64. This further confirms the estimate for the distance of the event obtained through the photometric redshift. Moreover, the rest frame X-ray light curve of GRB\,211227A is consistent with the typical values of the other SGRBs, if at $z$=0.228. However, if we put it at larger redshift (0.5, 1, 2), it is still consistent with the sample of short GRBs. Since the LGRBs fluxes are typically more luminous than SGRBs, this represents another test on the nature of the burst, from which we can safely classify GRB\,211227A as a short GRBs with extended emission (and therefore originated by a compact binary merger progenitor),
as also confirmed by its consistency with the short Amati relation in the previous section and from the galaxy features discussed in \S \ref{sec:211227_HG}.

\section{Conclusions}\label{conclusions}
GRB\,211106A and GRB\,211227A are two events that belong to the short GRB class that, right after their discovery, were proposed as possibly associated to bright, low-redshift host galaxies. To further investigate the properties of these two events and of their host galaxies, we performed multi-epoch and multi-band analysis of ground-based optical and near-infrared observations obtained with the ESO-VLT FORS2, HAWK-I, MUSE, and X-Shooter instruments. At the same time, we carried out an analysis of the available \textit{Swift}/XRT data for both bursts. We also carried out spectroscopic analysis of the GRB\,211106A and GRB\,211227A host galaxy candidates with the ESO-VLT MUSE and X-Shooter instruments, obtaining a redshift estimate of the former and studying in detail the properties of the latter. Finally, we investigated the kilonova detectability thresholds for both events and tested the GRBs properties by comparing them to those of a sample of short GRBs, namely the S-BAT4 sample \citep{Davanzo2014}, here extended up to December 2021.\\
From our analysis we obtained the following results:\\
\begin{itemize}
    \item We excluded an optical and a NIR counterpart for GRB\,211106A, from FORS2 and HAWK-I observations at 3$\sigma$ confidence level, down to $R \sim$ 26.7 mag at $t-t_0 \sim$ 2.9 d and $H \sim$ 23.6 mag at $t-t_0 \sim$ 4.8 d, respectively;
    
    \item By monitoring the GRB\,211106A field over more than 3 weeks, at the position of the radio/mm afterglow we detected an optical source with constant magnitude $R \sim 26.5$ mag from FORS2 observations. Its chance association probability with the burst is $P_{cc} \sim$ 0.13\%, so we consider it as best-candidate host galaxy of GRB\,211106A. We performed a SED fit, from which we derived a photometric redshift $z$ = 0.64;
    
    \item We investigated the possible presence of a kilonova associated with GRB\,211106A: we did not detect any source, as we would expect both in case of an AT2017gfo-like event and of a kilonova with different features. This is consistent with a host galaxy at $z$ = 0.64;
    
    \item We explored the possible presence of a supernova event at $z$ = 0.64 associated with GRB\,211106A: we found that, even in the case of significant extinction, SN1998bw-like and SN2006aj-like events should have been brighter than our limits, endorsing a compact binary merger progenitor scenario for the burst;
    
    \item We ruled out the presence of optical and NIR emission from GRB\,211227A from our FORS2 and X-Shooter images, placing 3$\sigma$ limits at $r \sim$ 25.4 mag, $z \sim$ 24.8 mag, $g \sim$ 25.4 mag  at $t-t_0 \sim$ 0.2 d, and $I \sim$ 25.8 mag at  at $t-t_0 \sim$ 1.2 d;

    \item We performed a spectroscopic analysis of a galaxy 3.7$''$ away from the GRB\,211227A best X-rays position, for which we derived $z$ = 0.228 and $P_{cc} \sim$ 1.13\%, making it the best candidate host galaxy. We investigated its properties in terms of metallicity, star forming rate, mass, and they all turned out to be consistent with typical values for short GRB hosts \citep{Berger14}, and significantly different from long GRB ones (see \S \ref{sec:211227_HG});

    \item An AT2017gfo-like kilonova should have been detected in our images, and our non-detection can be ascribed to a very faint emission. This may be produced either by a binary neutron star with low-mass dynamical ejecta or by a BH-NS binary with large mass inequality, low spin for the BH, and a stiff equation of state for the NS;
    
    \item We were able to exclude a SN presence at $z$ = 0.228 associated with GRB\,211227A thanks to the deep optical limits at $\sim$ 1.2 d, supporting the result of \cite{Luetal22};

    \item For both GRB\,211106A and GRB\,211227A it was not possible to detect an optical counterpart despite deep limits. They both fulfill the condition for their classification as dark bursts. This results can be ascribed for the former to a large intrinsic absorption, while for the latter to a low-density circumburst medium;
    
    \item Both GRB\,211106A and GRB\,211227A prompt-phase spectral properties, namely the peak energy $E_{peak}$ and the isotropic-equivalent energy $E_{iso}$, are consistent with short GRBs belonging to the S-BAT4 extended sample. In particular, they are significantly outside the 3$\sigma$ region in the $E_{peak} - E_{iso}$ correlation found for long GRBs \citep{AMATI_RELATION}, while they are consistent at 1$\sigma$ c.l. with the short Amati relation (\S \ref{sec:SBAT4});

    \item We placed all S-BAT4 SGRBs X-ray light curves in the common rest frame (2-10 keV) energy band, and the curves of the two GRBs are fully consistent with them, even if we consider a larger $z$ for GRB\,211227A, significantly distinguishing it from LGRBs.
\end{itemize}
The conclusions we obtained from the analysis of GRB\,211106A and GRB\,211227A and their candidate host galaxies, taken altogether, confirm the classification of the two events as short GRBs, namely originated by a compact binary merger progenitor. In particular, GRB\,211227A can be considered as a short GRB with extended emission, despite its T$_{90} >$ 2 s. Besides, the consistency of the prompt-phase spectral properties with the S-BAT4 sample and the X-ray rest frame light curves strengthens the association of the two bursts with the host galaxies at $z$ = 0.64 and $z$ = 0.228, respectively.\\
This work highlights the importance of performing deep and possibly rapid multiwavelength follow-up of single GRB events in order to avoid misleading host galaxy association and to constrain the GRB classification beyond the prompt emission duration.\\
On the other hand, the growing number of SGRBs detected will allow us to perform accurate sample studies, which can be used as tools to explore SGRBs properties in detail, as done in this work for GRB\,211106A and GRB\,211227A. Regarding this topic, a full discussion of the properties of the S-BAT4 sample will be presented in a forthcoming paper.

\begin{acknowledgements}
Based on observations collected at the European Organisation for Astronomical Research in the Southern Hemisphere under ESO programmes 106.21T6.015, 106.21T6.016, 106.21T6.019. The authors thank B. Sbarufatti, O.S. Salafia and G. Tagliaferri for useful suggestions and discussion. This work made use of data supplied by the UK Swift Science Data Centre at the University of Leicester. PDA, MF, MGB, SC, TS, CS acknowledge funding from the Italian Space Agency, contract ASI/INAF n. I/004/11/4”. PDA acknowledge support from PRIN- MIUR 2017 (grant 20179ZF5KS). AJL and DBM are supported by the European Research Council (ERC) under the European Union’s Horizon 2020 research and innovation programme (grant agreement No.~725246). The Cosmic Dawn Center is supported by the Danish National Research Foundation.
\end{acknowledgements}

\bibliographystyle{aa}
\bibliography{biblio}

\end{document}